\pdfoutput=1
\documentclass[fleqn,usenatbib]{mnras}
\usepackage[T1]{fontenc}
\usepackage{ae,aecompl}
\usepackage{graphicx}	
\usepackage{amsmath}	
\usepackage{amssymb}
\usepackage{xcolor}
\usepackage{pdflscape}
\usepackage{siunitx}
\def\lya{Ly$\alpha$~}
\def\HI{\hbox{H~$\scriptstyle\rm I$}}

\def\nHI{{\rm HI}}

\def\HeII{\hbox{He~$\scriptstyle\rm II$}}

\usepackage{dcolumn}
\newcolumntype{.}{D{.}{.}{2.7}}
\newcolumntype{e}{D{.}{.}{2.10}}
\newcolumntype{d}{D{.}{.}{2.3}}

\title[AGN luminosity function]{Evolution of the AGN UV
  luminosity function from redshift 7.5}

\author[Kulkarni et al.]
       {{Girish Kulkarni$^{1,2,3}$\thanks{Email: kulkarni@theory.tifr.res.in},
           G\'abor Worseck$^{4,6}$
           and Joseph F.~Hennawi$^{5,6}$} \\
         $^1$Institute of Astronomy,
         University of Cambridge, Madingley Road, Cambridge CB3 0HA,
         UK \\
         $^2$Kavli Institute of Cosmology,
         University of Cambridge, Madingley Road, Cambridge CB3 0HA,
         UK \\
         $^3$Department of Theoretical Physics,
         Tata Institute of Fundamental Research,
         Homi Bhabha Road, Mumbai 400005, India\\
         $^4$Institut f\"ur Physik und Astronomie, Universit\"at
         Potsdam, Karl-Liebknecht-Stra\ss e\ 24/25, D-14476 Potsdam,
         Germany \\
         $^5$Department of Physics, Broida Hall, UC Santa Barbara,
         Santa Barbara, CA 93106-9530, USA\\
         $^6$Max Planck Institute for Astronomy, K\"onigstuhl 17,
         D-69117 Heidelberg, Germany}
       \date{Accepted ---. Received ---; in original form ---}

\pubyear{2019}

\begin{document}
\label{firstpage}
\pagerange{\pageref{firstpage}--\pageref{lastpage}}
\maketitle

\begin{abstract}
  Determinations of the UV luminosity function of AGN at high
  redshifts are important for constraining the AGN contribution to
  reionization and understanding the growth of supermassive black
  holes.  Recent inferences of the luminosity function suffer from
  inconsistencies arising from inhomogeneous selection and analysis of
  data.  We address this problem by constructing a sample of more than
  80,000 colour-selected AGN from redshift $z=0$ to $7.5$ using
  multiple data sets homogenised to identical cosmologies, intrinsic
  AGN spectra, and magnitude systems.  Using this sample, we derive
  the AGN UV luminosity function from redshift $z=0$ to $7.5$.  The
  luminosity function has a double power law form at all redshifts.
  The break magnitude $M_*$ shows a steep brightening from $M_*\sim
  -24$ at $z=0.7$ to $M_*\sim -29$ at $z=6$.  The faint-end slope
  $\beta$ significantly steepens from $-1.7$ at $z<2.2$ to $-2.4$ at
  $z\simeq 6$.  In spite of this steepening, the contribution of AGN
  to the hydrogen photoionization rate at $z\sim 6$ is subdominant
  ($<3$\%), although it can be non-negligible ($\sim 10$\%) if these
  luminosity functions hold down to $M_{1450}=-18$.  Under reasonable
  assumptions, AGN can reionize \HeII\ by redshift $z=2.9$.  At low
  redshifts ($z<0.5$), AGN can produce about half of the hydrogen
  photoionization rate inferred from the statistics of \HI\ absorption
  lines in the IGM.  Our analysis also reveals important systematic
  errors in the data, which need to be addressed and incorporated in
  the AGN selection function in future in order to improve our
  results.  We make various fitting functions, codes, and data
  publicly available.
\end{abstract}


\section{Introduction}

The luminosity function of active galactic nuclei (AGN) and its
evolution over cosmological time scales has been a matter of central
interest of a large body of work over the last five decades
\citep{1968ApJ...151..393S, 1978A&A....68...17M, 1983ApJ...269..352S,
  1988ApJ...325...92K, 1988MNRAS.235..935B, 1993ApJ...406L..43H,
  1994ApJ...421..412W, 1995AJ....110...68S, 1995AJ....110.2553K,
  1995ApJ...438..623P, 2000MNRAS.317.1014B, 2001AJ....121...54F,
  2004AJ....128..515F, 2006AJ....131.2766R, 2007ApJ...654..731H,
  2009MNRAS.392...19C, 2010AJ....139..906W, 2011ApJ...728L..26G,
  2013ApJ...773...14R, 2013ApJ...768..105M, 2015AA...578A..83G,
  2015ApJ...798...28K, 2016ApJ...829...33Y, 2016ApJ...833..222J}.
Determination of the AGN luminosity function constrains models of the
build-up of supermassive black holes \citep{1982MNRAS.200..115S,
  2004ApJ...602..603Y, 2007ApJ...654..731H, 2008ApJ...679..118S,
  2009A&A...493...55E, 2000ApJ...531...42H, 2010MNRAS.401.2531A,
  2013ApJ...773...14R, 2014ApJ...787...73D, 2015MNRAS.448.3603D,
  2015MNRAS.452..575S, 2016MNRAS.462..190R}.  Due to the incidence of
supermassive black holes in most galaxies, the tight scaling relations
observed to exist between the mass of these black holes and properties
of their host galaxies, and the increasing consensus that AGN activity
feeds back on the host galaxy evolution, the AGN luminosity function
also constrains models of galaxy formation \citep{2006ApJ...650...42L,
  2008MNRAS.385.1846M}.  Finally, thanks to their brightness and high
Lyman-continuum (LyC) escape fractions, the luminosity function of AGN
determines their contribution to the ultraviolet background that
governs the temperature and ionization state of the intergalactic
medium (IGM), possibly even primarily driving hydrogen and helium
reionization 
\citep{2012ApJ...746..125H, 2015AA...578A..83G, 2017ApJ...847L..15O,
  2018MNRAS.474.2904P, 2019MNRAS.485...47P}.

There is a renewed interest in understanding the evolution of the UV
luminosity function of AGN, triggered by the discovery of 19
low-luminosity ($-18.9>M_{1450}>-22.6$) AGN candidates between
redshifts $z=4.1$ and $6.3$ by \citet{2015AA...578A..83G} using a
novel X-ray/NIR selection criterion.  This finding suggested that AGN
brighter than $M_{1450}\sim -18$ can account for all of the
metagalactic hydrogen photoionization rate inferred from the \lya
forest at $4<z<6$, despite the fact that these AGN are fainter than
the brightest star-forming galaxies at these redshifts.  (The UV
luminosity function of galaxies at $z=6$ has characteristic magnitude
$M^*_{1450}\sim -20$ \citep{2015ApJ...803...34B}.)  A significant
presence of AGN at high redshift ($z\sim 6$) and a dominant
contribution of AGN to reionization is appealing as the LyC escape
fraction of galaxies is uncertain, whereas for AGN it is close to
100\% \citep{2014ApJ...794...75S, 2015MNRAS.449.4204L,
  2016A&A...585A..48G}.  High-redshift galaxies down to rest-frame UV
magnitude $M_\mathrm{UV}=-12.5$ at $z=6$ \citep{2017ApJ...835..113L}
and redshifts up to $z=11.1$ \citep{2016ApJ...819..129O} have now been
reported.  But the escape of LyC photons has been measured in only a
few comparatively bright ($L>0.5L^*_\mathrm{galaxies}$) galaxies at
relatively low redshifts ($z < 4$).  The escape fraction in these
galaxies reveals a broad distribution from less than 2\% to more than
80\% \citep{2010ApJ...725.1011V, 2014Sci...346..216B,
  2015ApJ...810..107M, 2016A&A...585A..48G, 2016Natur.529..178I,
  2016ApJ...826L..24S, 2017MNRAS.468..389J, 2018ApJ...869..123S,
  2018arXiv180601741F, 2018A&A...619A.136M, 2018A&A...616A..30C,
  2018MNRAS.478.4851I, 2018MNRAS.474.4514I}, but the average escape
fraction is typically lower than 20\%.  Statistical constraints from
\HI\ column density measurements in gamma-ray burst (GRB) afterglow
spectra suggest an even lower escape fraction of 0.5\%
\citep{2007ApJ...667L.125C, 2009ApJS..185..526F, 2019MNRAS.483.5380T}.
This is a challenge for reionization models, which require an escape
fraction of about 20\% in galaxies as faint as $M_\mathrm{UV}=-13$
\citep{2016PASA...33...37F, 2015ApJ...802L..19R, 2016MNRAS.457.4051K}.
An enhanced incidence of high-redshift AGN may also be consistent with
the shallow bright-end slopes of the $z\sim 7$ UV luminosity function
of galaxies relative to a Schechter function
\citep{2012MNRAS.426.2772B, 2014MNRAS.440.2810B, 2014ApJ...792...76B,
  2015MNRAS.452.1817B} and the hard spectra of these galaxies
\citep{2015MNRAS.450.1846S, 2015MNRAS.454.1393S, 2017MNRAS.464..469S,
  2017ApJ...851...40L}.  Finally, AGN may also provide a natural
explanation for the large scatter in the \lya opacity between
different quasar sightlines close to redshift $z=6$
\citep{2015MNRAS.447.3402B, 2018MNRAS.479.1055B, 2018ApJ...864...53E}.
These opacity fluctuations extend to substantially larger scales
($\gtrsim 50\, h^{-1}$cMpc) than expected in galaxy-dominated
reionization models (\citealt{2015MNRAS.453.2943C}; although see
\citealt{2016MNRAS.460.1328D, 2015ApJ...813L..38D,
  2018ApJ...863...92B}).  AGN clustering can result in these
fluctuations naturally if there is a significant contribution
($\gtrsim 50\%$) of AGN to the ionising emissivity at $z=5$--$6$
\citep{2017MNRAS.465.3429C}.

Several counter-arguments against a higher incidence of AGN at high
redshift have also been made. \citet{2017MNRAS.468.4691D} and
\citet{2018MNRAS.473.1416M} pointed out that an enhanced number
density of AGN at $z\sim 6$ can lead to a \HeII\ reionization that
occurs much earlier than indicated by the observed evolution of
\HeII\ IGM opacity \citep{2011ApJ...733L..24W, 2016ApJ...825..144W}
and the evolution of the IGM temperature \citep{2011MNRAS.410.1096B,
  2014MNRAS.441.1916B} (see also \citealt{2017MNRAS.471..255K}).  For
instance, He~\textsc{ii} reionization is complete at $z=4.5$ in the
model of \citet{2015ApJ...813L...8M}, compared to $z=3$ in the
conventional scenario with sharply dropping AGN emissivity at $z>2$
\citep{2012ApJ...746..125H}.  Such early \HeII\ reionization will also
result in higher IGM temperatures due to the associated photoheating.
\citet{2017MNRAS.468.4691D} found that the temperature of the IGM at
mean density is twice as much in AGN-dominated reionization models as
the standard models at $z=3.5$--$5$, in conflict with constraints from
the \lya forest.  This inconsistency could be avoided by postulating a
reduced escape fraction of \HeII-ionizing photons in AGN, but it is
difficult to reconcile this with a unit escape fraction of
hydrogen-ionizing photons that is required to explain the Ly$\alpha$
opacity fluctuations.  Further arguments against AGN-dominated
reionization have been presented by \citet{2016MNRAS.459.2299F}, who
analysed the simulations of metal-line absorbers at $z\sim 6$.
\citet{2016MNRAS.459.2299F} find that in their cosmological radiation
hydrodynamical simulations AGN-dominated UV background results in too
many C~\textsc{iv} absorption systems relative to Si~\textsc{iv} and
C~\textsc{ii} at $z\sim 6$.  Finally, comparing the
\citet{2015AA...578A..83G} sample to X-ray-selected quasar data at
$z=0$--$6$, \citet{2017MNRAS.465.1915R} argued that the faint end of
the AGN UV luminosity function at $z\sim 6$ is probably shallower that
reported by \citet{2015AA...578A..83G}.  \citet{2017MNRAS.465.1915R}
argue that the apparent contradiction with the results of
\citet{2015AA...578A..83G} could be explained by contamination from
the host galaxies for faint AGN \citep[see
  also][]{2015MNRAS.453.1946G,2015MNRAS.448.3167W,2016MNRAS.463..348V}.

\begin{figure*}
  \begin{center}
    \includegraphics[width=\textwidth]{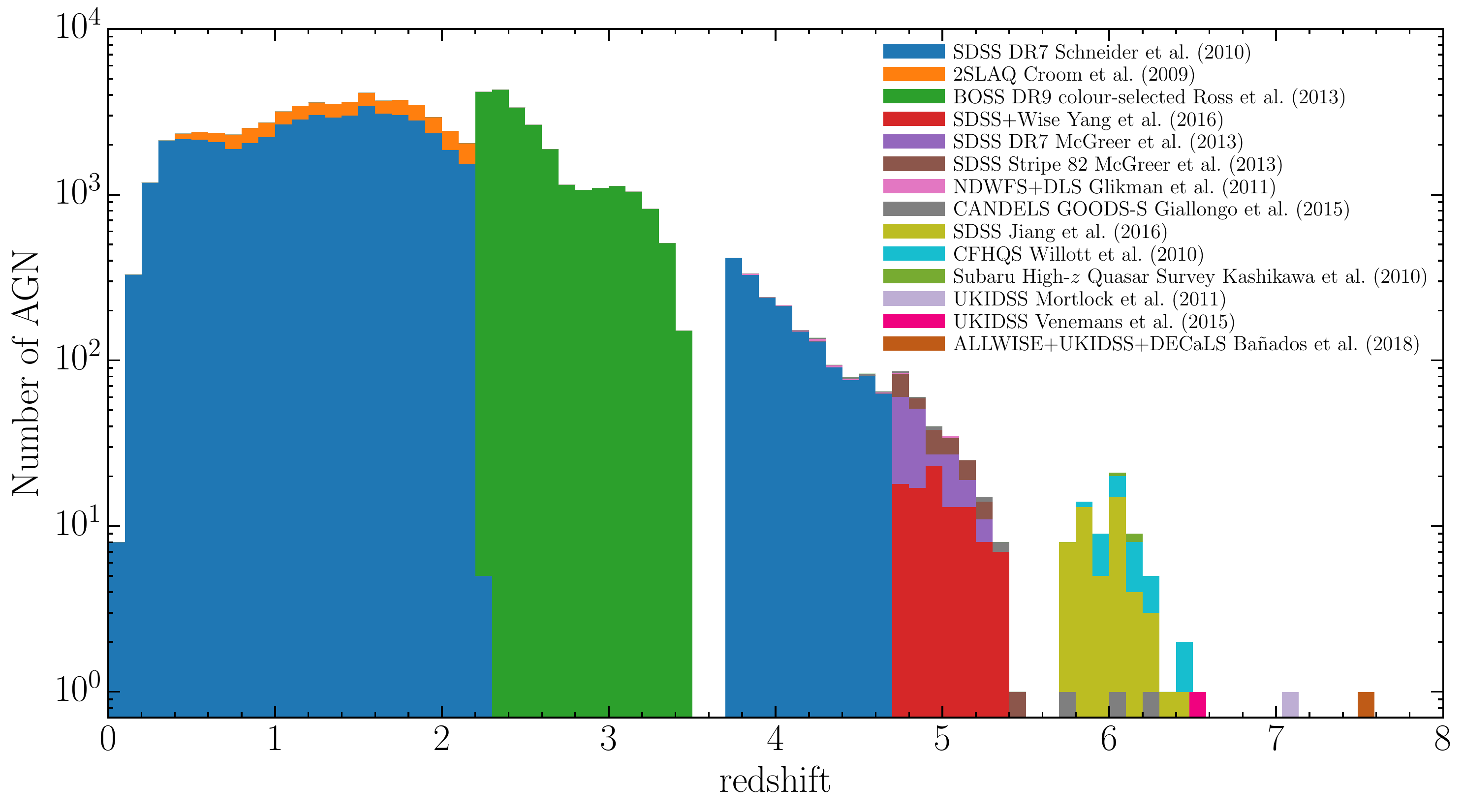}
    % data.py 
  \end{center}
  \caption{Redshift distribution of the 83,488 AGN considered in this
    analysis.  Shown here are the observed AGN numbers, without
    correcting for incompleteness.  Further details on each of these
    data sets are in Table~\ref{tab:samples} and
    Section~\ref{sec:sample}.}
  \label{fig:qsos}
\end{figure*}

The premier way towards a more robust understanding of the AGN
contribution to the UV background is to determine the evolution of the
AGN UV luminosity function across cosmic time.  To this end, numerous
measurements of the luminosity function at various redshifts have been
published in the last decade \citep{2009A&A...507..781S,
  2009MNRAS.392...19C, 2010AJ....139..906W, 2011ApJ...728L..26G,
  2012ApJ...755..169M, 2013ApJ...773...14R, 2013ApJ...768..105M,
  2015AA...578A..83G, 2016ApJ...833..222J, 2016ApJ...829...33Y,
  2016A&A...587A..41P, 2017ApJ...847L..15O, 2018PASJ...70S..34A,
  2018AJ....155..131M}.  However, many of these inferences suffer from
inconsistencies arising from inhomogeneous selection and analysis of AGN
data.  Some of the data sets analysed in these studies consist of
photometric samples, which is likely to increase sample contamination.
In some studies, distinct data sets binned differently in redshift and
magnitudes were inhomogeneously combined.  Some authors imposed
restricted priors on parameters while fitting luminosity function
models, e.g., by fixing luminosity function slopes to certain values,
which may bias the result.  Finally, some studies arbitrarily excluded
certain data sets.  This has resulted in a large scatter in the
inferred high-redshift hydrogen-ionizing AGN emissivity between
various recent studies.  For instance, there is an order of magnitude
scatter between various estimates of the hydrogen-ionizing AGN
emissivity at $z=5$--$6$ \citep{2011ApJ...728L..26G,
  2012ApJ...755..169M, 2015AA...578A..83G, 2018PASJ...70S..34A,
  2018AJ....155..131M, 2018MNRAS.474.2904P, 2017ApJ...847L..15O}.  Our
aim in this paper is to address this issue, by constructing an AGN
sample with robust redshift and completeness estimates and homogeneous
assumptions of the cosmology and intrinsic AGN spectrum.  After
constructing such a sample, we derive the UV luminosity function of
AGN from redshift $z=0$--$7.5$ and estimate the AGN contribution to
the UV background.  We discuss our sample construction in
Section~\ref{sec:sample}.  Our derived luminosity functions are
presented in Section~\ref{sec:lf}.  Section~\ref{sec:reion} presents
our inference of the AGN contribution to the \nHI-ionizing UV
background and \HeII\ reionization history.  We summarise our findings
in Section~\ref{sec:conc}.

We assume a flat cosmology with density parameters
$\left(\Omega_\mathrm{m},\Omega_\Lambda\right)=\left(0.3,0.7\right)$
and a Hubble constant $H_0=70$\,km\,s$^{-1}$\,Mpc$^{-1}$. Comoving
distances are given explicitly in comoving Mpc (cMpc). Magnitudes
are reported in the AB system \citep{1983ApJ...266..713O}, and
observed magnitudes are point spread function (PSF) magnitudes
\citep{2002AJ....123..485S} corrected for Galactic extinction
\citep{1998ApJ...500..525S} unless otherwise noted.
Our homogenised sample (Section~\ref{sec:sample}) uses absolute
monochromatic AB magnitudes at a rest frame wavelength of 1450\,\AA.

\section{Homogenised AGN Sample}
\label{sec:sample}

We started by compiling the samples of recent photometric rest-frame
UV-optical quasar surveys. The restriction to UV-optical surveys was
mainly driven by our science goal to characterise the UV luminosity
function of Type~1 quasars. X-ray-selected samples are less suited for
this purpose due to spectroscopic incompleteness and the $\sim
0.4$\,dex scatter in the conversion from X-ray to UV luminosity
\citep{2010A&A...512A..34L, 2015MNRAS.453.1946G, 2016ApJ...819..154L}
that contributes significantly to the error budget in the UV
luminosity function of X-ray-selected samples unless rest-frame UV
photometry is incorporated \citep{2015AA...578A..83G}.  The individual
surveys and their main characteristics are listed in
Table~\ref{tab:samples}.  Figure~\ref{fig:qsos} presents a redshift
histogram of the contributing surveys.

\subsection{Sample Selection}
\label{sect:samplesel}

We included surveys based on a set of simple criteria:
\begin{enumerate}
\item High spectroscopic completeness of the target sample.
\item Accurate rest-frame UV-optical CCD photometry.
\item Statistical power (sample size, coverage in $z$ and/or absolute magnitude).
\item A well-characterised selection function.
\end{enumerate}
As a prerequisite for a joint analysis of the QSO luminosity function
(QLF) we obtained the survey selection functions in electronic form,
either from the publication or by request from the authors. As a
reference for future surveys we make them electronically available
here in modified and homogenised form (see Section~\ref{sect:datahom}
and Appendix~\ref{sec:code})\footnote{These data and the code for
  deriving the luminosity functions are available on
  \url{https://github.com/gkulkarni/QLF}.  The data will also be made
  available on CDS upon acceptance of this manuscript for
  publication.}.

Due to their selection criteria and their statistical power specific
surveys contribute to distinct redshift ranges. At $z<2.2$ we
considered quasars from the SDSS DR7 quasar catalogue
\citep{2010AJ....139.2360S} and the 2SLAQ survey catalogue
\citep{2009MNRAS.392...19C}. We restricted the SDSS DR7 sample to the
48,664 $0.1<z<2.2$ quasars selected with the final SDSS quasar
selection algorithm \citep{2002AJ....123.2945R, 2006AJ....131.2766R}
from a survey area of 6248\,deg$^2$ \citep{2012ApJ...746..169S}. We
adopted the SDSS targeting photometry corrected for Galactic
extinction \citep{2010AJ....139.2360S}. To limit systematic
uncertainties in the correction for host galaxy light \citep[detailed
  in][]{2009MNRAS.392...19C} we restricted the 2SLAQ sample to 9365
$g<21.85$ $0.4<z<2.2$ quasars from its spectroscopic survey footprint
near the North Galactic Pole (NGP, 7027 quasars in $127.7$\,deg$^2$)
and the South Galactic Pole (SGP, 2338 quasars in
$64.2$\,deg$^2$). The small sample overlap between SDSS and 2SLAQ (102
quasars) has negligible impact on the QLF evaluation.

At $2.2<z<3.5$ we used a single sample of 23,301 uniformly
colour-selected quasars from 2236\,deg$^2$ in BOSS DR9
\citep{2013ApJ...773...14R} due to several improvements compared to
previous surveys. First, it covers a similar magnitude range as 2SLAQ
but with $>20$ times as many quasars. Second, although the SDSS DR7
sample provides better coverage of the bright end of the QLF at these
redshifts, its selection function is highly dependent on the assumed
incidence of (partial) Lyman limit systems in the IGM
\citep{2009ApJ...705L.113P, 2011ApJ...728...23W}. The BOSS DR9
selection function incorporates an improved treatment of the QSO SED
and the IGM, including stochasticity in the colour distribution due to
intervening Lyman-limit systems.  While the BOSS DR9 selection
function considers these improvements, the uncertainty in the QLF
remains dominated by assumptions in the selection function given the
large sample size \citep{2013ApJ...773...14R}.  Variability-selected
quasar samples circumvent this issue \citep{2013ApJ...773...14R,
  2013A&A...551A..29P, 2016A&A...587A..41P}, but may be affected by
(i) single-epoch imaging incompleteness at the faint end
\citep{2013ApJ...773...14R}, and (ii) uncertainties in the selection
function caused by the limited number of known $z\ga 3$ quasars not
selected by variability in the same footprint
\citep{2013A&A...551A..29P, 2016A&A...587A..41P}.

At $3.7<z<4.7$ we used a combination of SDSS DR7 \citep[1785 uniformly
  selected quasars from][]{2010AJ....139.2360S} and the NDWFS$+$DLS
survey \citep{2010ApJ...710.1498G,2011ApJ...728L..26G}. The lower cut
$z>3.7$ in SDSS limits the impact of systematic uncertainties in the
\citet{2006AJ....131.2766R} selection function
\citep{2009ApJ...705L.113P, 2011ApJ...728...23W}. We did not consider
the results from surveys for faint $z\sim 4$ quasars in the COSMOS
field \citep{2011ApJ...728L..25I, 2012ApJ...755..169M} due to
systematic errors in their selection functions\footnote{Both studies
  simulated quasar colours with a mean IGM attenuation curve
  \citep{1995ApJ...441...18M} that cannot account for stochastic Lyman
  continuum absorption, and therefore underpredicts the variance in
  quasar colours \citep{1999ApJ...518..103B, 2008MNRAS.387.1681I,
    2011ApJ...728...23W}. Modelling the colour variance in these
  surveys is essential, as most of the \citet{2011ApJ...728L..25I}
  quasars are near the edge of their colour selection region (see
  their Figure~1), and \citet{2012ApJ...755..169M} require modest
  attenuation of the $U$ band flux relative to the mid-infrared
  flux.}. Furthermore, 30 per cent of the \citet{2012ApJ...755..169M}
COSMOS sample have visually estimated photometric redshifts, and the
spectroscopic subsample reveals that 40 per cent of the visually
estimated redshifts are biased low
($z_\mathrm{spec}>z_\mathrm{est}+0.3$, see their Figure~9). These
unaccounted systematic redshift errors at least partly explain the
discrepancy in the $z\sim 4$ QLF between \citet{2011ApJ...728L..26G}
and \citet{2012ApJ...755..169M}, which justifies our preference for
the former sample that is 77 per cent spectroscopically complete at
$R$ magnitudes $<23.5$ \citep{2011ApJ...728L..26G}.

At $4.7\le z<5.5$ we combined several recent surveys, accounting for
sample overlap and updated selection functions. At the bright end of
the QLF we used the 99 quasars from the SDSS+WISE survey
\citep{2016ApJ...829...33Y} that have $M_{1450}\le -26.73$ in our
adopted cosmology.  For these 99 quasars selected from 14,555\,deg$^2$
we adopted the \citet{2016ApJ...829...33Y} selection function.  The
\citet{2016ApJ...829...33Y} sample partially overlaps with the
\citet{2013ApJ...768..105M} SDSS DR7 sample, so to avoid
double-counting quasars we used the latter sample only at
$M_{1450}>-26.73$, yielding 103 additional $4.7\le z<5.5$ quasars
selected in 6248\,deg$^2$. We used the $z\sim 5$ SDSS DR7 selection
function from \citet{2013ApJ...768..105M} that supersedes the one from
\citet{2006AJ....131.2766R} due to improved bandpass corrections and
IGM parameterization.  To these two bright-end samples we added the
faint-end sample from the \citet{2013ApJ...768..105M} SDSS Stripe~82
survey (59 uniformly selected $M_{1450}>-26.73$ $4.7\le z<5.5$ quasars
in 235\,deg$^2$) and two $4.7\le z<5.5$ quasars from
\citet{2011ApJ...728L..26G}, adopting the respective selection
functions.  We did not consider the limit on the $z\sim 5$ QLF by
\citet{2012ApJ...756..160I} due to systematic errors in their
selection function\footnote{\citet{2012ApJ...756..160I} underestimated
  the dispersion in rest-frame UV quasar colours with respect to SDSS
  at all redshifts (their Figure~4). Contrary to their claim,
  photometric errors have a small effect on the colour distribution of
  SDSS quasars given the statistical errors of $<0.03$\,mag in $gri$
  for 90 per cent of the SDSS DR7 bright quasar sample ($i<19.1$) and
  a relative calibration error of $\sim 1$ per cent
  \citep{2008ApJ...674.1217P}.}.

\begin{table*}
  \caption{AGN samples analysed in this work.}
  \label{tab:samples}
  \begin{tabular}{lcllrS}
    \hline
    Sample& $z$ range$^a$& Survey & Reference & Number & {Area} \\
    & & & & of quasars & {(deg$^2$)} \\
    \hline
    1 & 0.0--2.2 & SDSS DR7 & \citet{2010AJ....139.2360S} & 48664 & 6248.0 \\
    2$^b$ & 0.4--2.2 & 2SLAQ SGP & \citet{2009MNRAS.392...19C} & 2338 & 64.2 \\
    3$^b$ & 0.4--2.2 & --- NGP & \citet{2009MNRAS.392...19C} & 7027 & 127.7 \\
    4 & 2.2--3.5 & BOSS DR9 & \citet{2013ApJ...773...14R} & 23301 & 2236.0 \\
    5 & 3.7--4.7 & SDSS DR7 & \citet{2010AJ....139.2360S} & 1785 & 6248.0 \\
    6 & 3.6--5.2 & NDWFS & \citet{2011ApJ...728L..26G} & 12 & 1.71 \\
    7 & 3.8--5.3 & DLS & \citet{2011ApJ...728L..26G} & 12 & 2.05 \\
    8 & 4.7--5.4 & SDSS+WISE & \citet{2016ApJ...829...33Y} & 99 & 14555.0 \\
    9$^c$ & 4.7--5.5 & SDSS DR7 & \citet{2013ApJ...768..105M} & 103 & 6248.0 \\
    10$^c$ & 4.7--5.5 & --- Stripe 82 & \citet{2013ApJ...768..105M} & 59 & 235.0 \\
    11 & 5.7--6.5 & SDSS Main & \citet{2016ApJ...833..222J} & 24 & 11240.0 \\
    12 & 5.7--6.5 & --- Overlap & \citet{2016ApJ...833..222J} & 10 & 4223.0 \\
    13 & 5.7--6.5 & --- Stripe 82 & \citet{2016ApJ...833..222J} & 13 & 277.0 \\
    14 & 5.8--6.6 & CFHQS Deep & \citet{2010AJ....139..906W} & 1 & 4.47 \\
    15 & 5.8--6.6 & --- Very Wide & \citet{2010AJ....139..906W} & 16 & 494.0 \\
    16 & 5.8--6.5 & Subaru High-$z$ Quasar & \citet{2015ApJ...798...28K} & 2 & 6.5 \\
    17$^d$ &4.0--6.5 & CANDELS GOODS-S & \citet{2015AA...578A..83G} & 19 & 0.047 \\
    18$^e$ & 6.5--7.4 & UKIDSS & \citet{2011Natur.474..616M} & 1 & 3370.0 \\
    19$^e$ & 6.5--7.4 & UKIDSS & \citet{2015ApJ...801L..11V} & 1 & 3370.0 \\
    20$^e$ & 6.5--7.4 & ALLWISE+UKIDSS+DECaLS & \citet{2018Natur.553..473B} & 1 & 2500.0 \\
    \hline
  \end{tabular}\\
  \begin{minipage}{14.5cm}
    \textsuperscript{$a$}{Redshift range of the sample or, for small
      samples, approximate redshift range in which the survey is
      sensitive.}\\
    \textsuperscript{$b$}{Restricted to $z<2.2$.}\\
    \textsuperscript{$c$}{Restricted to $M_{1450}>-26.73$ quasars to
      avoid overlap with the \citet{2016ApJ...829...33Y} sample.}\\
    \textsuperscript{$d$}{Used only in Section~\ref{sec:global} and
      Appendix~\ref{sec:conv} due to lack of spectroscopic redshifts
      for majority of the sample.}\\
    \textsuperscript{$e$}{Used only in Section~\ref{sec:global} due to
      roughly estimated selection function.}
 \end{minipage}
\end{table*}

The SDSS colours of $5.1<z<5.5$ quasars are similar to those of M and
L dwarf stars, resulting in a low and uncertain completeness
\citep{2013ApJ...768..105M}. WISE mid-infrared selection performs
better \citep{2016ApJ...829...33Y}, but is restricted to the bright
end of the quasar population. Unlike \citet{2013ApJ...768..105M} we
include the 9 uniformly selected $M_{1450}>-26.73$ SDSS DR7 quasars
and the 10 SDSS Stripe~82 quasars at $z>5.1$, adopting their low
completeness. As we will show in Section~3.2, the resulting
QLF is consistent with those at lower and higher redshifts, indicating
that the \citet{2013ApJ...768..105M} selection functions are quite
reliable.

At $z\sim 6$ we combined the samples from all spectroscopic surveys
with a determined selection function as of June 2017.
\citet{2016ApJ...833..222J} recently compiled all quasars discovered
in several SDSS $z\sim 6$ surveys together with consistently derived
selection functions. Their uniform sample consists of 24 quasars from
the SDSS main survey (11,240\,deg$^2$), 10 additional quasars in
regions with two or more SDSS imaging scans (so-called overlap
regions, 4223\,deg$^2$), and 13 faint quasars from SDSS Stripe~82
(277\,deg$^2$). The CFHQS \citep{2010AJ....139..906W} provided a
uniform sample of 16 quasars in the Very Wide Survey (494\,deg$^2$)
and a single quasar in the Deep Survey ($4.47$\,deg$^2$). The one
quasar detected in both SDSS and CFHQS does not lead to underestimated
statistical errors in the QLF. Lastly, we included the two objects
from \citet{2015ApJ...798...28K}, one of which might be a galaxy due
to its narrow Ly$\alpha$ emission line (half width at half maximum
427\,km\,s$^{-1}$). With better photometry and additional spectroscopy
recently reported by \citet{2017ApJ...847L..15O} the
\citet{2015ApJ...798...28K} sample is complete.  Although we will
account for the slightly different redshift sensitivities for the
different surveys, we will quote a nominal redshift range $5.7<z<6.5$
for the combined $z\sim 6$ sample.

At the highest redshifts $z>6.5$ we considered ULAS~J1120$+$0641
\citep[$z=7.085$,][]{2011Natur.474..616M} and PSO~J$036.5078+03.0498$
\citep[$z=6.527$,][]{2015ApJ...801L..11V}, both discovered in UKIDSS imaging
(3370\,deg$^2$), in addition to the current highest-redshift quasar J1342$+$0928
at $z=7.54$ selected from a combination of UKIDSS, WISE and DECaLS
\citep[$\sim 2500$\,deg$^2$,][]{2018Natur.553..473B}. Although
currently only rough estimates exist concerning their selection
functions, these quasars provide constraints on the evolution of the
integrated quasar space density from $z\sim 6$ to $z\sim 7$.  We use
them in our secondary analysis in Section~\ref{sec:global}.  Although
we do not include the highly debated \citet{2015AA...578A..83G} sample
in our main analysis due to its rough selection function and lack of
spectroscopy for 17 of the 22 quasar candidates, we use it in
Section~\ref{sec:global} to constrain the faint end ($M_{1450}>-23$)
of the QLF at $z>4.1$. We restricted the \citet{2015AA...578A..83G}
sample to the 19/22 sources considered in their QLF.

\subsection{Sample Homogenisation}
\label{sect:datahom}

\begin{figure}
    \includegraphics[width=\columnwidth]{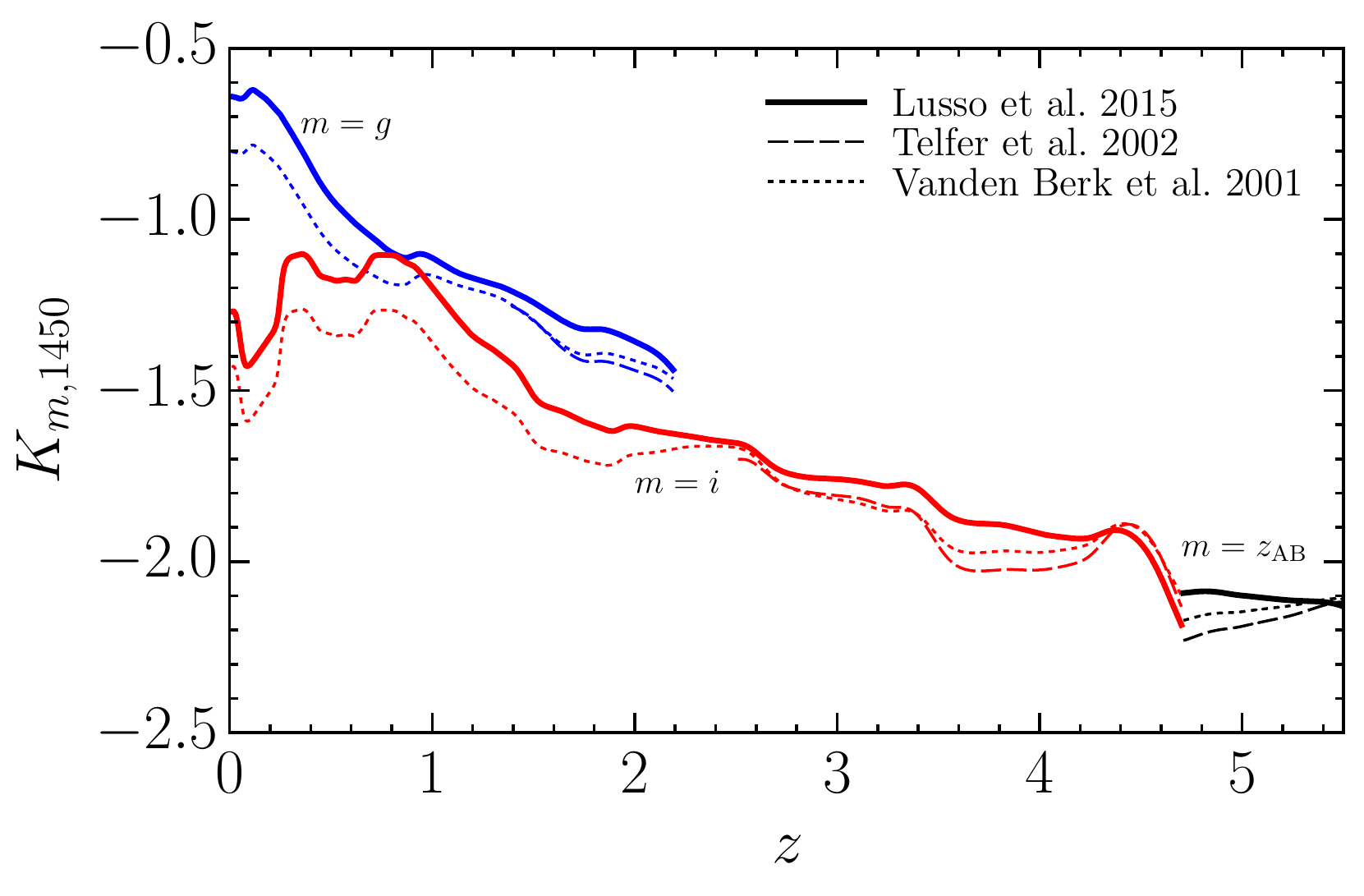}
  \caption{Bandpass corrections $K_{m,1450}$ from a broadband
    magnitude $m=\{g,i,z_\mathrm{AB}\}$ to the monochromatic AB
    magnitude at 1450\,\AA\ as a function of redshift $z$ for the
    \citet{2015MNRAS.449.4204L} quasar SED used in this work, and for
    two quasar composite spectra \citep{2001AJ....122..549V,
      2002ApJ...565..773T}.  The redshift range has been restricted to
    exclude the Ly$\alpha$ forest and to account for the different
    rest frame wavelength coverage of the spectra.}
  \label{fig:kcorr}
\end{figure}

For a joint fit of the QLF it is necessary to homogenise the different
survey samples in absolute magnitude, and to convert their selection
functions to the same absolute magnitude system. For the analysis of
the quasar UV emissivity and to be consistent with published work at
$z>3$ we chose to convert all samples and selection functions to the
absolute AB magnitude at a rest frame wavelength $\lambda=1450$\,\AA
\begin{equation}\label{eq:absmag}
  M_{1450}\left(z\right) = m-5\log{\left(\frac{d_L\left(z\right)}
    {\mathrm{Mpc}}\right)}-25-K_{m,1450}\left(z\right),
\end{equation}
with the luminosity distance
\begin{equation}
  d_L(z)=(1+z)\frac{c}{H_0}\int_0^z\frac{\mathrm{d}z^\prime}
  {\sqrt{\Omega_\mathrm{m}(1+z^\prime)^3+\Omega_\Lambda}}
  \label{eqn:dl}
\end{equation}
to a quasar at redshift $z$, the apparent magnitude $m$ in a filter used
in the survey, and the bandpass correction $K_{m,1450}\left(z\right)$
\citep{1956AJ.....61...97H, 1968ApJ...154...21O, 2000A&A...353..861W,
  2002astro.ph.10394H}. For the bandpass correction we used a
combination of the \citet{2015MNRAS.449.4204L} stacked quasar spectrum at
$\lambda<2500$\,\AA, and the \citet{2001AJ....122..549V} quasar
composite spectrum at longer wavelengths to cover the lowest
redshifts. The samples from SDSS and BOSS are defined in the SDSS $i$
band, while 2SLAQ is defined in the $g$ band. At $z>4.7$ we adopted
the SDSS $z$ band magnitude (in the following denoted $z_\mathrm{AB}$)
for SDSS DR7 quasars to avoid additional corrections due to the
Ly$\alpha$ forest. Figure~\ref{fig:kcorr} shows our bandpass
corrections for SDSS, BOSS and 2SLAQ as a function of redshift. We
ignored the luminosity dependence of the bandpass correction due to
the known anticorrelation of emission line equivalent width and
luminosity \citep{1977ApJ...214..679B}.  While the
\citet{2015MNRAS.449.4204L} spectrum is for luminous ($M_{1450}\simeq
-27.2$) quasars, UV composite spectra including fainter quasars
\citep{2002ApJ...565..773T, 2012ApJ...752..162S, 2014ApJ...794...75S}
give similar values, such that our bandpass corrections remain
applicable at $M_{1450}\la -24$. Empirical luminosity-dependent
bandpass corrections show a $\la 0.2$\,mag variation over $\sim 4$
orders of absolute magnitude depending on redshift and the filter, and
with a $\sim 0.2$\,mag intrinsic scatter due to individual quasar-to-quasar
variations \citep{2013ApJ...773...14R, 2013ApJ...768..105M,
  2013A&A...551A..29P}.

For the $z<2.2$ sample we corrected the SDSS $i$ and 2SLAQ $g$ band
magnitudes for host galaxy contamination following
\citet{2009MNRAS.392...19C}. Considering the different magnitude
limits of 2SLAQ and SDSS, the modelled host galaxy contamination is
small for $z>0.5$ quasars ($<0.1$\,mag in $g$, $<0.2$\,mag in $i$),
and is negligible at $z>0.8$.  In case the band defining the magnitude
limit of the survey undesirably overlaps with the Ly$\alpha$ forest
\citep{2010ApJ...710.1498G, 2011ApJ...728L..26G, 2013ApJ...768..105M}
we adopted their respective bandpass corrections to $M_{1450}$. In
particular, for the $z\sim 4$ sample of \citet{2010ApJ...710.1498G,
  2011ApJ...728L..26G} we recomputed $M_{1450}$ from the $R$ band
photometry to be consistent with the selection function defined in
$R$, and to avoid uncertainties in their spectrophotometry due to
incomplete spectral coverage. Since the \citet{2010ApJ...710.1498G,
  2011ApJ...728L..26G} $R$ band traces the rest frame UV, we assumed
negligible host galaxy contamination for their faint quasars. For the
remaining high-redshift surveys reporting $M_{1450}$ obtained by
various methods \citep{2010AJ....139..906W, 2011Natur.474..616M,
  2015ApJ...798...28K, 2015ApJ...801L..11V, 2016ApJ...829...33Y,
  2016ApJ...833..222J, 2018Natur.553..473B} we did not re-compute
$M_{1450}$, but applied appropriate shifts to correct to our adopted
cosmology.

The selection functions were treated similarly, i.e.\ the photometric
selection function of survey $j$ given in observed magnitudes
$f_{\mathrm{p},j}\left(m,z\right)$ \citep{2006AJ....131.2766R,
  2009MNRAS.392...19C, 2010ApJ...710.1498G, 2013ApJ...773...14R} was
transformed to our absolute magnitudes
$f_{\mathrm{p},j}\left(M_{1450},z\right)$ with
Equation~\ref{eq:absmag}, while the ones given in $M_{1450}$ were
adjusted to our cosmology. Note, however, that many surveys report
additional sources of incompleteness that require modifications to the
photometric selection functions.

For 2SLAQ we corrected for magnitude-dependent spectroscopic coverage
in the two survey areas \citep[$f_\mathrm{c,NGP}\left(g\right)$ and
  $f_\mathrm{c,SGP}\left(g\right)$; Figure~4
  in][]{2009MNRAS.392...19C} and spectroscopic redshift success
\citep[$f_\mathrm{s,2SLAQ}\left(g\right)$; Figure~6b
  in][]{2009MNRAS.392...19C} by multiplying them with the photometric
selection function, resulting in two area-specific 2SLAQ selection
functions
$f_\mathrm{NGP}\left(M_{1450},z\right)=f_\mathrm{p,2SLAQ}f_\mathrm{c,NGP}f_\mathrm{s,2SLAQ}$
and
$f_\mathrm{SGP}\left(M_{1450},z\right)=f_\mathrm{p,2SLAQ}f_\mathrm{c,SGP}f_\mathrm{s,2SLAQ}$
that are relevant for the QLF.  The $z<4.7$ SDSS photometric selection
function was modified to include known imaging incompleteness to
$f_{\mathrm{SDSS},z<4.7}=0.95f_{\mathrm{p,SDSS},z<4.7}$
\citep{2006AJ....131.2766R}. The BOSS colour-selected sample
\citep{2013ApJ...773...14R} contains quasars with
$f_\mathrm{c,BOSS}f_\mathrm{s,BOSS}\ge 0.85$, and we adopted
$f_\mathrm{BOSS}=\overline{f_\mathrm{c,BOSS}f_\mathrm{s,BOSS}}
f_\mathrm{p,BOSS}=0.962f_\mathrm{p,BOSS}$. \citet{2010ApJ...710.1498G}
presented two area-specific photometric selection functions due to
different filters employed, and more follow-up spectroscopy was
reported in \citet{2011ApJ...728L..26G}. We accounted for remaining
spectroscopic incompleteness at $R>23$, yielding the final selection
functions $f_\mathrm{NDWFS}$ and $f_\mathrm{DLS}$. The updated $z\sim
5$ SDSS photometric selection function \citep{2013ApJ...768..105M} was
modified to include imaging and spectroscopic incompleteness, yielding
$f_{\mathrm{SDSS},z\sim 5}=0.95^2f_{\mathrm{p,SDSS},z\sim 5}$. In the
deeper $z\sim 5$ SDSS Stripe~82 survey the spectroscopic
incompleteness is larger and magnitude-dependent \citep[Figure~14
  in][]{2013ApJ...768..105M}, resulting in $f_{\mathrm{S82},z\sim
  5}=0.95f_{\mathrm{s,S82},z\sim
  5}\left(i\right)f_{\mathrm{p,S82},z\sim 5}$. Likewise, imaging and
magnitude-dependent spectroscopic incompleteness was factored into the
\citet{2016ApJ...829...33Y} photometric selection function (their
Figures~5 and 7), resulting in
$f_\mathrm{SDSS+WISE}=0.97f_\mathrm{s,SDSS+WISE}
\left(z_\mathrm{AB}\right)f_\mathrm{p,SDSS+WISE}$.  We obtained a
rough estimate of the \citet{2015AA...578A..83G} selection function by
comparing the corrected and observed QLFs, i.e.\ taking
$f_\mathrm{GOODS-S}=\phi_\mathrm{obs}/\phi_\mathrm{corr}$ (see their
Table~3).  Finally, for the three $z>6.5$ quasars we assumed a
selection function of unity in a range of $z$ and $M_{1450}$ estimated
by the respective survey teams (private communication).

\section{Luminosity function}
\label{sec:lf}

After homogenising the samples and selection functions we are in a
position to compute the UV luminosity function of AGN. We begin our
analysis by computing binned estimates of the luminosity function as a
function of magnitude in several narrow redshift intervals.  We then
perform parametric maximum-likelihood fits of the luminosity function
in the individual redshift bins, examine the resulting parameters as a
function of redshift, and attempt a joint fit in magnitude and
redshift.  For simplicity we will use the notation $M\equiv M_{1450}$
in the following.

\subsection{Binned luminosity function estimates}
\label{sec:binnedlf}

In a magnitude bin $[M_\mathrm{min}, M_\mathrm{max})$, and redshift
  bin $[z_\mathrm{min}, z_\mathrm{max})$, we define the binned luminosity
function as \citep{2000MNRAS.311..433P}
\begin{equation}
  \phi \equiv \frac{N_\mathrm{QSO}}{V_\mathrm{bin}},
\end{equation}
where $N_\mathrm{QSO}$ is the number of quasars with magnitude
$M_\mathrm{min}\leq M<M_\mathrm{max}$ and redshift
$z_\mathrm{min}\leq z<z_\mathrm{max}$, and
\begin{equation}
  V_\mathrm{bin} = \int_{M_\mathrm{min}}^{M_\mathrm{max}}\mathrm{d}M
  \int_{z_\mathrm{min}}^{z_\mathrm{max}}\mathrm{d}z\, f(M, z)\,\frac{\mathrm{d}V}{\mathrm{d}z},
  \label{eqn:vi}
\end{equation}
is the effective volume of the bin. The inclusion of the survey selection function
$f(M,z)$ (Section~\ref{sect:datahom}) in Equation~(\ref{eqn:vi}) accounts for
what are sometimes called ``incomplete bins''
\citep{2006AJ....131.2766R}.  The comoving volume element $\mathrm{d}V/\mathrm{d}z$ is
given by
\begin{equation}
  \frac{\mathrm{d}V}{\mathrm{d}z}=\frac{\mathrm{d}V}{\mathrm{d}z\,\mathrm{d}\Omega}\times A\times\frac{4\pi}{41253},
\end{equation}
where $A$ is the survey area in deg$^2$, and 
\begin{equation}
  \frac{\mathrm{d}V}{\mathrm{d}z\,\mathrm{d}\Omega}=\frac{c}{H_0}\frac{d_L^2\left(z\right)}
       {\left(1+z\right)^2\left[\Omega_\mathrm{m}\left(1+z\right)^3+\Omega_\Lambda\right]^{1/2}}
  \label{eqn:dvdzdo}
\end{equation}
denotes the comoving volume element per unit solid angle
\citep{1999astro.ph..5116H}. The resulting luminosity function $\phi$ has units of
$\mathrm{cMpc}^{-3}\mathrm{mag}^{-1}$. 

\begin{figure*}
  \begin{center}
    \includegraphics[width=\textwidth,keepaspectratio]{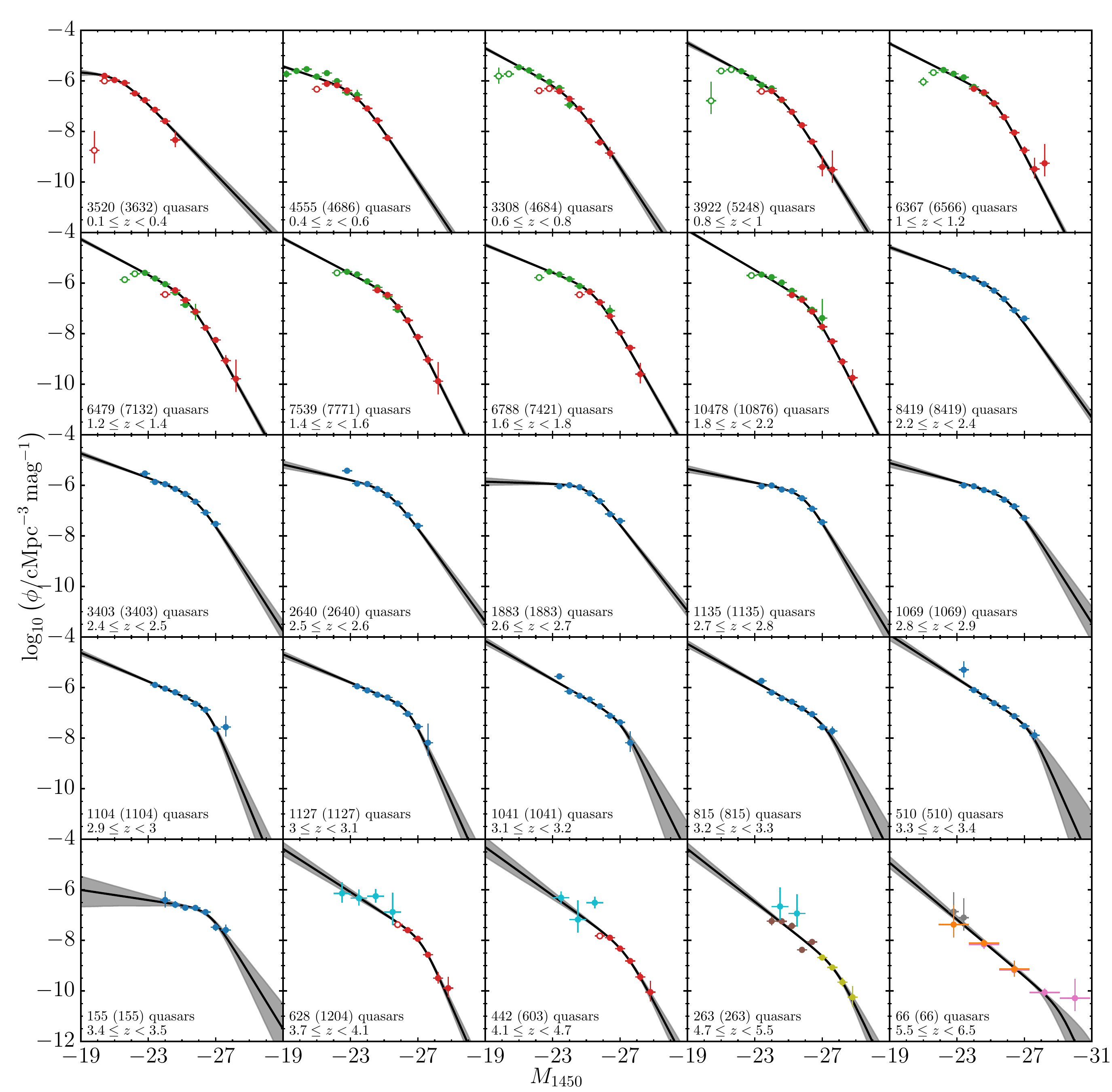}
    % mosaic.py 
  \end{center}
  \caption{Homogenised quasar luminosity functions at rest-frame
    wavelength $\lambda=1450$\,\AA\ in redshift bins from $z=0.1$ to
    $6.5$.  The symbols show our inferred binned luminosity functions
    from various data sets: \citet[red]{2010AJ....139.2360S},
    \citet[green]{2009MNRAS.392...19C}, \citet[dark
      blue]{2013ApJ...773...14R}, \citet[light
      blue]{2011ApJ...728L..26G}, \citet[yellow]{2016ApJ...829...33Y},
    \citet[DR7 brown]{2013ApJ...768..105M}, \citet[Stripe 82
      teal]{2013ApJ...768..105M}, \citet[pink]{2016ApJ...833..222J},
    \citet[orange]{2010AJ....139..906W}, and
    \citet[grey]{2015ApJ...798...28K}.  Open circles in corresponding
    colours indicate magnitude bins excluded due to incompleteness in
    the respective data sets.  The number of AGN before this selection
    is shown in parentheses; the selected number of AGN is shown
    outside parentheses.  In each redshift bin, the black curve shows
    our fiducial double power law model fit, which is represented by
    the median of the posterior probability distribution function.
    The grey shaded area shows the one-sigma uncertainty (68.26\%
    equal-tailed credible interval).  See Sections~\ref{sec:binnedlf}
    and \ref{sec:bins} for further details.}
  \label{fig:mosaic}
\end{figure*}

We evaluate the double integral in Equation~(\ref{eqn:vi}) by the Euler method,
i.e., by simply summing over the ``tiles'' of the selection function in $M$ and $z$
without interpolation.
This results in $V_i=0$ for a few quasars, which are subsequently removed
from our analysis\footnote{
  We note that the interpolation of sometimes coarse
  selection functions is not straightforward due to their strong gradients.
  The presence of objects with $V_i=0$ implies that the selection function has systematic errors.}.
In each bin we estimate the uncertainty in the luminosity function by
assuming Poisson statistics \citep{1986ApJ...303..336G} for the number
of quasars, i.e.\ assuming negligible uncertainty in the selection function.
While this is a reasonable approximation for small surveys, large surveys
with negligible Poisson errors are instead limited by rarely quantified systematic
errors due to implicit assumptions in their selection functions.
The resultant binned luminosity function estimates
are shown by the circles in Figure~\ref{fig:mosaic}.

From Figure~\ref{fig:mosaic} we see that the distribution of
luminosity function values in each redshift bin are suggestive of a
double power law form for the QLF.  We will fit such a form below.
However, in several redshift bins we note a suspicious decline of the
luminosity function at the faint limit of several surveys, for example
in the 2SLAQ sample at $z<2.2$, and in the SDSS sample at $z<1.8$ and
$z\sim 4$.  The inconsistency between the SDSS faint end and the
deeper 2SLAQ QLF indicates that the SDSS selection function is
systematically overestimated at its magnitude limit.  At $z\sim 4$ the
SDSS faint end QLF is inconsistent with the fainter
\citet{2011ApJ...728L..26G} QLF. We identify such discrepant bins by
eye, discard the contributing quasars from our analysis, and set their
selection function values to zero.  The discarded magnitude bins are
shown in Figure~\ref{fig:mosaic} by open circles.

\begin{figure*}
  \begin{center}
    \includegraphics[width=0.7\textwidth]{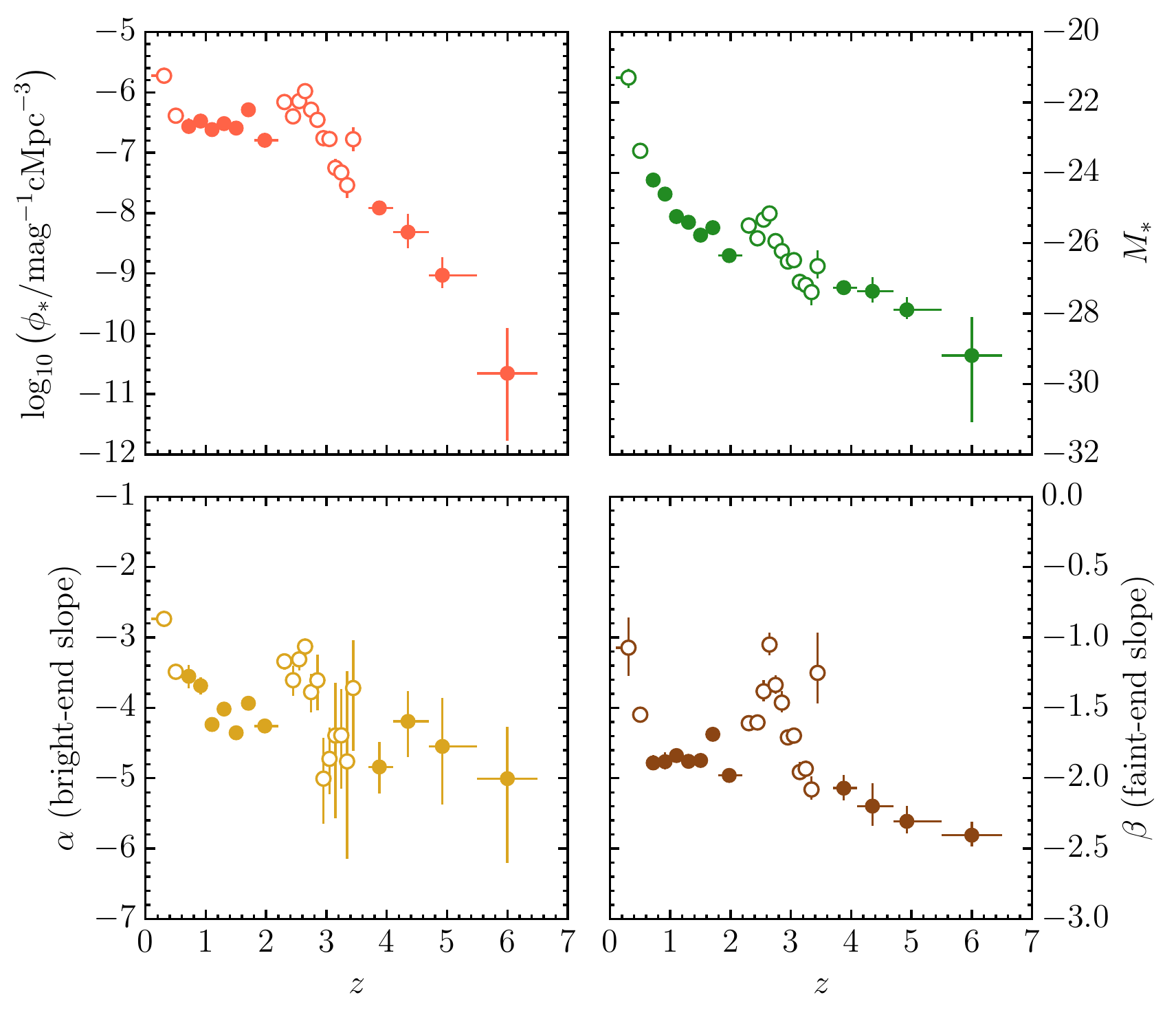}
    % summary_bins.py 
  \end{center}
  \caption{Redshift evolution of the four double power law parameters
    from the redshift bins shown in Figure~\ref{fig:mosaic}.  Vertical
    error bars show one-sigma (68.26\%) statistical uncertainties
    derived from the posterior distribution, whereas horizontal error
    bars show widths of the redshift bins.  We identify the general
    evolutionary trends of each of these parameters from the bins
    shown by the filled symbols.  The open symbols show bins that
    appear to be offset from these trends, likely due to 
    systematic errors.  The open circles at $2.2<z<3.5$ show the BOSS
    sample, while the bins at $z < 0.6$ contain AGN from the SDSS and
    2SLAQ data sets.}
  \label{fig:evoln}
\end{figure*}

\begin{table*}
  % bins_tabulate.py; Nqso and Nqso' added by hand
  \caption{Posterior median double power law luminosity function
    parameters and their $1\sigma$ (68.26\%) statistical errors in
    various redshift bins shown in Figure~\ref{fig:evoln}.  The
    luminosity function parameters $\phi_*$, $M_*$, $\alpha$, and
    $\beta$ are defined in Equation~(\ref{eqn:dpl}), with $\beta$
    denoting the faint-end slope.  Quasars in each bin have redshifts
    $z_\mathrm{min}\leq z < z_\mathrm{max}$ with a sample mean
    $\langle z\rangle$.  The total number of QSOs in each bin is given
    by $N_\mathrm{QSO}^\prime$.  The selected number of QSOs in each
    bin, after excluding faint QSOs due to excess incompleteness is
    given by $N_\mathrm{QSO}$.}
  \label{tab:bins}
  \begin{tabular}{lccrr....}
    \hline
    $\langle z\rangle$ &
    $z_\mathrm{min}$ &
    $z_\mathrm{max}$ &
    $N_\mathrm{QSO}$ &
    $N_\mathrm{QSO}^\prime$ &
    \multicolumn{1}{c}{$\log_{10}(\phi_*/$} &
    \multicolumn{1}{c}{$M_*$} &
    \multicolumn{1}{c}{$\alpha$} &
    \multicolumn{1}{c}{$\beta$} \\
    
    &
    &
    &
    &
    &
    \multicolumn{1}{c}{cMpc$^{-3}$mag$^{-1}$)} &
    &
    & \\
    \hline
    0.31$^a$ & 0.10 & 0.40 & 3520 & 3632 & -5.72^{+0.08}_{-0.12} & -21.30^{+0.24}_{-0.29} & -2.74^{+0.09}_{-0.11} & -1.07^{+0.22}_{-0.20} \\
    0.50$^a$ & 0.40 & 0.60 & 4555 & 4686 & -6.39^{+0.05}_{-0.05} & -23.38^{+0.10}_{-0.09} & -3.49^{+0.10}_{-0.10} & -1.55^{+0.04}_{-0.04} \\
    0.72 & 0.60 & 0.80 & 3308 & 4684 & -6.57^{+0.14}_{-0.12} & -24.21^{+0.22}_{-0.18} & -3.55^{+0.16}_{-0.17} & -1.89^{+0.06}_{-0.05} \\
    0.91 & 0.80 & 1.00 & 3922 & 5248 & -6.48^{+0.11}_{-0.10} & -24.60^{+0.16}_{-0.14} & -3.69^{+0.12}_{-0.13} & -1.88^{+0.07}_{-0.05} \\
    1.10 & 1.00 & 1.20 & 6367 & 6566 & -6.62^{+0.03}_{-0.03} & -25.24^{+0.05}_{-0.05} & -4.24^{+0.09}_{-0.09} & -1.84^{+0.02}_{-0.02} \\
    1.30 & 1.20 & 1.40 & 6479 & 7132 & -6.52^{+0.04}_{-0.04} & -25.41^{+0.06}_{-0.06} & -4.02^{+0.09}_{-0.09} & -1.88^{+0.03}_{-0.03} \\
    1.50 & 1.40 & 1.60 & 7539 & 7771 & -6.59^{+0.03}_{-0.03} & -25.77^{+0.04}_{-0.04} & -4.35^{+0.09}_{-0.09} & -1.87^{+0.02}_{-0.02} \\
    1.71 & 1.60 & 1.80 & 6788 & 7421 & -6.29^{+0.04}_{-0.04} & -25.56^{+0.06}_{-0.06} & -3.94^{+0.07}_{-0.08} & -1.69^{+0.03}_{-0.03} \\
    1.98 & 1.80 & 2.20 & 10478 & 1087& -6.79^{+0.03}_{-0.03} & -26.35^{+0.04}_{-0.04} & -4.26^{+0.07}_{-0.08} & -1.98^{+0.02}_{-0.02} \\
    2.30$^a$ & 2.20 & 2.40 & 8419 & 8419 & -6.16^{+0.07}_{-0.06} & -25.50^{+0.12}_{-0.11} & -3.34^{+0.11}_{-0.12} & -1.61^{+0.04}_{-0.04} \\
    2.45$^a$ & 2.40 & 2.50 & 3403 & 3403 & -6.40^{+0.08}_{-0.08} & -25.86^{+0.14}_{-0.13} & -3.61^{+0.21}_{-0.22} & -1.60^{+0.05}_{-0.04} \\
    2.55$^a$ & 2.50 & 2.60 & 2640 & 2640 & -6.15^{+0.08}_{-0.08} & -25.33^{+0.17}_{-0.16} & -3.31^{+0.15}_{-0.16} & -1.38^{+0.08}_{-0.07} \\
    2.65$^a$ & 2.60 & 2.70 & 1883 & 1883 & -5.98^{+0.05}_{-0.06} & -25.16^{+0.13}_{-0.13} & -3.13^{+0.11}_{-0.12} & -1.05^{+0.08}_{-0.08} \\
    2.75$^a$ & 2.70 & 2.80 & 1135 & 1135 & -6.29^{+0.07}_{-0.07} & -25.94^{+0.14}_{-0.13} & -3.78^{+0.26}_{-0.29} & -1.34^{+0.07}_{-0.06} \\
    2.85$^a$ & 2.80 & 2.90 & 1069 & 1069 & -6.46^{+0.12}_{-0.10} & -26.22^{+0.23}_{-0.18} & -3.61^{+0.37}_{-0.43} & -1.46^{+0.08}_{-0.07} \\
    2.95$^a$ & 2.90 & 3.00 & 1104 & 1104 & -6.76^{+0.07}_{-0.06} & -26.52^{+0.11}_{-0.09} & -5.01^{+0.58}_{-0.64} & -1.71^{+0.05}_{-0.04} \\
    3.05$^a$ & 3.00 & 3.10 & 1127 & 1127 & -6.77^{+0.08}_{-0.07} & -26.48^{+0.11}_{-0.10} & -4.72^{+0.44}_{-0.50} & -1.70^{+0.06}_{-0.05} \\
    3.15$^a$ & 3.10 & 3.20 & 1041 & 1041 & -7.25^{+0.15}_{-0.12} & -27.10^{+0.21}_{-0.18} & -4.39^{+0.75}_{-1.18} & -1.96^{+0.07}_{-0.05} \\
    3.25$^a$ & 3.20 & 3.30 & 815 & 815 &  -7.33^{+0.12}_{-0.13} & -27.19^{+0.19}_{-0.21} & -4.39^{+0.66}_{-0.76} & -1.93^{+0.06}_{-0.05} \\
    3.34$^a$ & 3.30 & 3.40 & 510 & 510 &  -7.54^{+0.22}_{-0.22} & -27.39^{+0.29}_{-0.37} & -4.76^{+1.29}_{-1.38} & -2.08^{+0.09}_{-0.07} \\
    3.44 & 3.40 & 3.50 & 155 & 155 &  -6.78^{+0.20}_{-0.20} & -26.65^{+0.44}_{-0.34} & -3.72^{+0.68}_{-0.89} & -1.25^{+0.29}_{-0.22} \\
    3.88 & 3.70 & 4.10 & 628 & 1204 & -7.92^{+0.12}_{-0.10} & -27.26^{+0.14}_{-0.12} & -4.84^{+0.36}_{-0.38} & -2.07^{+0.10}_{-0.09} \\
    4.35 & 4.10 & 4.70 & 442 & 603 &  -8.32^{+0.31}_{-0.26} & -27.37^{+0.39}_{-0.32} & -4.19^{+0.43}_{-0.50} & -2.20^{+0.16}_{-0.14} \\
    4.92 & 4.70 & 5.50 & 263 & 263 &  -9.03^{+0.29}_{-0.22} & -27.89^{+0.36}_{-0.26} & -4.55^{+0.69}_{-0.82} & -2.31^{+0.11}_{-0.09} \\
    6.00 & 5.50 & 6.50 & 66 & 66 &  -10.66^{+0.75}_{-1.12} & -29.19^{+1.09}_{-1.89} & -5.00^{+0.73}_{-1.20} & -2.40^{+0.09}_{-0.08} \\
    \hline
  \end{tabular}\\
  \begin{minipage}{13.0cm}
    \textsuperscript{$a$}{Redshift bin not considered in the joint QLF fit due to systematic errors (open circles in Figure~\ref{fig:evoln}).}
  \end{minipage}
\end{table*}

\subsection{Double power law fits}
\label{sec:bins}

In each redshift bin, we model the QLF as a double power law
\citep[e.g.][]{1988MNRAS.235..935B}
\begin{equation}
  \phi(M) =
  \frac{\phi_*}{10^{0.4(\alpha+1)(M-M_*)}+10^{0.4(\beta+1)(M-M_*)}}
  \label{eqn:dpl}
\end{equation}
with four free parameters: (i) the amplitude $\phi_*$, (ii) the break
magnitude $M_*$, (iii) the bright-end slope $\alpha$, and (iv) the
faint-end slope $\beta$.  By assuming broad, uniform priors, we obtain
posterior probability distributions for these parameters using the
Markov Chain Monte Carlo technique \citep[MCMC, e.g.,][]{jaynes}.  The
joint posterior probability distribution of the model parameters is
then written as
\begin{multline}
  p(\phi_*, M_*, \alpha, \beta | \{M_i, z_i\}) \propto \\ p(\phi_*, M_*,
  \alpha, \beta)p(\{M_i, z_i\} | \phi_*, M_*, \alpha, \beta),
\end{multline}
where the constant of proportionality is independent of the luminosity
function parameters, and $\{M_i, z_i\}$ denotes the magnitudes and
redshifts of quasars falling in a redshift bin $[z_\mathrm{min},
  z_\mathrm{max})$.  We use a uniform prior distribution $p(\phi_*,
  M_*, \alpha, \beta)$ and, following the standard practice of
  unbinned luminosity function estimates
  \citep[e.g.,][]{2001AJ....121...54F}, assume that the likelihood
\begin{equation}
  \mathcal{L}\equiv p(\{M_i, z_i\} | \phi_*, M_*, \alpha, \beta)
\end{equation}
is given by $\phi(M)$ itself with suitable normalisation.  The
negative logarithm of the likelihood $S\equiv -2\ln\mathcal{L}$ can
then be written as
\begin{multline}
  S = -2\sum_{i=1}^{N_\mathrm{QSO}}\ln\phi(M_i, z_i)\\+2\int_{M_\mathrm{min}}^{M_\mathrm{max}}\mathrm{d}M
  \int_{z_\mathrm{min}}^{z_\mathrm{max}}\mathrm{d}z\, \phi(M,z) f(M, z)\,\frac{\mathrm{d}V}{\mathrm{d}z},
  \label{eqn:S}
\end{multline}
where the integral over magnitude is on the surveyed range of $M$.  We
use the \texttt{emcee} code \citep{2013PASP..125..306F} for MCMC.

The above likelihood can also be understood as the limit of the
Poisson likelihood in luminosity and redshift bins
\citep{1983ApJ...269...35M, 2001AJ....121...54F}.  We can write the
probability of observing $n_{ij}$ quasars in the $(M_i, z_j)$ bin as
the Poisson distribution
\begin{equation}
  \mathcal{L}=\prod_{i,j}\frac{e^{-\mu_{ij}}\mu_{ij}^{n_{ij}}}{n_{ij}!},
  \label{eqn:lhood}
\end{equation}
where 
\begin{equation}
  \mu_{ij}= \int_{M_i}^{M_{i+1}}\mathrm{d}M\int_{z_j}^{z_{j+1}}\mathrm{d}z\, \phi(M,z) f(M, z)
  \,\frac{\mathrm{d}V}{\mathrm{d}z},
\end{equation}
is the average number of quasars expected in the $(M_i, z_j)$ bin
given the luminosity function $\phi(M,z)$.  In the limit of
infinitesimal bins, $n_{ij}=0$ or $1$, and Equation~(\ref{eqn:lhood})
can be simplified to obtain Equation~(\ref{eqn:S}).

Our estimates for the double power law luminosity function are shown
in Figure~\ref{fig:mosaic} for 25 redshift bins.  We adopt the
posterior median as our fiducial model fit, and the 68.26\%
equal-tailed credible interval as the uncertainty on $\phi$.  (We
adopt similar definitions for the derived quantities.)  The resultant
parameter values are listed in Table~\ref{tab:bins}.\footnote{Full
  posterior distributions for the luminosity functions themselves
  should be obtained by running our publicly available code.  This is
  also true for other models and derived quantities such as
  emissivities that we discuss in this paper.} Consistent with
previous studies, the double power law model provides an excellent
description of the luminosity function model over almost the complete
range of redshifts spanned by the data.  It is only in the highest
redshift bin ($z=5.5$--$6.5$) that the data seem to favour a single
power law.  In this bin, the resultant posterior distribution of the
break magnitude $M_*$ is bimodal with favoured values at the faint
($M_*>-18$) and bright end of the data ($M_*<-30$).  While in the
literature $z\sim 6$ quasars have been assumed to lie on the bright
end of the luminosity function \citep[e.g.,][]{2016ApJ...833..222J}, a
comparison with the luminosity function at lower redshifts ($z<5.5$)
suggests that these AGN should instead be understood to describe the
faint-end of a double power law.  As we discuss below, $M_*$ gets
progressively brighter with redshift.  Therefore, after inspecting the
data at lower redshifts, we use restricted priors in the
$z=5.5$--$6.5$ redshift bin in order to avoid bimodal distributions.
In this bin, we restrict the bright-end slope $\alpha$ to values less
than $-4$, which is equivalent to forcing $M_*$ to be at the bright
end of the data.  Other parameters continue to have wide uniform
priors.  This also illustrates the importance of analysing the
evolution of the QLF with redshift.

The redshift evolution of the four double power law parameters is
shown in Figure~\ref{fig:evoln} and tabulated in Table~\ref{tab:bins}.
We find interesting evolutionary trends in each of the four
parameters.  The break magnitude $M_*$ evolves by more than eight
magnitudes from redshift $z=0$ to $7$.  The amplitude of the
luminosity function $\phi_*$ evolves moderately from $z=0$ to $z\sim
3$ and then drops by more than five orders of magnitude to about
$10^{-12}$\,cMpc$^{-3}$mag$^{-1}$ at $z\sim 7$.  The bright end slope
$\alpha$ has significant scatter but still shows a trend towards more
negative values, i.e., steeper luminosity function bright ends, at
high redshifts. Finally, the faint end slope also shows signs of
increasing steepness towards high redshifts, but with a marked
discontinuity at $2.2\le z<3.5$.

Discontinuities and scatter in the QLF parameters over short redshift
intervals in Figure~\ref{fig:evoln} reveal further likely systematic
errors in the survey selection functions.  Quasars at $2.2\le z<3.5$
taken solely from BOSS \citep{2013ApJ...773...14R} follow the redshift
trends in $\phi_*$ and $M_*$, but with significant scatter in $\Delta
z=0.1$ intervals that is much larger than the statistical error.  The
discontinuity at $z=2.2$ indicates a mismatch of BOSS and
SDSS$+$2SLAQ. The most striking feature, however, is the apparent
rapid redshift evolution of the faint-end slope revealed in the BOSS
sample, which is also highlighted in Figure~\ref{fig:mosaic}. Given
the relatively smooth evolution of all QLF parameters at lower and
higher redshifts it is unlikely that the QLF evolution at $2.2\le
z<3.5$ indicated by BOSS is physical.  Rather it reflects the
systematics limit of the large BOSS sample induced by a fixed
selection function that critically depends on the assumed quasar
spectral energy distribution, the IGM parameterization, and the
modeled photometric errors at the targeted magnitude limit of the
single-epoch SDSS imaging
\citep{2011ApJ...728...23W, 2012ApJS..199....3R, 2013ApJ...773...14R}.
Consequently, we exclude all BOSS quasars from further analysis.

The imperfect match between SDSS and 2SLAQ (Figure~\ref{fig:mosaic})
causes low-level systematics, as evidenced by the apparent
discontinuities in $\alpha$ at $z<2.2$ and the jump in $\beta$ at
$z=1.8$ in Figure~\ref{fig:evoln}.  At $z<0.6$ the faint-end slope
shows a sharp increase which we attribute to remaining uncertainties
in the correction for host galaxy light and potentially missed AGN in
extended sources. We exclude $z<0.6$ quasars from further
consideration. Due to different selection function parameters
inter-survey systematics are definitely present at $z>3.5$ as well,
but Poisson errors of the limited samples dominate.

\subsection{Evolution of the luminosity function}
\label{sec:global}

\begin{figure*}
  \begin{center}
    \includegraphics[width=0.7\textwidth]{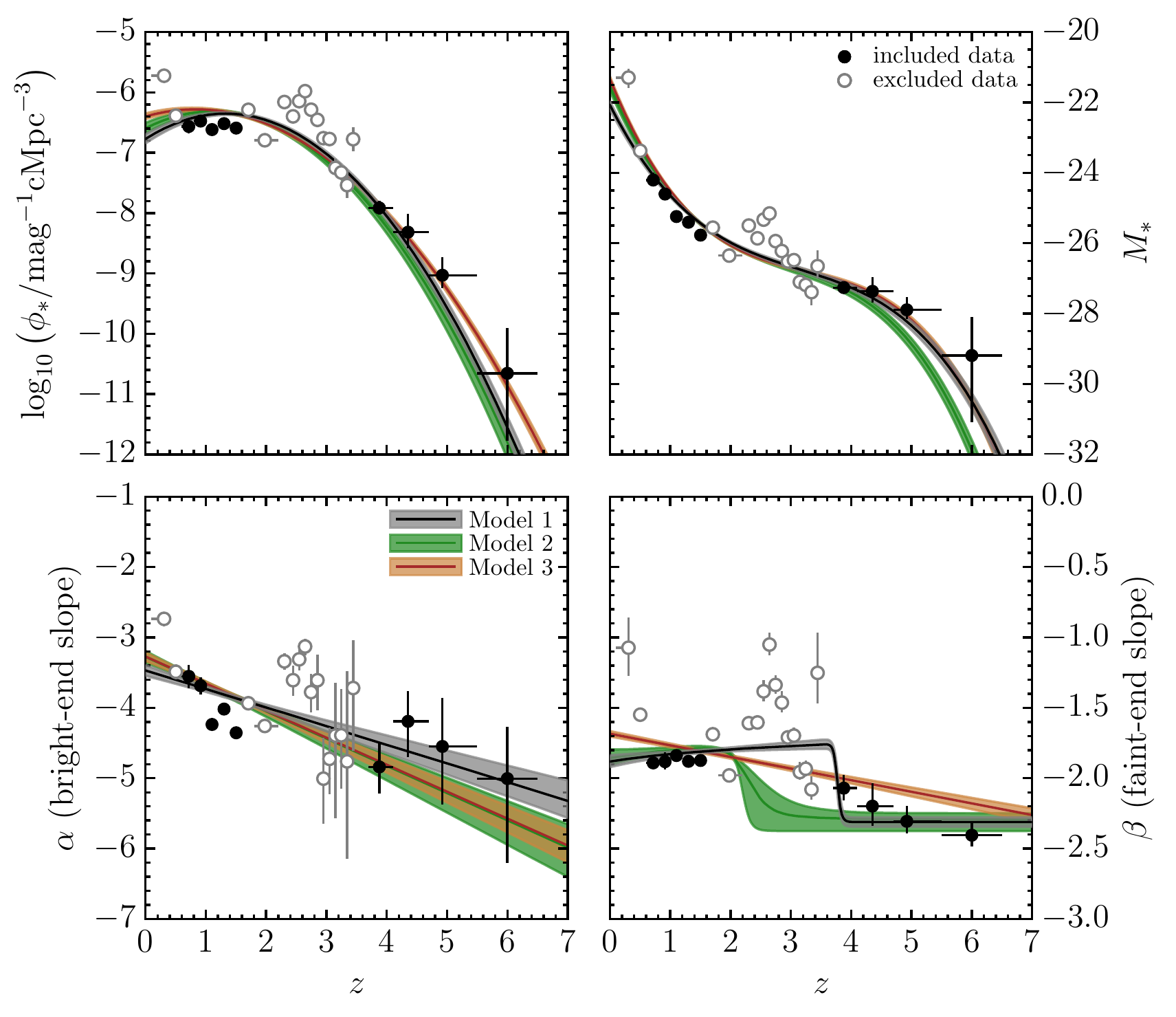}
    % summary_fromFile.py 
  \end{center}
  \caption{Luminosity function parameter evolution in the global
    models.  The symbols show the posterior median values of
    parameters with one-sigma (68.26\%) uncertainties in redshift bins
    from Figure~\ref{fig:evoln}.  Redshift bins deemed to be affected
    by systematics and removed from the global analysis are shown by
    the grey open circles.  In each panel, the solid curves and shaded
    regions show the three derived global models with one-sigma
    uncertainties.}
  \label{fig:evoln_global}
\end{figure*}

After excluding the redshift bins that are most obviously affected by
systematic errors, the remaining redshift bins (filled symbols in
Figure~\ref{fig:evoln}) are consistent with a smooth redshift
evolution of the luminosity function parameters that may be described
by parametric models \citep{1976A&A....53...15M, 1983ApJ...269..352S,
  1988ApJ...325...92K, 1988MNRAS.235..935B, 1993ApJ...406L..43H,
  1994ApJ...421..412W, 1995AJ....110...68S, 1995AJ....110.2553K,
  1995ApJ...438..623P, 2000MNRAS.317.1014B, 2001AJ....121...54F,
  2006AJ....131.2766R, 2007A&A...472..443B, 2009MNRAS.399.1755C,
  2013ApJ...773...14R, 2013A&A...551A..29P}.  Such descriptions have
also been developed in the literature for the X-ray
\citep[e.g.,][]{2015MNRAS.451.1892A} and so-called bolometric
luminosity functions \citep[e.g.,][]{2007ApJ...654..731H}.  Such
``global'' models of the QLF evolution are useful as
they give a continuous description of the luminosity function.  This
allows one to reduce the bias introduced by binning the data in
arbitrary redshift bins.  By potentially allowing for extrapolations
beyond the redshift range spanned by the data, such models are
valuable for understanding of the physics behind the luminosity
function.  Ideally, one would want to use physically meaningful
parameters that govern the formation and evolution of the AGN
population.  Unfortunately, such physical parameterisation is yet to
be developed.  We therefore set up an empirical parameterisation to
describe the evolution of the four parameters of the double power law
model in Equation~(\ref{eqn:dpl}) as 
\begin{align}
  &\log_{10}\phi_*(z) = F_0(\{c_{0,j}\}, z)\nonumber\\
  &M_*(z) = F_1(\{c_{1,j}\}, z)\nonumber\\
  &\alpha(z) = F_2(\{c_{2,j}\}, z)\nonumber\\
  &\beta(z) = F_3(\{c_{3,j}\}, z),
  \label{eqn:global}
\end{align}
where the $\{c_{n,j}\}$ are the new model parameters, and the $\{F_j\}$
are functions that vary smoothly with redshift $z$.  The joint
posterior probability distribution of these parameters can be now
written as
\begin{equation}
  p(\{c_{n,j}\} | \{M_i, z_i\}) \propto p(\{c_{n,j}\})p(\{M_i, z_i\} | \{c_{n,j}\}),
\end{equation}
where the likelihood 
\begin{equation}
  \mathcal{L}\equiv p(\{M_i, z_i\} | \{c_{n,j}\}),
\end{equation}
is now given by $\phi(M,z)$ with suitable normalisation.  Note that
$\phi(M,z)$ is given by Equation~(\ref{eqn:dpl}), but now the four
parameters in that equation are redshift-dependent.  The negative
logarithm of the likelihood $S\equiv -2\ln\mathcal{L}$ is a
straightforward generalisation of Equation~(\ref{eqn:S}).  We consider
models in which the evolution of the four double power law parameters
is modelled independently as in Equations~(\ref{eqn:global}), an
approach sometimes termed as ``flexible double power law''
\citep{2015MNRAS.451.1892A}.  We present three such models in this
paper.  These are shown in Figure~\ref{fig:evoln_global}.  The three
models differ in the way they describe the evolution of the faint-end
slope $\beta$ and in the selection of the AGN data.  The models are
defined as follows.

\begin{itemize}

\item
  In Model 1 we assume that the functions $F_0$, $F_1$ and $F_2$ from
  Equation~\eqref{eqn:global} are Chebyshev polynomials in
  $\left(1+z\right)$, written as
  \begin{equation}
    F_i\left(1+z\right)=\sum_{j=0}^{n_i}c_{i,j}T_j\left(1+z\right)
    \label{eqn:cbs}
  \end{equation}
  for $i\in\{0,1,2\}$, where $c_{i,j}$ are the parameters from
  Equation~\eqref{eqn:global} and $T_j\left(1+z\right)$ are Chebyshev
  polynomials of the first kind. We try successively higher orders of
  Chebyshev polynomials in order to arrive at a good fit with the
  data.  As we discuss below, we find that $\phi_*$, $M_*$ and
  $\alpha$ prefer quadratic, cubic, and linear evolutions in
  $\left(1+z\right)$, respectively.  For the faint-end slope $\beta$
  we adopt a double power law to account for a possible break at
  $z\sim 3$ that is currently not covered with credible data with
  well-characterised selection function.  Thus we write
  \begin{equation}
    F_3\left(1+z\right)=c_{3,0}+\frac{c_{3,1}}{10^{c_{3,3}\zeta}+10^{c_{3,4}\zeta}},
    \label{eqn:beta}
  \end{equation}
  where
  \begin{equation}
    \zeta = \log_{10}\left(\frac{1+z}{1+c_{3,2}}\right),
  \end{equation}
  thus resulting in a five-parameter model with parameters $c_{3,i}$.
  The parameters $c_{3,3}$ and $c_{3,4}$ thus determine the low and
  high redshift slopes of this evolution, with a break at redshift
  $c_{3,2}$.  This is similar to the model of
  \citet{2007ApJ...654..731H}, who also favoured a broken power law
  model for the evolution of the faint-end slope of the bolometric
  luminosity function of quasars.  Model 1 thus has 14 parameters.
  Excluding those deemed to be dominated by systematic errors, as
  discussed in the previous section, all of the remaining AGN from
  Table~\ref{tab:samples} are included while fitting this model.

\item Model 2 parameterises the luminosity function evolution in the
  same way as Model 1, so that the faint-end slope evolution is
  described by a double power law while the evolution of the other
  parameters $\phi_*$, $M_*$ and $\alpha$ is described by,
  respectively, quadratic, cubic, and linear polynomials in $(1+z)$.
  The total number of parameters is 14.  However, while fitting this
  model, we exclude the samples from \citet{2015AA...578A..83G},
  \citet{2011Natur.474..616M}, \citet{2015ApJ...801L..11V}, and
  \citet{2018Natur.553..473B} from the analysis (samples 17--20 from
  Table~\ref{tab:samples}).  As discussed in
  Section~\ref{sect:samplesel}, these samples have approximate
  selection functions.  Removing them allows us to understand the
  effect this has on the favoured evolution model.

\item In Model 3, we again exclude samples from
  \citet{2015AA...578A..83G}, \citet{2011Natur.474..616M},
  \citet{2015ApJ...801L..11V}, and \citet{2018Natur.553..473B} from
  the analysis (samples 17--20 from Table~\ref{tab:samples}).  We also
  continue to describe the evolution of $\phi_*$, $M_*$ and $\alpha$
  by quadratic, cubic, and linear polynomials in $(1+z)$,
  respectively.  But in this model, the evolution of the faint-end
  slope $\beta$ is also assumed to be linear in $(1+z)$.  This model
  thus has just 11 parameters.
\end{itemize}

\begin{figure*}
  \begin{center}
    \includegraphics[width=\textwidth,keepaspectratio]{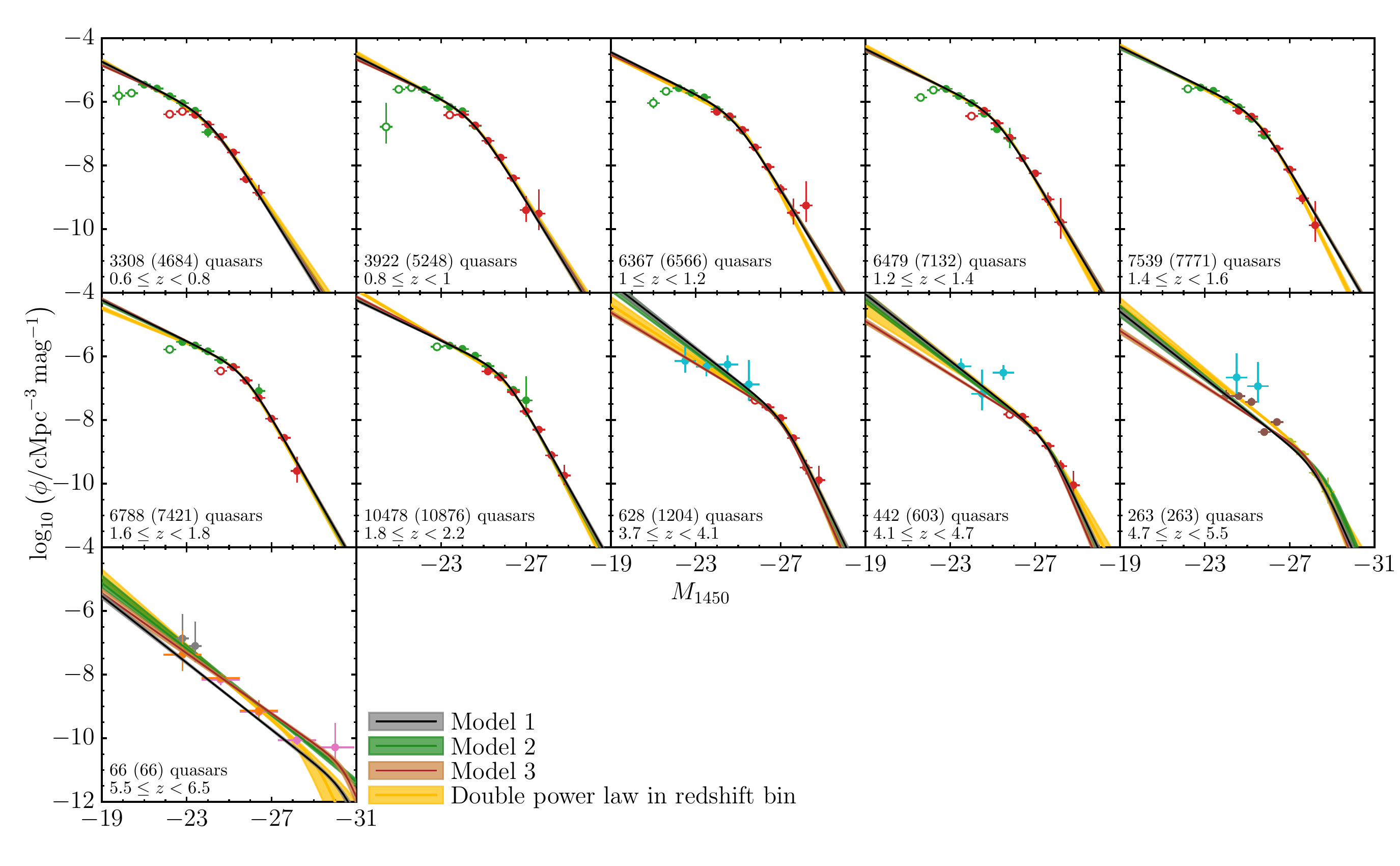}
    % mosaic_selected.py
  \end{center}
  \caption{Luminosity function estimates from $z=0.6$ to $6.5$.
    Similar to Figure~\ref{fig:mosaic}, the symbols show our inferred
    binned luminosity functions.  In each redshift bin, yellow curves
    show our fiducial double power law luminosity function model in
    that redshift bin.  Other curves show the three global evolution
    models.  Shaded regions show the one-sigma (68.26\%)
    uncertainties.}
  \label{fig:mosaic_global}
\end{figure*}

\begin{figure}
  \begin{center}
    \includegraphics[width=\columnwidth,keepaspectratio]{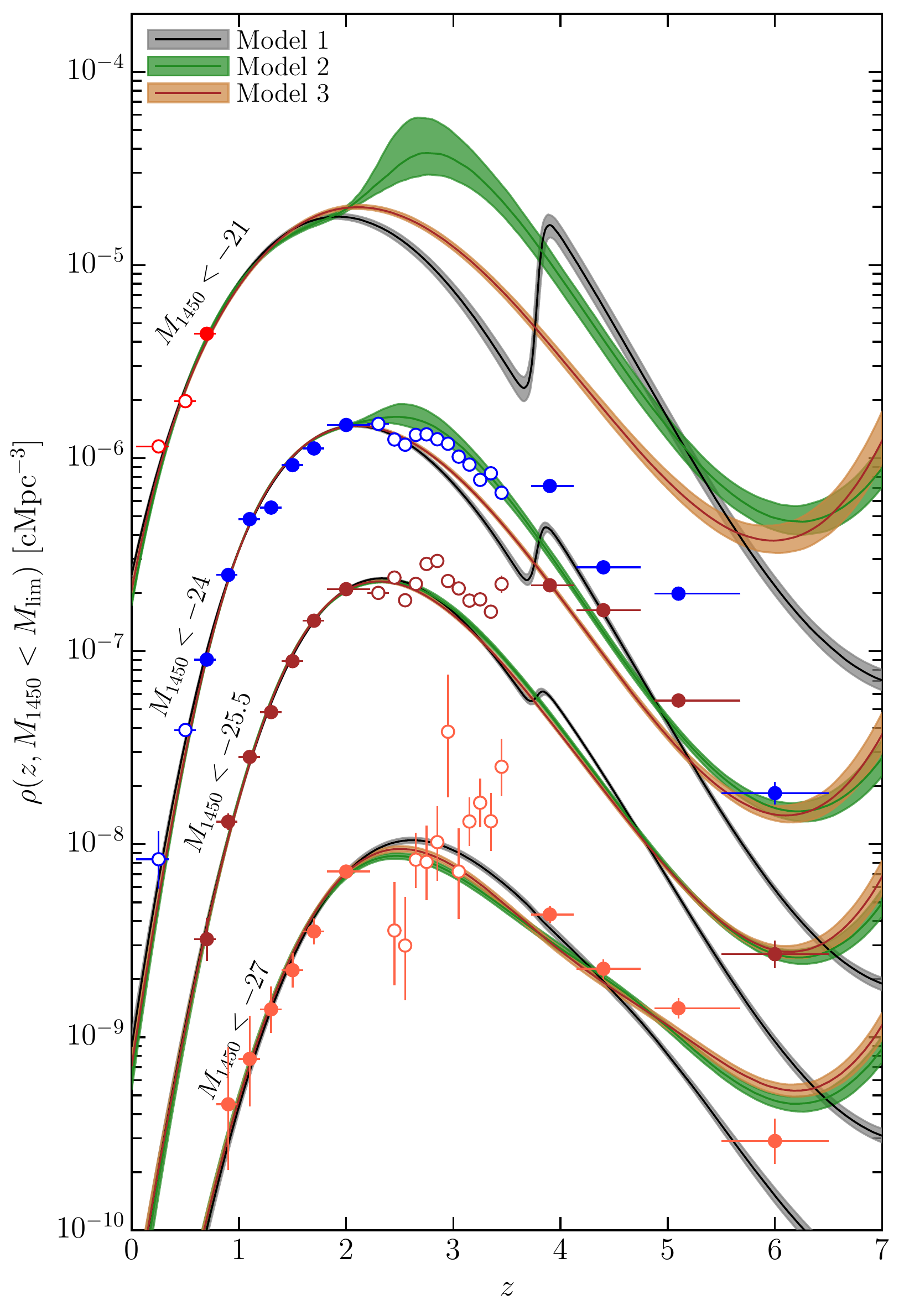}
    % rhoqso_data.plot_using_npz
  \end{center}
  \caption{AGN number density evolution in the global model, when the
    luminosity function is integrated (from top to bottom) down to
    $M_\mathrm{1450}=-21, -24, -25.5$ and $-27$.  Points show the
    observed number density evolution, with vertical error bars
    denoting one-sigma (68.26\%) uncertainties.  Open circles show the
    same in redshifts bins affected by systematics.  Curves show the
    estimated density from the global models, with the accompanying
    shaded regions denoting the one-sigma uncertainty. }
  \label{fig:rhoqso}
\end{figure}

Figure~\ref{fig:mosaic_global} shows the three global models in
comparison with the binned fits from the previous section.  The shaded
regions show the one-sigma (68.26\%) uncertainty.  The symbols show
the luminosity function binned in luminosity and redshift, and the
yellow shaded regions show the posterior distribution of the double
power law luminosity functions in various redshift bins, as in
Figure~\ref{fig:mosaic}.  Bins containing data with large systematic
error are excluded from Figure~\ref{fig:mosaic_global}.  All three
global models are in good agreement with the luminosity functions in
redshift bins, although Model~1 reproduces best the apparent rapid
evolution in the faint-end slope $\beta$ at $z>4$.  Given the
systematic errors in the data, a quantitative model selection is of
limited use.

Figure~\ref{fig:evoln_global} shows the parameter evolution in the
global models, by comparing it with the results from the fits in
individual redshift bins shown in Figure~\ref{fig:evoln}.  All three
models capture the steepening of the faint end of the luminosity
function towards higher redshifts.  The derived form of Model~1 is
shown by the black curves in Figure~\ref{fig:evoln_global}.  The
accompanying grey shaded area depicts the one-sigma (68.26\%)
uncertainty.  The model is in excellent agreement with the results of
the fits in redshift bins discussed in the previous section.
The deviation of the BOSS quasars at $z=2$--$4$ from the smooth
evolution is again strikingly apparent, as is the deviation of the
SDSS and 2SLAQ quasars at $z<0.6$.  This is an indirect justification
for the data selection discussion previously in
Section~\ref{sec:bins}.  Unfortunately, this model suffers from a
remarkably sharp break in the evolution of the faint-end slope $\beta$
at about $z\sim 3.5$.  As seen in Figure~\ref{fig:evoln_global}, the
data seem to require this break, although it seems unlikely that such
a sharp break at this redshift would be physical.  Model 1 thus serves
to emphasize the necessity of better quality data at these redshifts.
Models 2 and 3 are shown in Figure~\ref{fig:evoln_global} by the green
and orange curves, respectively.  The corresponding parameter values
are tabulated in Table~\ref{tab:global}.

The evolution of the comoving number density of quasars is shown in
Figure~\ref{fig:rhoqso} when the luminosity function is integrated
down to different limits.  Symbols show the number densities derived
from the data.  Solid curves and shaded regions show the global
models.  This number density evolution again highlights the systematic
error in data at $z\sim 3$.  The number density of AGN down to the
limit of the deepest spectroscopic surveys ($M_{1450}<-21$) is about
few times $10^{-6}$ cMpc$^{-3}$ at its peak.  At higher redshifts,
data go down to only about $M_{1450}=-24$. This density rapidly
increases with redshift at low redshifts and then drops with redshift
gradually at high redshifts.  Figure~\ref{fig:rhoqso} also shows the
familiar downsizing feature in which the number density of faint AGN
peaks at lower redshifts than that of the bright AGN
\citep{2004ApJ...605..625H, 2005MNRAS.360L..39N, 2006AJ....131.2766R,
  2006A&A...451..443M, 2009A&A...493...55E, 2013A&A...558A..89K,
  2010MNRAS.401.2531A, 2015MNRAS.451.1892A}.  While the number density
of AGN with $M_*<-27$ peaks at $z\sim 2.5$, the number density of AGN
with $M_*<-24$ peaks at $z\sim 2$.  However, the difference between
our three models is dramatically evident in Figure~\ref{fig:rhoqso}.
Model~1 prefers a decrease in the number density of faint quasars at
$z\sim 3$ followed by an increase at higher redshift.  This is caused
by the rapid steepening of the faint-end slope in this model at this
redshift (Figure~\ref{fig:evoln_global}).  Figure~\ref{fig:rhoqso}
reveals another property of these models: when extrapolated, the AGN
number density diverges in all three models at high redshifts.  This
results from the steep faint-end slope at high redshifts combined with
the rapid brightening of the break luminosity.  While no data exist at
redshift $z>7.5$, this divergent behaviour is shared by previous
models in the literature \citep{2007ApJ...654..731H}.

\begin{table}
  \caption{Derived luminosity function evolution models.  These
    parameters are defined in Equations~(\ref{eqn:cbs}) and
    (\ref{eqn:beta}).  See Section~\ref{sec:global} for further
    details and the redshift range of validity of these models.
    Errors indicate one-sigma (68.26\%) uncertainties. }
  \label{tab:global}
  \begin{tabular}{p{1.0cm}eee}
    \hline 
    Param. &
    \multicolumn{1}{c}{Model 1} &
    \multicolumn{1}{c}{Model 2} &
    \multicolumn{1}{c}{Model 3} \\
    \hline 
    $c_{0,0}$&-7.798^{+0.145}_{-0.157} &   -7.432^{+0.192}_{-0.211} &   -6.942^{+0.086}_{-0.086} \\
    $c_{0,1}$&1.128^{+0.085}_{-0.081} &    0.953^{+0.122}_{-0.107} &    0.629^{+0.046}_{-0.045} \\
    $c_{0,2}$&-0.120^{+0.005}_{-0.006} &   -0.112^{+0.007}_{-0.008} &   -0.086^{+0.003}_{-0.003} \\ 
    \\      
    $c_{1,0}$&-17.163^{+0.219}_{-0.226} &  -15.412^{+0.288}_{-0.318} &  -15.038^{+0.156}_{-0.150} \\
    $c_{1,1}$&-5.512^{+0.127}_{-0.124} &   -6.869^{+0.183}_{-0.169} &   -7.046^{+0.100}_{-0.101} \\
    $c_{1,2}$&0.593^{+0.011}_{-0.010} &    0.778^{+0.016}_{-0.015} &    0.772^{+0.013}_{-0.013} \\
    $c_{1,3}$&-0.024^{+0.00035}_{-0.00039} &   -0.032^{+0.001}_{-0.001} &   -0.030^{+0.001}_{-0.001} \\ 
    \\       
    $c_{2,0}$&-3.223^{+0.127}_{-0.121} &   -2.959^{+0.127}_{-0.143} &   -2.888^{+0.097}_{-0.093} \\
    $c_{2,1}$&-0.258^{+0.047}_{-0.051} &   -0.351^{+0.054}_{-0.052} &   -0.383^{+0.039}_{-0.041} \\
    \\       
    $c_{3,0}$&-2.312^{+0.034}_{-0.032} &   -2.264^{+0.038}_{-0.036} &   -1.602^{+0.029}_{-0.028} \\
    $c_{3,1}$&0.559^{+0.049}_{-0.045} &    0.530^{+0.054}_{-0.049} &    -0.082^{+0.009}_{-0.009} \\
    $c_{3,2}$&3.773^{+0.017}_{-0.016} &    2.379^{+0.118}_{-0.085} &  \multicolumn{1}{c}{---} \\        
    $c_{3,3}$&141.884^{+31.521}_{-8.832} & 12.527^{+7.349}_{-3.618} & \multicolumn{1}{c}{---} \\        
    $c_{3,4}$&-0.171^{+0.101}_{-0.116} &   -0.229^{+0.135}_{-0.150} & \multicolumn{1}{c}{---} \\
    \hline
  \end{tabular}
\end{table}

\begin{figure*}
  \begin{center}
    \begin{tabular}{cc}
      \includegraphics[width=0.47\textwidth]{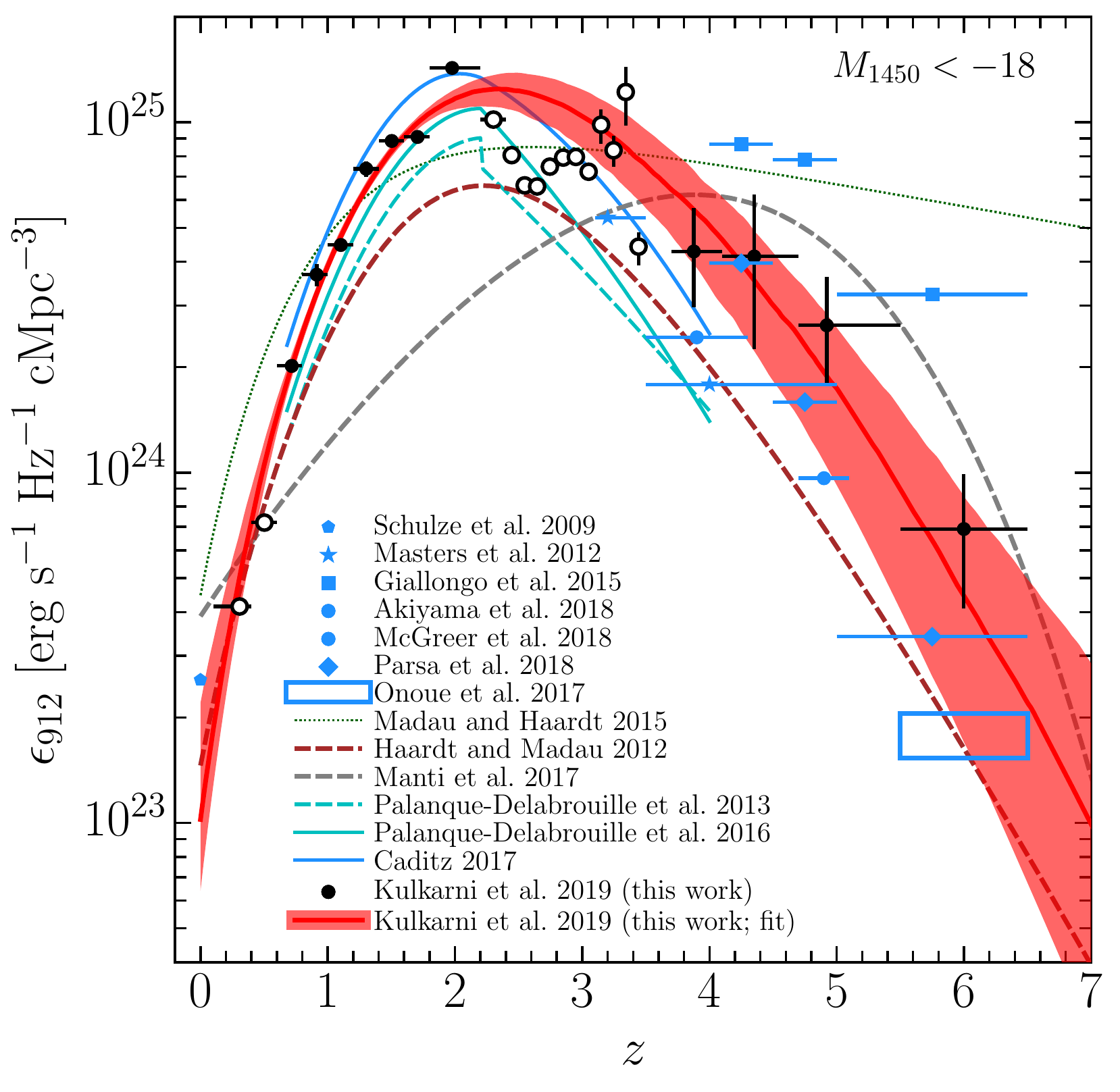} & 
      \includegraphics[width=0.47\textwidth]{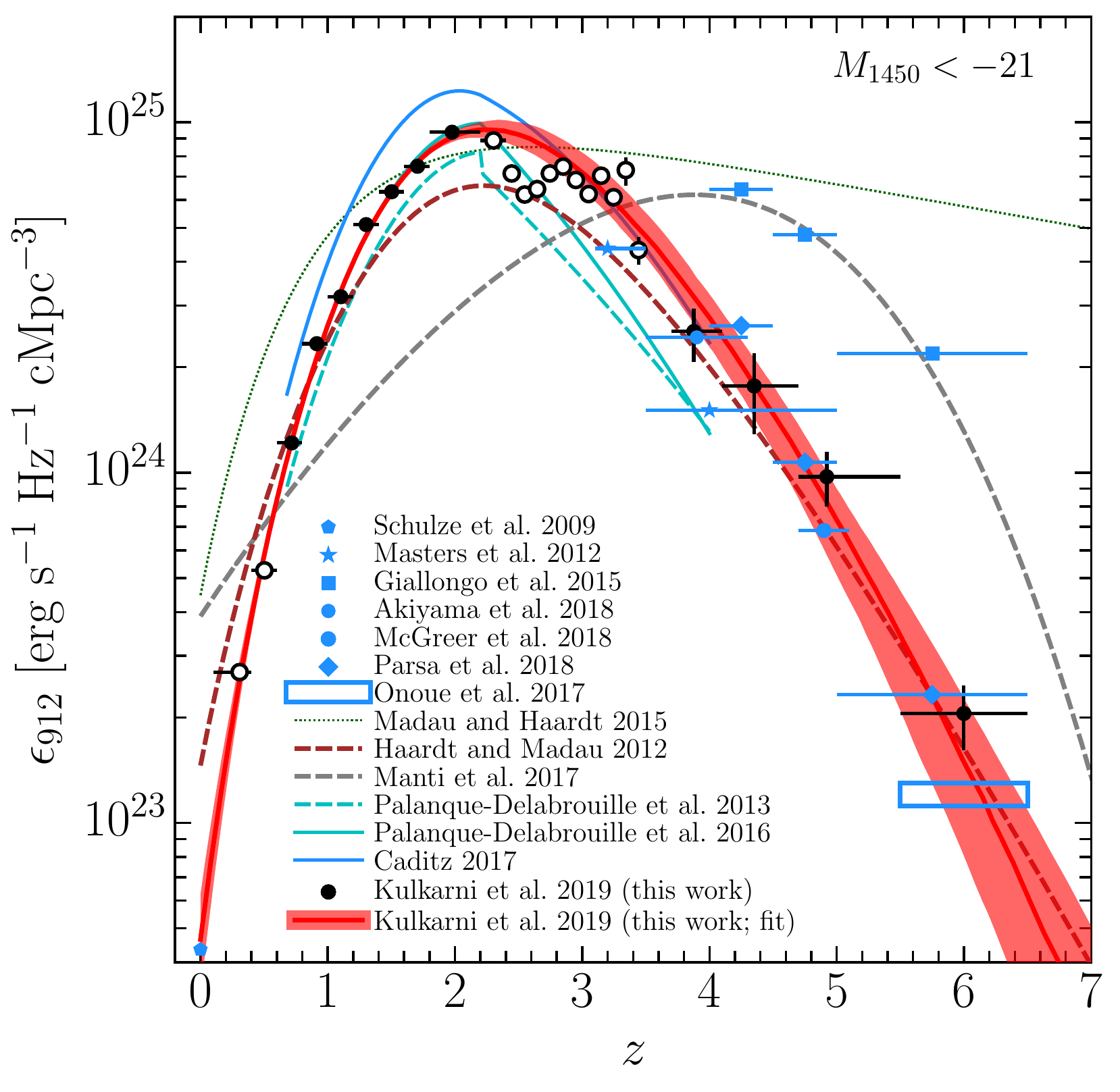} \\
      % gammapi.py -- draw_emissivity_18()
      % gammapi.py -- draw_emissivity_21()
    \end{tabular}
  \end{center}
  \caption{The 912~\AA\ emissivity of AGN down to a limiting magnitude
    for a luminosity function integration limit of $M_{1450}=-18$
    (left panel) and $M_{1450}=-21$ (right panel).  Black filled
    circles with one-sigma (68.26\%) error bars in both panels show
    the emissivity determinations in redshift bin deemed to have low
    systematic errors.  Open circles show emissivities for redshift
    bins that we remove from analysis due to high systematic errors.
    Solid red curves in both panels show the derived posterior median
    emissivity evolution model, with the shaded area showing the
    one-sigma (68.26\%) uncertainty.  Also shown for comparison in
    both panels are models by \citet[pentagon
      symbol]{2009A&A...507..781S}, \citet[star]{2012ApJ...755..169M},
    \citet[square]{2015AA...578A..83G},
    \citet[triangle]{2018PASJ...70S..34A},
    \citet[circle]{2018AJ....155..131M},
    \citet[diamond]{2018MNRAS.474.2904P}, \citet[open
      rectangle]{2017ApJ...847L..15O}, \citet[dotted green
      curve]{2015ApJ...813L...8M}, \citet[dashed
      brown]{2012ApJ...746..125H}, \citet[dashed
      grey]{2017MNRAS.466.1160M}, \citet[dashed
      blue]{2013A&A...551A..29P}, \citet[solid
      cyan]{2016A&A...587A..41P}, and \citet[solid
      blue]{2017A&A...608A..64C}.}
  \label{fig:e912_2}
\end{figure*}

\section{The AGN contribution to reionization}
\label{sec:reion}

We now discuss the contribution of AGN to the hydrogen and helium
reionization in our luminosity function model.  We first derive the
redshift evolution of the 912\,\AA\ emissivity of AGN, and then use
this to estimate the contribution of AGN to the \ion{H}{i}
photoionization rate between $z=0$ and $z=7$, as well as the redshift
evolution of the average \ion{He}{iii} fraction in the IGM for a
quasar-driven \ion{He}{ii} reionization.

\subsection{Quasar emissivity at the hydrogen Lyman limit}
\label{sec:e912}

For simplicity we assume that all quasars have a universal UV SED,
parameterised as a power law $f_\nu\propto\nu^{\alpha_\nu}$
with a break at 912\,\AA,
\begin{equation}
  f_\nu\propto\begin{cases}
  \nu^{-0.61} & \text{if}~\lambda\ge 912~\text{\AA},\\
  \nu^{-1.70} & \text{if}~\lambda<912~\text{\AA}                
  \end{cases}
  \label{eqn:sed}
\end{equation}
as derived from a stacked spectrum of 53 luminous ($M_{1450}\simeq
-27$) $z\simeq 2.4$ quasars \citep{2015MNRAS.449.4204L}, and
consistent with recent composite spectra of low-$z$ quasars with a
wide range in luminosity \citep{2012ApJ...752..162S,
  2014ApJ...794...75S}.  The extreme UV SED of faint ($M_{1100}\approx
-22$) AGN may be significantly harder
\citep[$\alpha_\nu=-0.56$,][]{2004ApJ...615..135S}, but the exact
value of the slope critically depends on the total rest-frame
wavelength coverage, the adopted continuum windows, the correction for
IGM line blanketing, and the ability to distinguish
low-equivalent-width emission lines from the underlying continuum
\citep{2014ApJ...794...75S,2015MNRAS.449.4204L,2016ApJ...817...56T}.
For the computation of the hydrogen Lyman limit emissivity the choice
of the spectral index at $\lambda<912$\,\AA\ is inconsequential,
whereas for the calculation of the \ion{H}{i} photoionization rate in
the IGM and the \ion{He}{ii} reionization history other sources of
uncertainty dominate (see below).  We note that all composite AGN
spectra are consistent with an AGN Lyman limit escape fraction of
unity (see also \citealt{2018A&A...613A..44G} for a recent sample of
faint $z\sim 4$ AGN).  We therefore adopt an escape fraction of unity
for Lyman continuum photons.

With our adopted SED the specific comoving volume emissivity of
quasars at 1450\,\AA\ can be written as
\begin{equation}
\epsilon_{1450}\left(z\right)=\int\limits_{-\infty}^{M_{1450}^\mathrm{lim}}\mathrm{d}M_{1450}\,\phi\left(M_{1450},z\right)10^{-0.4\left(M_{1450}-51.60\right)},
\label{eqn:epsilon}
\end{equation}
which depends on the adopted magnitude limit $M_{1450}^\mathrm{lim}$
and on the faint-end slope of the QLF if $M_{1450}^\mathrm{lim}\gg
M_*$.  The 912\,\AA\ emissivity is then given by
\begin{equation}
  \epsilon_{912}=\epsilon_{1450}\times\left(\frac{912}{1450}\right)^{0.61}.
  \label{eqn:epsilon_912}
\end{equation}
The black circles in Figure~\ref{fig:e912_2} show the comoving
912\,\AA\ emissivity obtained from the individual QLF fits
(Table~\ref{tab:bins}) for $M_{1450}<-18$ (left panel) and
$M_{1450}<-21$ (right panel), respectively. Redshift bins that were
removed from the analysis due to large systematic errors in the QLF
parameters are shown as open circles, with the error bars representing
68.26\% equal-tailed credible intervals.  Our derived emissivity
values at 912\,\AA\ and at 1450\,\AA\ are listed in
Table~\ref{tab:emissivity_bins} for reference.  The emissivity peaks
between $z=2$ and 3 at $\epsilon_{912}\simeq 10^{25}\,\mathrm{erg\,
  s^{-1}\, Hz^{-1}\, cMpc^{-3}}$ depending on the magnitude limit, and
decreases rapidly towards lower and higher redshifts.  Systematic
errors in the faint-end slope derived from BOSS data are more
pronounced for $M_{1450}<-18$ due to extrapolation of the QLF.

To account for redshift effects in the calculation of the \ion{H}{i}
photoionization rate, a continuous function
$\epsilon_{912}\left(z\right)$ is required. As our parametric QLF
models from the previous section suffer from non-monotonic or
divergent AGN number densities, we do not use them to derive the
corresponding $\epsilon_{912}\left(z\right)$, but instead fit the
individual emissivity values derived from credible data with a
five-parameter functional form used by \citet{2012ApJ...746..125H}
\begin{equation}
  \epsilon_{912}\left(z\right)=\epsilon_0(1+z)^a\frac{\exp(-bz)}{\exp(cz)+d}\quad,
  \label{eqn:e912fit}
\end{equation}
assuming a Gaussian likelihood for the emissivity values.  For
$M_{1450}<-18$ we obtain
\begin{multline}
  \epsilon_{1450}\left(z\right)=\left(10^{24.72}\mathrm{erg\, s^{-1}\, Hz^{-1}\, cMpc^{-3}}\right)\left(1+z\right)^{8.42}\\\times\frac{\exp(-2.1z)}{\exp(1.09z)+38.56}\quad,
  \label{eqn:e912_18}
\end{multline}
 and for $M_{1450}<-21$ we get
 \begin{multline}
  \epsilon_{1450}\left(z\right)=\left(10^{23.91}\mathrm{erg\, s^{-1}\, Hz^{-1}\, cMpc^{-3}}\right)\left(1+z\right)^{8.26}\\\times\frac{\exp(-1.3z)}{\exp(1.62z)+13.6}\quad,
  \label{eqn:e912_21}
\end{multline}
where the parameter values are median values of marginalised posterior
distributions.  The 912\,\AA\ emissivity is then given by
Equation~(\ref{eqn:epsilon_912}).  The resultant curves and the
corresponding one-sigma uncertainties are shown in
Figure~\ref{fig:e912_2}.  Table~\ref{tab:gamma2} provides the derived
emissivities at 912\,\AA\ and 1450\,\AA\ together with their derived
errors at $0<z<15$, extrapolating at $z<0.6$ and at $z>6.5$.

\subsection{Comparison to the literature}

Before proceeding to derive estimates of the quasar contribution to
the IGM \ion{H}{i} photoionization rate from the fitted
912\,\AA\ emissivity, it is instructive to compare our results to
recent estimates from the literature\footnote{We do not compare to
  \citet{2017MNRAS.466.1160M} due to an error in their analysis.
  Integration of the 912\,\AA\ emissivity to
  $M_{1450}^\mathrm{lim}=-19$ with their double power law QLF
  parameterization as a function of redshift yields values 13--75 per
  cent higher than those implied by their Equation 9.  This also
  explains the striking discrepancy to literature values at $z<3$
  (their Figure 4).}.  The various blue symbols in
Figure~\ref{fig:e912_2} show emissivity values that we have computed
from other recent QLF determinations in narrow redshift ranges.  The
various curves show emissivities derived from parametric QLF model
fits over larger redshift ranges, or fits to
$\epsilon_{912}\left(z\right)$. All QLFs using different magnitude
systems have been converted to $M_{1450}$ with the
\citet{2015MNRAS.449.4204L} SED, and all QLFs have been adjusted to
our cosmology.  For the $z\simeq 0$ QLF reported by
\citet{2009A&A...507..781S} we convert their $B_J$ magnitudes in the
Vega system to our AB magnitudes as $M_{1450,\mathrm{AB}}=M_{B_J,
  \mathrm{Vega}}+0.59$.  Lyman limit emissivities have been
consistently derived using Equation~\eqref{eqn:sed} and for our
adopted magnitude limits whenever possible.  We note that due to
strong covariance in the QLF parameters it is not straightforward to
compute statistical errors of $\epsilon$ from given QLF fits, and we
refrained from estimating them from the fits in the literature.  The
statistical errors of our binned emissivities are based on the
posterior of the QLF fits from unbinned data, and thus naturally
account for the covariance in the QLF parameters.  In contrast,
\citet{2015MNRAS.451L..30K} derived the majority of their values by
refitting binned QLFs with fixed QLF slopes, and they calculated
errors by propagating the errors in the QLF slopes, some of which they
had assumed or inflated from the cited literature.  These incorrect
procedures affect the emissivity values derived by
\citet{2015MNRAS.451L..30K}.

The brown dashed curve in Figure~\ref{fig:e912_2} shows the AGN
912\,\AA\ emissivity model adopted by \citet{2012ApJ...746..125H},
which is based on the bolometric luminosity function from
\citet{2007ApJ...654..731H}.  The bolometric emissivity derived by
\citet{2007ApJ...654..731H} converges for luminosities $L>0$ due to
their shallow faint-end QLF slope, which should yield a higher
912\,\AA\ emissivity for both our adopted magnitude limits. We
attribute much of the discrepancy to the conversion from bolometric to
912\,\AA\ emissivity assumed by \citet{2007ApJ...654..731H}.
Figure~\ref{fig:e912_2} also shows the emissivity curve from
\citet{2015ApJ...813L...8M} that was inspired by recent QLF fits
including the highly debated \citet{2015AA...578A..83G} results.  The
emissivity adopted by \citet{2015ApJ...813L...8M} exceeds our fits by
more than a factor of two at $z\la 1$ and $z\ga 4$. Moreover, we
stress that our emissivities have been derived for fixed magnitude
limits $M_{1450}^\mathrm{lim}$ at all redshifts
(Equation~\eqref{eqn:epsilon}), whereas other authors have adopted
magnitude limits
$M_{1450}^\mathrm{lim}\left(z\right)=M_*\left(z\right)+5$ that vary
with break magnitude and redshift
\citep{2015AA...578A..83G,2015ApJ...813L...8M,2015MNRAS.451L..30K,2019MNRAS.485...47P}.
We deem the latter convention to be physically unfounded, because (i)
the customary QLF double power-law parameterization thus far lacks a
deeper physical meaning, (ii) $M_*$ decreases by more than five
magnitudes with redshift (Figure~\ref{fig:evoln}, see also
\citealt{2013ApJ...768..105M} and \citealt{2016ApJ...829...33Y}), and
(iii) the various QLF fits at the same redshift are highly discordant
(Appendix~\ref{sec:qlfliterature}).  Inhomogeneous magnitude limits
lead to artificial scatter in the derived emissivities if the
faint-end slope of the QLF is not sufficiently shallow. Our fixed
magnitude limits bracket a reasonable range of QLF extrapolations
beyond the range covered by current data, whereas a varying limit
$M_*+5$ includes feeble $M_{1450}\simeq -18$ AGN at $z<0.6$, but
excludes verified $M_{1450}\simeq -24$ quasars at $z\simeq 6$
(Figure~\ref{fig:evoln}).

\citet{2013A&A...551A..29P,2016A&A...587A..41P} presented two
variability-selected quasar samples that are not included in our
analysis, and therefore provide a valuable cross-check.  In
Figure~\ref{fig:e912_2} we show the emissivities computed from their
parametric model fits to binned QLFs at
$0.68<z<4$. \citet{2013A&A...551A..29P} fitted pure luminosity
evolution models to binned QLFs from their data set and the one by
\citet{2009MNRAS.399.1755C}, but with a discontinuity in the QLF
slopes at $z=2.2$ that cause artificial discontinuities in the QSO
number density and the emissivity. The good agreement with our
inferences at $z<2.2$ is partially due to sample
overlap. \citet{2016A&A...587A..41P} presented an independent sample
of 13876 variability-selected quasars. They fitted their binned QLFs
with a pure luminosity evolution model at $z<2.2$ and a luminosity and
density evolution model at higher redshifts, imposing continuity at
$z=2.2$. The emissivity computed from their QLF is in good agreement
with our results at $z<2.2$, but is systematically lower at higher
redshifts. \citet{2017A&A...608A..64C} corrected an apparent error in
the bandpass correction applied by \citet{2016A&A...587A..41P} that
results in a higher QLF at $z>3$.  However, his higher inferred
$\phi_*$ at $z=0$ causes a 30--60 per cent higher emissivity at
$z<2.2$ compared to our inferences. Since neither
\citet{2013A&A...551A..29P,2016A&A...587A..41P} nor
\citet{2017A&A...608A..64C} fitted the QLF in narrow redshift bins
from unbinned quasar data, it remains unclear whether there is a
systematic difference between colour-selected and variability-selected
samples at $z<2.2$. As variability-selected samples to not probe the
faint end of the QLF at $z>3$, inferences of the high-redshift quasar
number density and emissivity are highly uncertain at present.

We also calculate the 912\,\AA\ emissivities from various
determinations of the UV QLF in narrow redshift ranges
\citep{2009A&A...507..781S,2012ApJ...755..169M,2015AA...578A..83G,2017ApJ...847L..15O,2018PASJ...70S..34A,2018AJ....155..131M,2018MNRAS.474.2904P}.
All these QLF determinations are not fully independent from ours due
to partial sample overlap, mostly from SDSS at the bright end. We use
both QLF determinations from \citet{2017ApJ...847L..15O} with and
without including a faint X-ray-selected $z\sim 6$ AGN candidate
\citep{2018MNRAS.474.2904P}, and indicate the emissivities with a box
in Figure~\ref{fig:e912_2}.  For $M_{1450}<-21$ our
$\epsilon_{912}\left(z\right)$ parameterization and the individual
values are generally consistent with those computed from other QLFs in
narrow redshift ranges, apart from the results by
\citet{2015AA...578A..83G} that have been disputed by several studies
\citep[][see Appendix~\ref{sec:conv} for further
  discussion]{2015MNRAS.453.1946G,2016MNRAS.463..348V,2017MNRAS.465.1915R,2018AJ....155..131M,2018MNRAS.474.2904P}.
As detailed in Section~\ref{sect:samplesel}, the
\citet{2012ApJ...755..169M} QLF at $z\sim 4$ -- and hence the derived
emissivity -- is underestimated due to systematic error in their
photometric redshifts, while their $z\sim 3.2$ results likely suffer
from systematic uncertainty in the SDSS selection function
\citep{2011ApJ...728...23W,2012ApJS..199....3R,2013ApJ...773...14R}.

Integration to $M_{1450}=-18$ results in larger discrepancies due to
extrapolation of the QLF with an uncertain faint-end slope.
\citet{2018PASJ...70S..34A} obtained a very flat faint-end slope
$\beta=-1.30\pm 0.05$ that is inconsistent with our determinations at
all redshifts. Since they selected only point sources, they might be
missing a significant fraction of AGN at $M_{1450}>-23.5$.  In
addition, only $4.6$ per cent of their AGN have spectroscopic
redshifts, such that their correction for contamination is very rough.
The large difference in the inferred $M_{1450}<-18$ emissivity at
$z\simeq 5$ with respect to \citet{2018AJ....155..131M} is due to
their shallower faint-end slope $\beta=-1.97\pm 0.09$ compared to our
result at this redshift ($\beta=-2.30^{+0.11}_{-0.08}$). We attribute
the difference in the faint-end slope to the fixed bright-end slope
$\alpha=-4$ in \citet{2018AJ....155..131M} that is larger than our
measurements at $z>3.5$ (Figure~\ref{fig:evoln}).  Similarly,
\citet{2017ApJ...847L..15O} fixed the bright-end slope to
$\alpha=-2.8$ which had been determined by \citet{2016ApJ...833..222J}
from a single power-law fit to the QLF.  Our
bright-end slopes are inconsistent with such high values at all
redshifts, indicating that single power-law fits to a limited range in
magnitude yield only approximate estimates of $\alpha$, biasing the
other derived QLF parameters, quasar number densities and
emissivities.  Full double-power law fits to unbinned data over a wide
magnitude range are required.

\begin{figure*}
  \begin{center}
    % rtg2.draw_g_paper() 
    \includegraphics[scale=0.65]{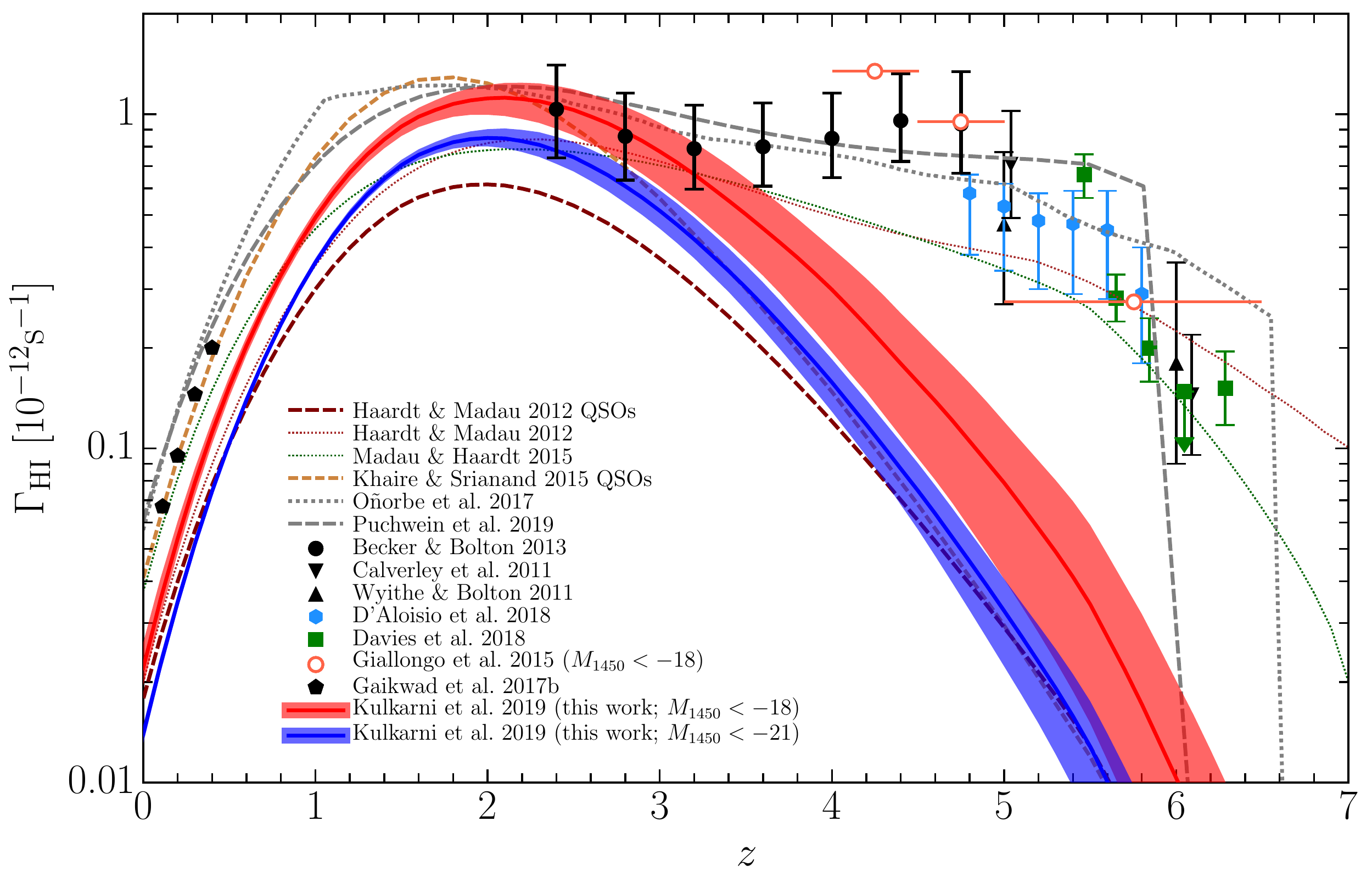}
  \end{center}
  \caption{AGN contribution to the hydrogen photoionization rate, when
    the AGN luminosity function is integrated down to $M_{1450}=-21$
    (blue curve and shaded region) and $M_{1450}=-18$ (red curve and
    shaded region).  The shaded regions show the one-sigma (68.26\%)
    uncertainty.  Also shown are the photoionization rate measurements
    by \citet[filled circles]{2013MNRAS.436.1023B}, \citet[inverted
      triangles]{2011MNRAS.412.2543C},
    \citet[triangles]{2011MNRAS.412.1926W}, \citet[blue
      hexagons]{2018MNRAS.473..560D} and \citet[green
      squares]{2018ApJ...855..106D}, and
    \citet[pentagons]{2017MNRAS.467.3172G}, and models of
    \citet[dotted brown curve]{2012ApJ...746..125H}, the QSO
    contribution in this model (dashed brown),
    \citet[dotted green]{2015ApJ...813L...8M}, the QSO
    contribution from the model of \citet[dashed
      orange]{2015MNRAS.451L..30K}, \citet[dotted
      grey]{2017ApJ...837..106O}, and \citet[dashed
      grey]{2019MNRAS.485...47P}.  The photoionization rate derived
    from the luminosity function fits of \citet{2015AA...578A..83G}
    are shown by the red open circles.}
  \label{fig:gammapi}
\end{figure*}

\subsection{Hydrogen photoionization rate}
\label{sec:gammahi}

Our calculation of the AGN contribution to the UV background follows
previous work on UV background synthesis models
\citep[e.g.][]{1996ApJ...461...20H,2012ApJ...746..125H}.  The main
quantity of interest is the \ion{H}{i} photoionization rate of the UV
background
\begin{equation}
  \Gamma_\ion{H}{i}\left(z\right)=\int_{\nu_{912}}^\infty\mathrm{d}\nu
  \frac{4\pi J_\nu\left(\nu,z\right)}{h\nu} \sigma_\ion{H}{i}\left(\nu\right)\quad,
  \label{eqn:gammapi}
\end{equation}
where $h$ is Planck's constant, $\sigma_\ion{H}{i}(\nu)$ is the
\ion{H}{i} photoionization cross-section \citep{1997MNRAS.292...27H,
  2011piim.book.....D}, and
\begin{multline}
  J_\nu(\nu, z)=\frac{c}{4\pi}\int_{z}^\infty\mathrm{d}z^\prime\frac{\left(1+z\right)^3}{H\left(z^\prime\right)\left(1+z^\prime\right)}\epsilon_\nu\left(\nu_\mathrm{em},z^\prime\right)\\
  \times\exp{\left[-\tau_\mathrm{eff}\left(\nu, z, z^\prime\right)\right]}
  \label{eqn:flux}
\end{multline}
is the angle- and space-averaged specific intensity of the UV
background. In the above equation
$H\left(z^\prime\right)=H_0(\Omega_\mathrm{m}\left(1+z^\prime\right)^3+\Omega_\Lambda)^{1/2}$
is the Hubble parameter, and
$\epsilon_\nu\left(\nu_\mathrm{em},z^\prime\right)$ is the comoving
emissivity of all \ion{H}{i} Lyman continuum sources at redshift
$z^\prime>z$ and emitted frequency
$\nu_\mathrm{em}=\nu\left(1+z^\prime\right)/\left(1+z\right)>\nu_{912}$.
In practice, we adopt an upper limit $z^\prime_\mathrm{max}=7$ when
integrating Equation~\eqref{eqn:flux}.  Due to the short mean free
path at high redshifts $z^\prime_\mathrm{max}=7$ is sufficient to give
converged result.  We do not require severe extrapolation of our
emissivity models to high redshifts.  Assuming that quasars are the
only ionizing sources and considering Equation~\eqref{eqn:sed} we have
\begin{equation}
  \epsilon_\nu\left(\nu_\mathrm{em},z^\prime\right) = \epsilon_{912}\left(z^\prime\right)\left(\frac{\nu_\mathrm{em}}{\nu_{912}}\right)^{-1.70}\quad,
  \label{eqn:epsilon_freq}
\end{equation}
with $\epsilon_{912}\left(z^\prime\right)$ obtained as described in
Section~\ref{sec:e912}.  For Poisson-distributed absorbers with an
\ion{H}{i} column density distribution
$f\left(N_\ion{H}{i},z^{\prime\prime}\right)=\partial^2n/\left(\partial
N_\ion{H}{i}\partial z^{\prime\prime}\right)$, the effective optical
depth to \ion{H}{i} Lyman continuum photons travelling between
redshifts $z^\prime$ and $z$ is \citep{1980ApJ...240..387P}
\begin{equation}
  \tau_\mathrm{eff}\left(\nu,z,z^\prime\right) = \int_{z}^{z^\prime}\mathrm{d}z^{\prime\prime}\int_0^\infty
  \mathrm{d}N_\ion{H}{i} f\left(N_\ion{H}{i},z^{\prime\prime}\right)\left[1-e^{-\tau_1}\right]\quad,
   \label{eqn:taueff}
\end{equation}
where $\tau_1\approx
N_\ion{H}{i}\sigma_\ion{H}{i}\nu\left(1+z^{\prime\prime}\right)/\left(1+z\right)$
is the Lyman continuum optical depth through an individual
absorber\footnote{The helium content of the absorber can be ignored
  due to the small \ion{H}{i} photoionization cross section at
  $\lambda<304$\,\AA.}.  For the \ion{H}{i} column density
distribution $f\left(N_\ion{H}{i},z\right)$ we adopt the piecewise
power-law parametrization by \citet{2012ApJ...746..125H} that is
consistent with $f\left(N_\ion{H}{i},z\right)$ measurements at
$z<3.5$, and roughly reproduces both the \ion{H}{i} Ly$\alpha$
effective optical depth \citep[but not in detail---see][]{2015MNRAS.450.4081P,2017MNRAS.464..897B,2017ApJ...837..106O}
and the measured mean free path to \ion{H}{i} Lyman limit photons at
$z<5.5$ \citep{2009ApJ...705L.113P,2014MNRAS.445.1745W}.
\citet{2019MNRAS.484.4174K} suggest that variations in
$f\left(N_\ion{H}{i},z\right)$ result in modest (10--40 per cent)
changes in $\Gamma_\ion{H}{i}$ at $z<3$. We note, however, that at
$z\ga 3.5$ all UV background synthesis models are based on brazen
extrapolations of $f\left(N_\ion{H}{i},z\right)$, whose detailed shape
for (partial) Lyman limit systems is not well constrained at these
redshifts \citep{2010ApJ...718..392P}.

Equations~\eqref{eqn:flux} and \eqref{eqn:taueff} assume that sources
and absorbers are uncorrelated, and that $J_\nu(\nu, z)$ is spatially
uniform, i.e.\ that the mean free path to \ion{H}{i} Lyman continuum
photons is much larger than the average distance between the sources
\citep[e.g.][]{1999ApJ...514..648M,2004MNRAS.350.1107M,2009ApJ...703.1416F,2012ApJ...746..125H}.
Obviously, these assumptions do not hold for rare sources and during
\ion{H}{i} reionization.  At $z\simeq 5$ our (extrapolated) AGN number
densities suggest an average distance of $\approx 70$\,cMpc between
$M_{1450}<-21$ AGN (Figure~\ref{fig:rhoqso}), which is comparable to
the mean free path \citep[$83\pm 10$\,cMpc,][]{2014MNRAS.445.1745W}.
Consequently, if only $M_{1450}<-21$ AGN contribute to the emissivity
then the UV radiation field at $z\simeq 5$ must fluctuate.  If the QLF
reaches to fainter magnitudes, or if star-forming galaxies contribute
to the emissivity, the UV background remains uniform to higher
redshifts, and the assumptions of standard UV background synthesis
models remain valid.

The red and blue curves in Figure~\ref{fig:gammapi} show the inferred
AGN \ion{H}{i} photoionization rate as a function of redshift for our
integration limits $M_{1450}=-18$ and $-21$, respectively. For
comparison we also plot the predictions of recent UV background
synthesis models for AGN
\citep{2012ApJ...746..125H,2015ApJ...813L...8M,2015MNRAS.451L..30K}
and AGN$+$galaxies \citep{2012ApJ...746..125H,2019MNRAS.485...47P}, as
well as inferences from the \ion{H}{i} Ly$\alpha$ forest
\citep{2011MNRAS.412.1926W,2013MNRAS.436.1023B,2017MNRAS.467.3172G,2018MNRAS.473..560D},
the Ly$\alpha+\beta$ forest \citep{2018ApJ...855..106D}, and the
quasar proximity effect \citep{2011MNRAS.412.2543C}.  Where necessary,
literature values have been rescaled by a few per cent to adjust to
our cosmology.

Differences in the \ion{H}{i} photoionization rates inferred from UV
background synthesis models mostly arise from obvious differences in
the emissivities of AGN and galaxies, but also due to differences in
the parametrization of the IGM and the AGN SED.  For an integration
limit of $M_{1450}=-21$ the AGN contribution to the hydrogen
photoionization rate falls short of 100\% across the redshift range.
It is marginally consistent with the measured photoionization rate at
$z=2.4$.  The photoionization rate in our model for $M_{1450}=-21$ has
the same evolution but a higher amplitude as the QSO contribution to
the \HI\ photoionization rate in the model of
\citet{2012ApJ...746..125H}.  For both of our integration limits, the
photoionization rate peaks at $z\sim 2$.  For the integration limit of
$M_{1450}=-18$, AGN can provide all the flux necessary to explain the
observed \lya forest between $z=2.4$ and $3.2$, with a contribution
from other sources necessary only at higher redshifts.  However, the
photoionization rate contributed by AGN falls short of the inference
of \citet{2017MNRAS.467.3172G} from the low-redshift ($z<0.6$)
\lya\ data.  We discuss this low-redshift evolution in greater detail
in the next section.

An important conclusion from Figure~\ref{fig:gammapi} is that the AGN
contribution to hydrogen reionization is likely subdominant, although
it can be non-negligible if faint AGN down to $M_{1450}=-18$ emit
hydrogen-ionizing photons with our assumed SED and a unit escape
fraction.  At $z=6.1$, AGN with $M_{1450}<-18$ contribute about $10\%$
of the required \HI\ ionizing flux.  The contribution of
$M_{1450}<-21$ at this redshift is $\sim 3\%$ relative to the
measurements \citep{2011MNRAS.412.2543C, 2018MNRAS.473..560D,
  2018ApJ...855..106D}.  At $z=6$, our determinations are lower than
those by \citet{2015AA...578A..83G} by almost an order of magnitude.
This difference arises from the difference in the inferred
emissivities in our models relative to \citet{2015AA...578A..83G}, as
discussed in the previous section.  The photoionization rate evolution
in the model of \citet{2015ApJ...813L...8M} agrees with our
determination at low redshifts ($z<0.5$) for $M_{1450}<-18$ but is
much higher at $z>4$, as expected from the higher emissivities assumed
by these authors.  At $3<z<6$ our model photoionization rates are
understandably lower than those in the models of
\citet{2017ApJ...837..106O} and \citet{2019MNRAS.485...47P} as these
authors include contribution to the photoionization rate from galaxies
in their models.  The differences in our model from that of
\citet{2015MNRAS.451L..30K} are a result of the inhomogeneous
redshift-dependent integration limits used by these authors and their
refitting of the results of \citet{2009MNRAS.392...19C} and
\citet{2013A&A...551A..29P}.

\subsection{Photon underproduction at $z=0$?}

It is instructive to closely examine if the corresponding hydrogen
photoionization rate is consistent with the \HI\ column density
distribution function measured from the \lya\ forest
\citep{2016ApJ...817..111D} at low redshifts ($z<0.5$).
\citet{2014ApJ...789L..32K} argued that in order to match the
\HI\ column density distribution function observed by
\citet{2016ApJ...817..111D} at these redshifts, hydrodynamical
cosmological simulations require a hydrogen photoionization rate that
is a factor of five larger than that in the UV background model of
\citet{2012ApJ...746..125H}.  Several recent studies have addressed
this `photon underproduction crisis' \citep{2015MNRAS.451L..30K,
  2015ApJ...811....3S, 2017MNRAS.467.3172G, 2017MNRAS.467.4802F,
  2017MNRAS.466..838G, 2017MNRAS.467L..86V}.  On the one hand, these
studies emphasised the uncertainty in the \citet{2012ApJ...746..125H}
UVB model at these redshifts due to the lack of certainty in the UV
photon emissivities of galaxies and AGN \citep{2015MNRAS.451L..30K,
  2015ApJ...811....3S}.  On the other hand, they noted the uncertainty
in the results of the cosmological simulations at these redshifts, due
to effects such as AGN feedback and limited numerical resolution
\citep{2015ApJ...811....3S, 2017MNRAS.467L..86V, 2017MNRAS.471.1056N,
  2017ApJ...837..106O, 2017MNRAS.466..838G, 2017MNRAS.467.3172G,
  2017ApJ...835..175G}.  A general conclusion of these studies was
that the discrepancy between the photoionization rate required by the
observed \HI\ column density distribution and predicted by the UVB
model of \citet{2012ApJ...746..125H} is likely to be smaller than that
found by \citet{2014ApJ...789L..32K}.

\begin{figure}
  \begin{center}
    % rtg2.draw_g_puc()
    \includegraphics[width=\columnwidth,keepaspectratio]{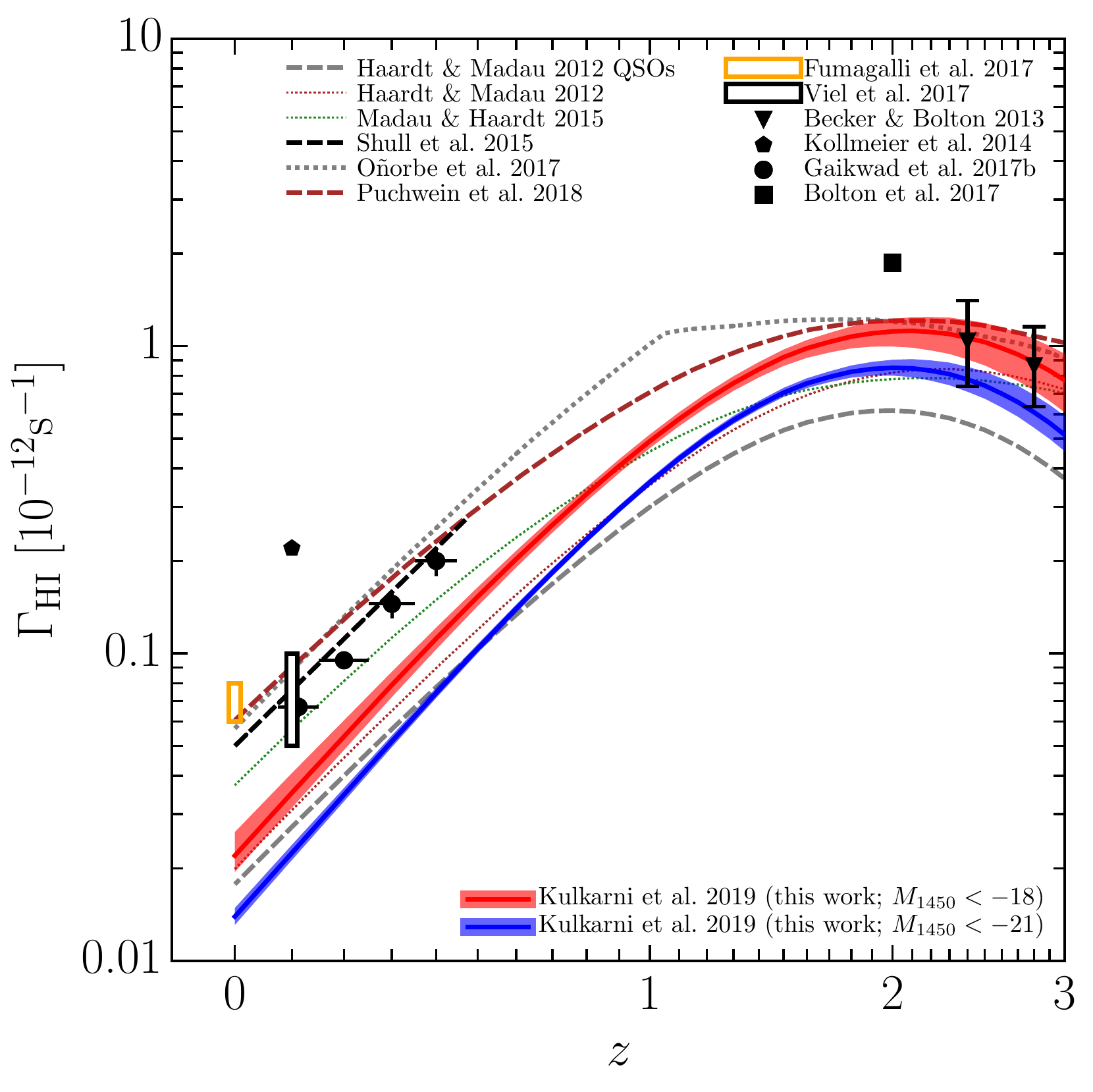}
  \end{center}
  \caption{Evolution of the hydrogen photoionization rate at low
    redshifts, when the AGN luminosity function is integrated down to
    $M_{1450}=-21$ (blue curve and shaded region) and $M_{1450}=-18$
    (red curve and shaded region).  Shaded regions show the one-sigma
    (68.26\%) uncertainty.  Also shown are the models and inferences
    from \citet[dotted brown curve]{2012ApJ...746..125H}, the QSO
    contribution in this model (dashed grey), \citet[dotted
      green]{2015ApJ...813L...8M}, \citet[dashed
      black]{2015ApJ...811....3S}, the QSO contribution from the model
    of \citet[dashed orange]{2015MNRAS.451L..30K}, \citet[dotted
      grey]{2017ApJ...837..106O}, \citet[dashed
      brown]{2019MNRAS.485...47P}, \citet[yellow
      box]{2017MNRAS.467.4802F}, \citet[black
      box]{2017MNRAS.467L..86V}, \citet[inverted
      triangle]{2013MNRAS.436.1023B},
    \citet[pentagon]{2014ApJ...789L..32K}, and
    \citet[circle]{2017MNRAS.467.3172G}. Note that we use the 
    \HI\ column density distribution model from
    \citet{2012ApJ...746..125H} to derive the photoionization rate.
    \label{fig:puc}}
\end{figure}

Figure~\ref{fig:puc} shows the evolution of the \HI\ photoionization
rate due to AGN in our model for luminosity function integration
limits of $M_{1450}=-21$ and $-18$ at redshifts $z<3$.  Note that we
use the 912\,\AA\ emissivity from Figure~\ref{fig:e912_2}, derived by
fitting the model from Equation~(\ref{eqn:e912fit}) to the
emissivities obtained from luminosity functions in various redshift
bins.  While doing this, as discussed above, redshift bins that were
interpreted as being affected by systematic errors were ignored.  As a
result, the emissivity model used in calculating the photoionization
rate is an extrapolation at $z<0.6$.  However, it is interesting to
note that the extrapolated emissivities at these redshifts are
consistent with the emissivities calculated from our luminosity
function fits to the $z<0.6$ data for either of our chosen integration
limits.  This can be seen in Figure~\ref{fig:e912_2}.  At $z\sim 0$,
the comoving 912\,\AA\ emissivity in our model is lower by almost a
factor of 2 than that derived from the luminosity function inferred at
this redshift by \citet{2009A&A...507..781S} for the $M_{1450}<-18$
case.  For the $M_{1450}<-21$ integration, our comoving emissivity is
higher than the emissivity in the model of \citet{2009A&A...507..781S}
by about 50\%.  This results from the very faint $M_*$ ($\sim -19$)
and steep faint-end slope ($\beta=-2$) obtained by
\citet{2009A&A...507..781S}.  (See Appendix~\ref{sec:qlfliterature}.)

We find that the low-redshift \HI\ photoionization rate in our model
is approximately equal to that in the \citet{2012ApJ...746..125H} UVB
model for both of our integration limits.  At $z=0.1$, this rate is
smaller by a factor of 2 than the photoionization rate derived by
\citet{2017MNRAS.467.3172G} from the \HI\ column density distribution
measurements of \citet{2016ApJ...817..111D}.  The model of
\citet{2015ApJ...813L...8M} also results in a higher ionisation rate
than our $M_{1450}<-18$ inference at $z<0.7$.
\citet{2015ApJ...811....3S} compared the \HI\ column density
distribution measurements to cosmological simulations with an enhanced
photoionization rate relative to the \citet{2012ApJ...746..125H} model
to find that $\Gamma_\mathrm{HI}=4.6\times
10^{-14}(1+z)^{4.4}\,\mathrm{s}^{-1}$ produces the observed
\HI\ column densities.  This photoionization rate was achieved in the
simulations of \citet{2015ApJ...811....3S} by a combination of quasars
and galaxies (with an escape fraction of hydrogen-ionizing photons
assumed to be $f_\mathrm{esc}=0.05$).  As seen in
Figure~\ref{fig:puc}, this photoionization rate is in closer agreement
with the \citet{2015ApJ...813L...8M} model.  The photoionization rate
estimate by \citet{2017MNRAS.467.4802F} from the H$\alpha$ surface
brightness of a $z\sim 0$ galaxy observed by VLT/MUSE is somewhat
higher than most other models shown in Figure~\ref{fig:puc}.  However,
it is possible that the \citet{2017MNRAS.467.4802F} estimate is an
upper limit, as the contribution of local sources to the
photoionization rate is neglected in their modelling.  The requirement
of an enhanced photoionization rate at $z\sim 0$ relative to our
inference is also confirmed by the simulations presented by
\citet{2017MNRAS.467L..86V}, \citet{2017ApJ...837..106O}, and
\citet{2019MNRAS.485...47P}.  But note that, as discussed above in
Section~\ref{sec:e912}, a direct comparison of our results with the
models of \citet{2015ApJ...813L...8M}, \citet{2015MNRAS.451L..30K},
and \citet{2019MNRAS.485...47P} is difficult because of the
inhomogenous redshift-dependent integration limits adopted by these
authors.

Before considering the deficit in the photoionization rate obtained
from AGN at low redshifts relative to measurements from the
\lya\ forest as a photon underproduction crisis, it is worthwhile to
recall various assumptions entering our derivation.  With a spectral
index of $-1.7$, our ionisation rate estimate falls short of the
\lya\ forest measurements by a factor of $\sim 2$ with an integration
limit of $M_{1450}=-18$ on the luminosity function.
\citet{2015ApJ...813L...8M} implicitly assume an integration limit of
$M_{1450}=-14$ at $z\sim 0$ (see also \citealt{2015MNRAS.451L..30K,
  2019MNRAS.485...47P}) as this limit corresponds to $M_*+5$ for the
\citet{2009A&A...507..781S} luminosity function measurement.  It is
unclear if the assumption of unit LyC escape fraction is valid for
such faint AGN, but we find that even with this extremely faint
integration limit, the photoionization rate deficit is reduced by only
about 25\% for a spectral index of $-1.7$.  While this result does not
change significantly if we assume a spectral index of $-1.4$ as
suggested by \citet{2014ApJ...794...75S}, assuming a much harder
spectrum with a spectral index of $-0.56$ \citep{2004ApJ...615..135S}
for faint ($M_{1450}>-23$) AGN completely alleviates the deficit for
an integration limit of $M_{1450}=-18$.  One should also note that our
ionization rate model ignores the large uncertainties in the
\HI\ column density distribution at the Lyman limit
\citep{2011ApJ...736...42R, 2017ApJ...849..106S, 2017MNRAS.466..838G}.
Alternatively, it may be possible to balance the ionization rate
deficit by the contribution from low-redshift LyC-leaking galaxies
\citep{2016Natur.529..178I, 2018MNRAS.474.4514I, 2018MNRAS.478.4851I}

\subsection{Helium reionization}

\begin{figure}
  \begin{center}
    % qhe.py 
    \includegraphics[width=\columnwidth,keepaspectratio]{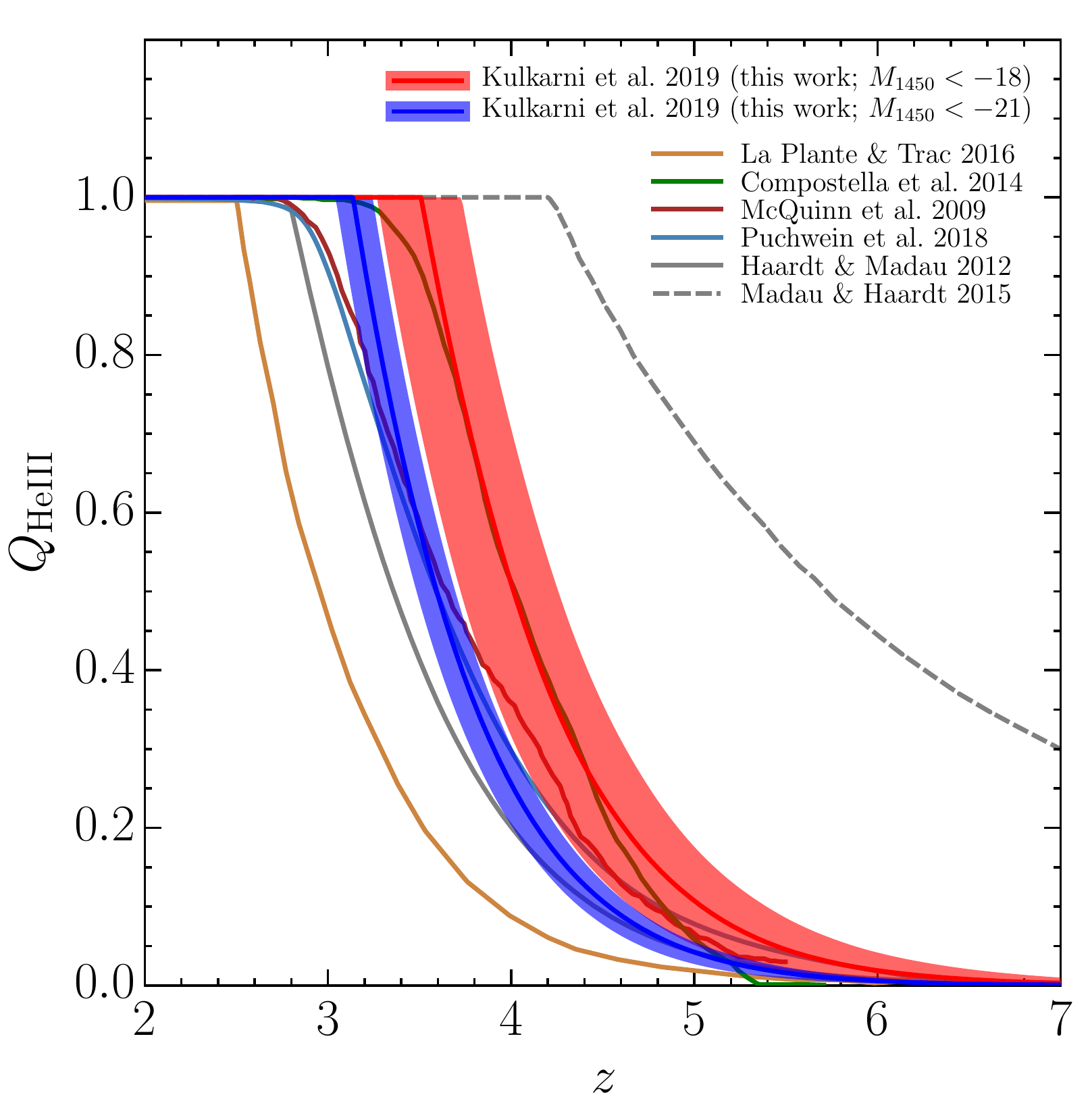}
  \end{center}
  \caption{Redshift evolution of the volume-averaged \ion{He}{iii}
    fraction $Q_\ion{He}{iii}$ considering AGN with $M_{1450}<-21$
    (blue) and $M_{1450}<-18$ (red) in our calculation of the emission
    rate of ionizing photons.  The shaded regions show the one-sigma
    (68.26\%) confidence interval resulting from the uncertainty in
    the emissivity.  The other curves show previous determinations of
    $Q_\ion{He}{iii}\left(z\right)$ based on
    Equation~\eqref{eqn:qHedot} \citep{2012ApJ...746..125H,
      2015ApJ...813L...8M, 2016ApJ...828...90L, 2019MNRAS.485...47P}
    and on cosmological radiative transfer simulations of
    quasar-driven \ion{He}{ii} reionization
    \citep{2009ApJ...694..842M, 2014MNRAS.445.4186C}.  Note that
    Equation~\eqref{eqn:qHedot} ignores the presence of \ion{He}{ii}
    Lyman limit systems impeding reionization at
    $Q_\ion{He}{iii}\rightarrow 1$, requiring further analytic
    modelling \citep{2017ApJ...851...50M} or numerical simulation
    \citep{2019MNRAS.485...47P}.
  \label{fig:qhe}}  
\end{figure}

We now consider the implications of our AGN luminosity function models
for \ion{He}{ii} reionization. The time evolution of the
volume-averaged \ion{He}{iii} fraction $Q_\ion{He}{iii}$ is given by
\citep[e.g.][]{2012ApJ...746..125H}
\begin{equation}
  \frac{\mathrm{d}Q_\ion{He}{iii}}{\mathrm{d}t}=\frac{\dot{n}_\mathrm{ion,4}}{\langle n_\mathrm{He}\rangle}-\frac{Q_\ion{He}{iii}}{\langle t_\mathrm{rec,He}\rangle}\quad,
  \label{eqn:qHedot}
\end{equation}
where $\dot n_\mathrm{ion,4}$ is the emission rate of $\ge 4$\,Ry
photons per unit proper volume, $\langle n_\mathrm{He}\rangle$ is the
average proper helium number density, and $\langle
t_\mathrm{rec,He}\rangle$ is the average recombination time scale for
\ion{He}{iii}.  With our quasar emissivity model from
Section~\ref{sec:e912} the emission rate can be written as
\begin{equation}
\dot
n_\mathrm{ion,4}\left(z\right)=-\frac{4^{\alpha_\nu}}{h\alpha_\nu}\left(1+z\right)^3\epsilon_{912}\left(z\right)\quad,
\end{equation}
with the quasar spectral index $\alpha_\nu=-1.7$ and the comoving
912\,\AA\ emissivity $\epsilon_{912}\left(z\right)$ is obtained from
the fits shown in Figure~\ref{fig:e912_2}.  The recombination time
scale in Equation~\eqref{eqn:qHedot} is given by
\begin{equation}
  \langle t_{\rm rec}\rangle=[(1+2\chi) \langle n_\mathrm{H}\rangle \alpha_\mathrm{B}\,C]^{-1}\quad,
  \label{eq:trec}
\end{equation}
where $\langle n_\mathrm{H}\rangle=1.881\times
10^{-7}(1+z)^3$~cm$^{-3}$ is the average proper hydrogen number
density, $\alpha_\mathrm{B}$ is the Case~B \ion{He}{iii} recombination
coefficient \citep{1997MNRAS.292...27H}, and $\chi=0.079$ is the
cosmic number fraction of helium for a cosmic helium mass fraction of
$Y_\mathrm{He}=0.24$.  We assume that the clumping factor $C$ for
helium is the same as for hydrogen, for which
\citet{2012ApJ...747..100S} obtained
\begin{equation}
  C = 2.9\left(\frac{1+z}{6}\right)^{-1.1}
\end{equation}
over the redshift range $5<z<9$. 

Equation~\eqref{eqn:qHedot} neglects \ion{He}{ii} Lyman limit systems
that considerably delay the end of \ion{He}{ii} reionization
\citep{2009MNRAS.395..736B, 2017ApJ...851...50M}.  In the absence of
these self-shielded systems, $Q_\ion{He}{iii}$ can continue to
increase beyond unity and the mean free path of \ion{He}{ii}-ionizing
photons diverges.  We set $Q_\ion{He}{iii}=1$ when this happens.  This
can be corrected by accounting for the \ion{He}{ii} column density
distribution and filtering the \ion{He}{ii}-ionizing radiation field
through it. Unfortunately, the \ion{He}{ii} column density
distribution is itself uncertain, as it depends on the relative
contributions of quasars and galaxies to the UV background
\citep[e.g.][]{2012ApJ...746..125H,2019MNRAS.485...47P}.  We consider
the simplified treatment of $Q_\ion{He}{iii}$ accurate enough for the
purpose of this work, while noting that the redshift of \ion{He}{ii}
reionization is likely overestimated in any such model.

The resultant \ion{He}{ii} reionization histories are shown in
Figure~\ref{fig:qhe} for our two considered magnitude limits, with the
shaded regions showing the one-sigma uncertainty in $Q_\ion{He}{iii}$
resulting from the uncertainty in the quasar emissivity alone,
i.e.\ using a fixed quasar spectral energy distribution and a fixed
redshift evolution of the clumping factor.  In our model, \ion{He}{ii}
reionization starts around the earliest quasars at $z>6$ and finishes
at $2.9\la z\la 3.7$ depending on the extent of the faint-end QLF.
Considering the limitations of the modelling discussed above, both
\ion{He}{ii} reionization histories are consistent with recent
\ion{He}{ii} Ly$\alpha$ effective optical depth measurements
supporting substantial progression of \ion{He}{ii} reionization by
$z\simeq 3.4$ \citep{2016ApJ...825..144W}, and an end of the process
at $z\simeq 2.7$ \citep{2011ApJ...733L..24W,2016ApJ...825..144W} with
the build-up of a quasi-homogeneous \ion{He}{ii}-ionizing background
\citep{2014MNRAS.437.1141D,2017MNRAS.465.2886D}.

Figure~\ref{fig:qhe} also shows previous solutions of
Equation~\eqref{eqn:qHedot} with different parameterizations of the
quasar emissivity, spectral energy distribution and clumping factor
\citep{2012ApJ...746..125H, 2015ApJ...813L...8M, 2016ApJ...828...90L,
  2019MNRAS.485...47P}, as well as the results from cosmological
radiative transfer simulations of \ion{He}{ii} reionization
\citep{2009ApJ...694..842M,2014MNRAS.445.4186C}.  The differences of
our results to the ones by \citet{2012ApJ...746..125H} and
\citet{2015ApJ...813L...8M} mostly result from differences in the
adopted quasar emissivity (Figure~\ref{fig:e912_2}).  The large quasar
emissivity adopted by \citet{2015ApJ...813L...8M} results in an early
completion of \ion{He}{ii} reionization at $z\approx 4$, which is
inconsistent with the measured strong \ion{He}{ii} absorption at
$2.7<z<3$ \citep{2016ApJ...825..144W, 2018MNRAS.473.1416M,
  2019MNRAS.485...47P} if the \ion{He}{ii}-ionizing background is not
fluctuating on large scales due to rare clustered and/or short-lived
quasars that result in a spatially varying mean free path of
\ion{He}{ii}-ionizing photons \citep{2010ApJ...714..355F,
  2014MNRAS.440.2406M, 2014MNRAS.437.1141D, 2017MNRAS.465.2886D}.  The
quasar emissivity considered by \citet{2012ApJ...746..125H} falls
below our determinations at the relevant redshifts
(Figure~\ref{fig:e912_2}), which results in a delayed completion of
\ion{He}{ii} reionization at $z\simeq 2.8$.

\citet{2016ApJ...828...90L} varied the parameters of
Equation~\eqref{eqn:qHedot} to estimate the uncertainty in
$Q_\ion{He}{iii}\left(z\right)$, in particular the
QLF\footnote{\citet{2016ApJ...828...90L} erroneously varied the
  $z>3.5$ QLF parameters independently, neglecting their covariance
  and the survey data.}.
Considering the limitations of Equation~\eqref{eqn:qHedot},
\ion{He}{ii} reionization finishes too late ($z\sim 2.5$) in their
fiducial model, which is probably due to their considered QLFs and
varying magnitude limits their emissivity calculations.
\citet{2019MNRAS.485...47P} adopted a somewhat higher quasar
emissivity than \citet{2012ApJ...746..125H} that is comparable to our
fit for $M_{1450}<-21$ at the redshifts of interest.  Instead of their
solution to Equation~\eqref{eqn:qHedot} we show their results of
one-cell simulations with gas at cosmic mean density (see Section~3.3
and Appendix~C of \citealt{2019MNRAS.485...47P} for details).  In
these simulations the \ion{He}{ii} number density is correctly
calculated as \ion{He}{ii} is gradually ionized by a radiation field
resulting from one-dimensional radiative transfer calculations through
an inhomogeneous IGM.  This accounts for \ion{He}{ii} Lyman limit
systems in a simple way, resulting in a smooth transition of
$Q_\ion{He}{iii}$ to unity.

In Figure~\ref{fig:qhe} we also plot $Q_\ion{He}{iii}\left(z\right)$
from two numerical simulations of quasar-driven \ion{He}{ii}
reionization \citep{2009ApJ...694..842M,2014MNRAS.445.4186C} that
broadly reproduce the measured \ion{He}{ii} effective optical depths
\citep{2016ApJ...825..144W}.  In these simulations the outputs of
$N$-body or hydrodynamic simulations are post-processed with radiative
transfer around quasars sourced according to a specific model, and the
effects of \ion{He}{ii} Lyman limit systems are approximately captured
with sub-grid filtering methods \citep{2009ApJ...694..842M} or with
adaptive mesh refinement \citep{2014MNRAS.445.4186C}.  Their different
timing of \ion{He}{ii} reionization is mostly due to the adopted
quasar model (QLF, spectral energy distribution, quasar lifetime,
anisotropic emission, halo mass range) and stochasticity in the $z\ga
3.5$ quasar number density in the limited simulation volumes.  Thus,
while current simulations do not reproduce the measured scatter in the
\ion{He}{ii} effective optical depths at $2.7<z<3.5$ in detail
\citep{2016ApJ...825..144W}, this may reflect limitations of the
modelling other than the QLF \citep{2017MNRAS.468.4691D}.  Our
$Q_\ion{He}{iii}\left(z\right)$ for $M_{1450}<-18$ agrees reasonably
well with the simulation by \citet{2014MNRAS.445.4186C}, in which
\ion{He}{ii} reionization is accomplished by the $z<5$ quasar
population evolved assuming pure density or pure luminosity evolution
to match the QLF by \citet{2011ApJ...728L..26G} for $M_{1450}<-19.5$
\citep[see also][]{2013MNRAS.435.3169C}.  Future efforts in numerical
modelling of \ion{He}{ii} reionization should investigate quasar
models which follow the QLF evolution in detail, and which are
consistent with the measured redshift evolution and scatter of the
\ion{He}{ii} effective optical depth, as well as the measured redshift
evolution of the IGM temperature-density relation
\citep[e.g.][]{2011MNRAS.410.1096B, 2014MNRAS.441.1916B,
  2018MNRAS.474.2871R, 2018ApJ...865...42H}.

\section{Summary and Conclusions}
\label{sec:conc}

We have analysed the evolution of the AGN UV luminosity function from
redshift $z=0$ to $7.5$ using a combined sample of 83,488 mostly
UV-optical colour-selected AGN from 12 data sets, homogenised with
respect to the assumed cosmology, magnitude system and bandpass
correction (if possible).  The vast majority of them (83,469 AGN from
11/12 samples) have spectroscopic redshifts, and for all but 22 AGN
the selection functions have been characterised.  After restricting
the sample due to persisting incompleteness at the faint end we arrive
at 77,659 spectroscopically confirmed $0<z<7.5$ AGN extending to
absolute AB magnitudes $M_{1450}\simeq -19$ at $z\simeq 0.5$ and to
$M_{1450}\sim -22$ at $z>3.5$, respectively.  To facilitate
comparisons to future data sets we make the homogenised AGN sample and
the homogenised selection functions publicly available.

Binning the $1450$\,\AA\ AGN luminosity function in narrow redshift
ranges we find that it is excellently described by the customary
double power law in magnitude at all redshifts
(Figure~\ref{fig:mosaic}).  Its four parameters significantly evolve
with redshift (Figure~\ref{fig:evoln}): (i) The break magnitude $M_*$
of the AGN luminosity function shows a steep brightening from
$M_*\simeq -24$ at $z\simeq 0.7$ to $M_*\sim -29$ at $z\simeq 6$.
(ii) The corresponding amplitude $\phi_*$ drops by a factor $\sim
20,000$ from $\phi_*\simeq 4\times 10^{-7}$~mag$^{-1}$cMpc$^{-3}$ at
$z<2.2$ to $\phi_*\sim 2\times 10^{-11}$~mag$^{-1}$cMpc$^{-3}$ at
$z\simeq 6$.  (iii) The faint-end slope $\beta$ significantly
decreases from $\simeq -1.7$ at $z<2.2$ to $\simeq -2.4$ at $z\simeq
6$, resulting in a steepening of the luminosity function.  (iv) The
bright-end slope $\alpha$ also shows a moderate decrease with
redshift, but is less constrained at the highest redshifts because of
the bright break magnitude.  In contrast to several previous studies,
our fits are based on unbinned homogenised AGN data and their
selection functions, and the confidence intervals fully account for
covariance in the luminosity function parameters.  From the continuity
of the luminosity function at lower redshifts we argue that its
apparent single power law description at $z\simeq 6$ can be
interpreted as the faint end of the double power law whose break
magnitude $M_*$ is at the bright end of the $z\simeq 6$ quasar
population.

Similar continuity arguments let us scrutinize the value of several
photometrically-selected AGN samples in determinations of the
luminosity function.  Our analysis has revealed systematic errors in
the survey selection functions caused by their fixed and simplified
assumptions, i.e.\ regarding the parameterisation of the IGM and the
AGN SED, and the treatment of photometric errors at the survey
magnitude limit.  Several of these systematic errors are easily
identified from artificial faint-end drops of the luminosity function
(Figure~\ref{fig:mosaic}), an apparently zero accessible volume for
discovered AGN, or a mismatch between the observed and the simulated
AGN color distribution.  In large samples the precision of the AGN
luminosity function is limited by unaccounted systematic errors in the
survey selection functions, which lead to (i) unphysical variations in
the luminosity function parameters (i.e.\ for BOSS in
Figure~\ref{fig:evoln}) and (ii) inter-survey systematics in combined
samples that are further amplified by heterogeneous selection function
parameter choices amongst the surveys.

With only partially credible data it is challenging to describe the
redshift evolution of the AGN luminosity function with a viable
parametric model.  We have developed three such models
(Figure~\ref{fig:evoln_global}), finding that our fourteen-parameter
Model 1 describes the redshift evolution of the luminosity function
rather well.  However, this model prefers a break in the faint-end
slope at $z\simeq 3.5$, which causes an unphysical discontinuity in
the cumulative AGN number density (Figure~\ref{fig:rhoqso}).  Our
other two models do not have these features, but they do not match the
measured faint-end slope at $z>4$ as well as Model 1 does.

With our determinations of the luminosity function we have revisited
the question of the contribution of AGN to reionization and the UV
background.  We have made the first determinations of the AGN
912\,\AA\ emissivity with homogeneous faint-end limits at all
redshifts, whose statistical uncertainties have been properly
calculated from the posterior distributions of the luminosity function
(Figure~\ref{fig:e912_2}). Our parametric fits yield a peak in the
912\,\AA\ emissivity at $z\approx 2.4$, with a decline by $\simeq 2$
orders of magnitude to $z\simeq 0$ and $z\simeq 7$, respectively. At
$z>4$ our determined emissivities are lower by a factor 2--10 than
recently claimed by \citet{2015AA...578A..83G}. At the same time, the
912\,\AA\ emissivity of $M_{1450}<-18$ AGN exceeds the model
considered by \citet{2012ApJ...746..125H} by a factor $>2$ at $z>3$,
suggesting a somewhat higher contribution of AGN to the high-$z$ UV
background.

Having derived the AGN \ion{H}{i} photoionization rate by filtering
the \ion{H}{i}-ionizing AGN emissivity through the IGM \ion{H}{i}
column density distribution, we find that while at $z=2$--$3$
$M_{1450}<-18$ AGN are almost sufficient to explain measurements from
the Ly$\alpha$ forest, additional UV sources are required at higher
redshifts (Figure~\ref{fig:gammapi}).  Boldly extrapolating the
\ion{H}{i} column density distribution to $z=6$, we estimate that
$M_{1450}<-18$ AGN contribute to the \ion{H}{i} photoionization rate
only at the $\sim 5$ per cent level.  This indicates a minor
contribution of such AGN to \ion{H}{i} reionization.  At $z<0.5$
$M_{1450}<-18$ AGN fall short by a factor of $\sim 2$ to explain UV
background measurements (Figure~\ref{fig:puc}), but we hesitate to
claim a `photon underproduction crisis' \citep{2014ApJ...789L..32K},
because (a) the apparent tension may be alleviated either by harder
extreme-UV SEDs in faint ($M_{1450}>-23$) AGN
\citep{2004ApJ...615..135S} or by known $-18\la M_{1450}\la -14$ AGN
at these redshifts \citep{2009A&A...507..781S} albeit their Lyman
continuum escape fraction is unknown, and (b) current UV background
synthesis models do not account for the large uncertainty in the
low-redshift column density distribution of (partial) \ion{H}{i} Lyman
limit systems \citep{2011ApJ...736...42R, 2017ApJ...849..106S}.
Helium reionization is accomplished at $z\simeq 3.5$ by $M_{1450}<-18$
AGN (Figure~\ref{fig:qhe}), but should be delayed by \ion{He}{ii}
Lyman limit systems \citep{2009MNRAS.395..736B, 2017ApJ...851...50M},
requiring further modelling and detailed numerical simulations of the
\ion{He}{ii} reionization process.

There are several promising paths forward in the characterization of
the AGN luminosity function, some of which may substantially advance
on current UV-optical broadband color-selected samples that have
dominated the field for the past 20 years.  These include, but are not
limited to, comprehensive surveys for faint high-$z$ quasars selected
by broadband photometry \citep{2016ApJ...828...26M,
  2017ApJ...839...27W} or variability \citep{2017MNRAS.464.1693H},
infrared-selected highly complete surveys at $z\sim 3$
\citep{2017ApJ...851...13S, 2018AJ....155..110Y}, multi-band
narrow-band surveys targeting AGN at all epochs
\citep{2014arXiv1403.5237B}, slitless spectroscopic surveys with
upcoming space telescopes \citep{2011arXiv1110.3193L,
  2013arXiv1305.5422S}, and next-generation X-ray surveys
\citep[e.g.][]{2013A&A...558A..89K}. However, we caution that future
progress based on massive surveys requires a detailed quantification
of their selection functions and the incorporation of systematic
errors into AGN luminosity function measurements, in a similar manner
to recent surveys for lensed high-redshift galaxies that are limited
by the accuracy of the lensing mass models \citep{2017ApJ...843..129B,
  2018ApJ...854...73I, 2018MNRAS.479.5184A}.  For a better
characterization of the AGN contribution to the UV background one will
need to (a) perform spectroscopic surveys for faint AGN at all
redshifts and determine their characteristic UV SED, (b) carefully
assess the impact of the host galaxy on the survey selection and the
ionizing power, and (c) measure precisely the amplitude and shape of
the column density distribution of (partial) \ion{H}{i} Lyman limit
systems to inform UV background synthesis models at $z<2$ and $z>4$.

Using the homogenised data and code that we make publicly available
(Appendix \ref{sec:code}), the luminosity functions presented in this
paper can be easily compared with theoretical models or future data
sets.  These luminosity functions update previous determinations by
including new data, by analysing unbinned data, and extending the
analysis to higher redshifts.  Our public code can be used either to
reproduce all of our analysis starting from the QSO catalogues, or to
quickly evaluate our luminosity function fits at desired redshifts and
luminosities.

\section*{Acknowledgements}

We thank Eilat Glikman, Linhua Jiang, Nobunari Kashi\-kawa, Ian
McGreer, Nick Ross, Chris Willott, and Jinyi Yang for sharing data and
especially for sharing their quasar selection functions.  It is a
pleasure to acknowledge useful discussions with James Aird, Eduardo
Ba\~nados, Manda Banerji, Tirthankar Roy Choudhury, George Efstathiou,
Xiaohui Fan, Andrea Ferrara, Prakash Gaikwad, Enrico Garaldi,
Francesco Haardt, Martin Haehnelt, Paul Hewett, David Hogg, Vikram
Khaire, Sergey Koposov, Alex Krolewski, Donald Lynden-Bell, Roberto
Maiolino, Richard McMahon, Daniel Mortlock, Ewald Puchwein, Gordon
Richards, Alberto Rorai, Bram Venemans and Stephen Warren.  We
  thank the anonymous referee for their comments. GK acknowledges
support from ERC Advanced Grant 320596 `The Emergence of Structure
During the Epoch of Reionization'.

\appendix

\section{Posterior distributions}

\begin{figure*}
  \begin{center}
    % corner.corner() with bins.lfs[-4]
    \includegraphics[width=0.65\textwidth]{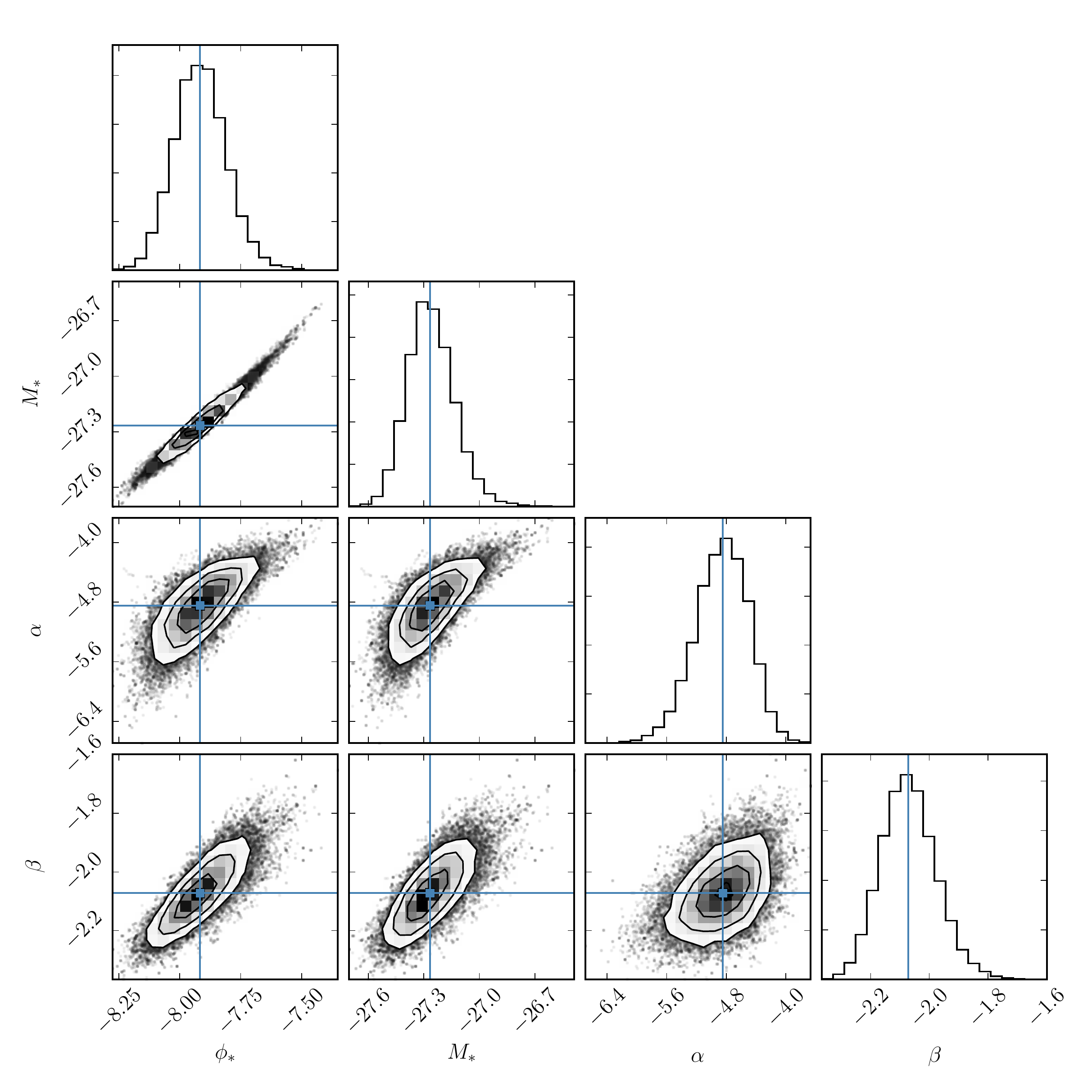}
  \end{center}
  \caption{Posterior distributions of the four double-power-law
    parameters in the $3.7\leq z < 4.1$ redshift bin.  The blue
    squares indicate median values.  Similar behaviour of the
    posterior distributions is seen in all other redshift bins defined
    in Section~\ref{sec:bins}.
    \label{fig:corner}}
\end{figure*}

Figure~\ref{fig:corner} shows marginalised one-dimensional and
two-dimensional posterior probability distribution functions (PDFs) of
the four parameters, $\phi_*$, $M_*$, $\alpha$ and $\beta$, of the
double-power-law luminosity function in the $3.7\leq z < 4.1$ redshift
bin.  The procedure adopted for fitting this model is described in
Section~\ref{sec:bins}.  Figure~\ref{fig:corner} illustrates that the
four parameters are well-constrainted.  This figure also shows the
degeneracies between the parameters.  There is a relatively strong
correlation between the amplitude of the luminosity function $\phi_*$
and the break magnitude $M_*$.  The faint-end slope $\beta$ is
positively correlated with the other three parameters.  Similar
behaviour of the posterior distributions is seen in all the redshift
bins defined in Section~\ref{sec:bins}.  In our highest-redshift bin
($5.5\leq z < 6.5$), we impose a prior $\alpha < -4$, which changes
the posterior distributions.  However, various parameter correlations
remain qualitatively unchanged.

\section{Comparison with other luminosity function determinations}
\label{sec:qlfliterature}
\begin{figure*}
  \begin{center}
    % summary_grand.py
    \includegraphics[width=\textwidth]{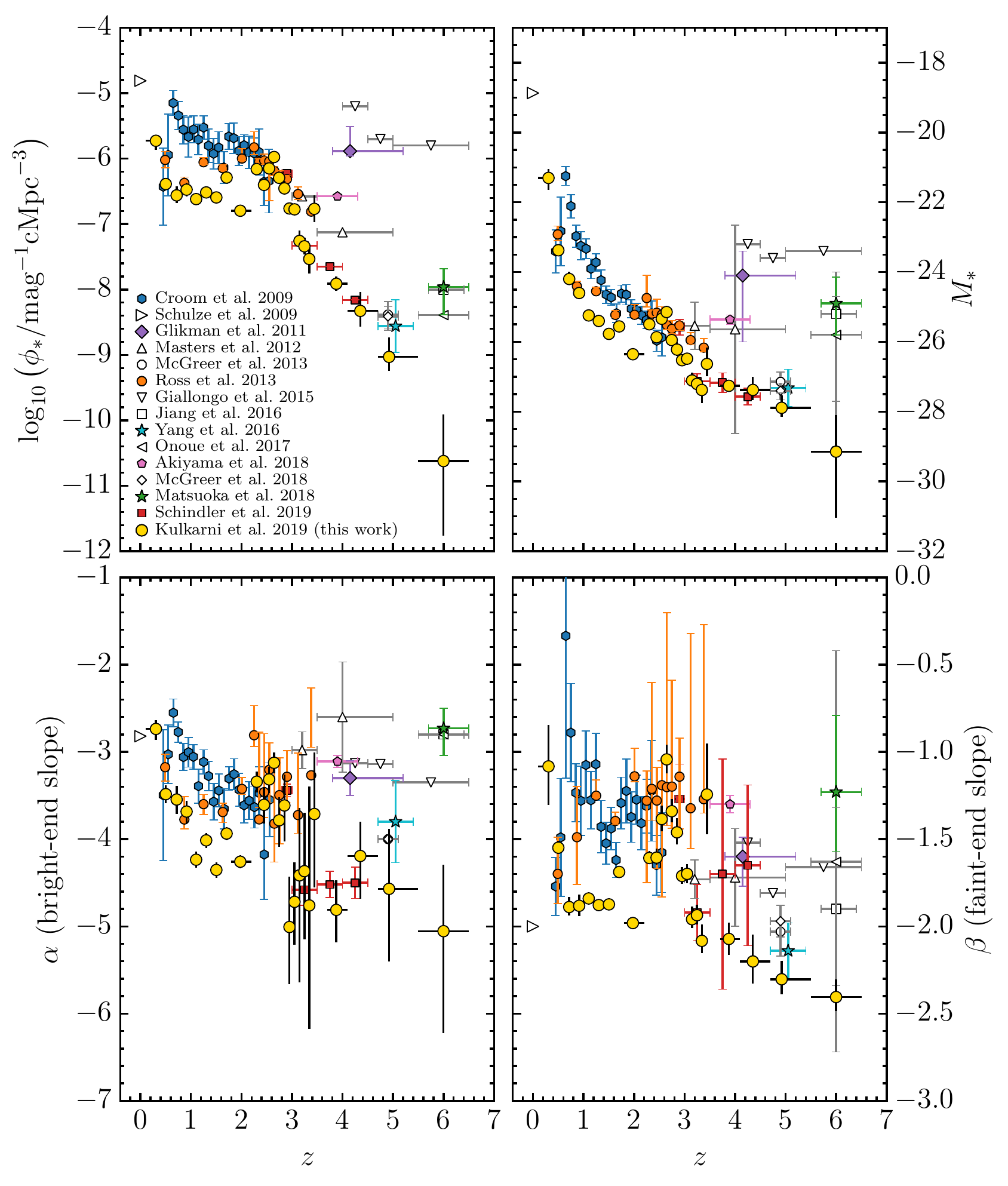}
  \end{center}
  \caption{A comparison of our inferred double power law parameter
    values with those reported in the literature.  Yellow points show
    our determinations from Figure~\ref{fig:evoln}.  (Our statistical
    errors are often smaller than the symbol size.)
    Other data points show values from
    \citet[blue hexagons]{2009MNRAS.399.1755C},
    \citet[open rightward triangles]{2009A&A...507..781S},
    \citet[purple diamonds]{2011ApJ...728L..26G},
    \citet[open triangles]{2012ApJ...755..169M},
    \citet[open octagon]{2013ApJ...768..105M},
    \citet[orange circles]{2013ApJ...773...14R},
    \citet[open downward triangles]{2015AA...578A..83G},
    \citet[open square]{2016ApJ...833..222J},
    \citet[cyan star]{2016ApJ...829...33Y},
    \citet[open leftward triangles]{2017ApJ...847L..15O},
    \citet[magenta pentagons]{2018PASJ...70S..34A},
    \citet[open diamonds]{2018AJ....155..131M},    
    \citet[green star]{2018ApJ...869..150M}, and 
    \citet[red squares]{2019ApJ...871..258S}.
    Inferences that fixed values of some parameters are shown using
    open symbols.  Where necessary, values from the literature have
    been converted to our cosmology.
    \label{fig:params_grand}}
\end{figure*}

Figure~\ref{fig:params_grand} compares the parameters of the double
power law luminosity function from our analysis of
Section~\ref{sec:bins} to values reported in the literature.  There is
significant disagreement between various inferences.  The break
magnitude and the luminosity function amplitudes are particularly
severely discordant towards the higher-redshift end of
Figure~\ref{fig:params_grand}, although the disagreement is
appreciable at other redshifts too.  The slopes are in conflict at all
redshifts. In general, our break luminosity is brighter and the
faint-end slope is steeper than other determinations.

Several results from the literature shown in
Figure~\ref{fig:params_grand} make restrictive assumptions while
fitting double power law models to data.  For example,
\citet{2013ApJ...768..105M} and \citet{2018AJ....155..131M} fix the
bright-end slope $\alpha$ to $-4$ at $z\sim 5$.
\citet{2016ApJ...833..222J} and \citet{2017ApJ...847L..15O} fix the
bright-end slope $\alpha$ to $-2.8$ at $z\sim 6$.
\citet{2015AA...578A..83G} fix the faint-end slope and the break
luminosity in their highest-redshift bin ($z=5.0$--$6.5$).  This
biases the inference of other parameters and will also underestimate
the uncertainties.

The choice of data sets is also often different.
For instance, \citet{2015AA...578A..83G} do not include the SDSS
Stripe 82 data from \citet{2013ApJ...768..105M} and the AGN samples of
\citet{2010AJ....139..906W} and \citet{2015ApJ...798...28K} in their
analysis.  (We discuss the results of \citet{2015AA...578A..83G} in
greater detail in Appendix~\ref{sec:conv}.)

Our disagreement with \citet{2018PASJ...70S..34A} is discussed above
in Section~\ref{sec:reion}.  \citet{2018PASJ...70S..34A} report a very
flat faint-end slope at $z\sim 4$ that is inconsistent with our
inference at all redshifts.  As discussed in Section~\ref{sec:reion},
this could be because of missing faint ($M_{1450} > -23.5$) AGN
population due to point source selection and partly also because of
photometric redshift errors.  The differences with
\citet{2012ApJ...755..169M} and \citet{2009MNRAS.399.1755C} are less
straightforward to understand.  The uncertainty reported by
\citet{2012ApJ...755..169M} on $\phi_*$ spans negative values.  (We
discuss other issues with this data set in
Section~\ref{sect:samplesel} above.)  At lower redshifts, our fits to
the $z<2$ luminosity function have steeper faint ends than those
reported by \citet{2009MNRAS.399.1755C}.  Upon closer inspection, we
found that our binned luminosity functions differ systematically from
those of \citet{2009MNRAS.399.1755C} in one or two faintest bins at
all redshifts while being in agreement in all other bins.  One
possible reason behind this discrepancy is that these authors made
specific interpolation choices while integrating over the completeness
map.

It can also been seen that there is inconsistency in many $z\sim 6$
QLF parameterisations in the literature.  To account for QLF evolution
across the redshift bin, many $z=6$ analyses assume that $\phi_*$
evolves as $\phi_*(z)=\phi_*(z=6)\times 10^{k(z-6)}$ where $k$ is
negative, historically estimated to be $-0.47$
\citep{2001AJ....121...54F}.  \citet{2016ApJ...833..222J} assume
$k=-0.7$ but their $\phi_*$ at $z=6$ is higher than that in all $z=5$
fits.  The same holds for \citet{2017ApJ...847L..15O}.  It is
interesting to note that the $\phi_*(z=6)$ inferred by
\citet{2016ApJ...829...33Y} is factor of 10 smaller than the inference
by \citet{2016ApJ...833..222J}.  Recent analysis by
\citet{2018ApJ...869..150M} using the SHELLQs survey gives parameter
values that are in agreement with \citet{2016ApJ...833..222J}.  The
SHELLQs AGN are not included in our analysis, but the parameter
inference by \citet{2018ApJ...869..150M} shown in
Figure~\ref{fig:params_grand} suggests that smooth redshift evolution
in any of the four double-power-law parameters is precluded.  If the
$z\sim 4$--$5$ data are correct, then the luminosity function
evolution is highly non-monotonic.  This raises fundamental issues on
the interpretation of QLFs and the validity of parametric models. At
this point we can only speculate whether the QLF suddenly changes from
$z=5$ to $6$ or whether there are data issues.

At high redshifts, our parameter values are closest to those reported
by \citet{2016ApJ...829...33Y} at $z=4.7$--$5.4$ and by
\citet{2019ApJ...871..258S} at $z=2.8$--$5$.

Figure~\ref{fig:params_grand} suggests that it is incorrect to fix QLF
parameters while fitting QLF models and one should only fit models if
data actually span the magnitude range.  In order to emphasize this,
we have shown inferences that used fixed parameters as open symbols.

\section{Comparison with G15}
\label{sec:conv}
\begin{figure*}
  \begin{center}
    % summary_giallongo.py
    \includegraphics[width=0.7\textwidth]{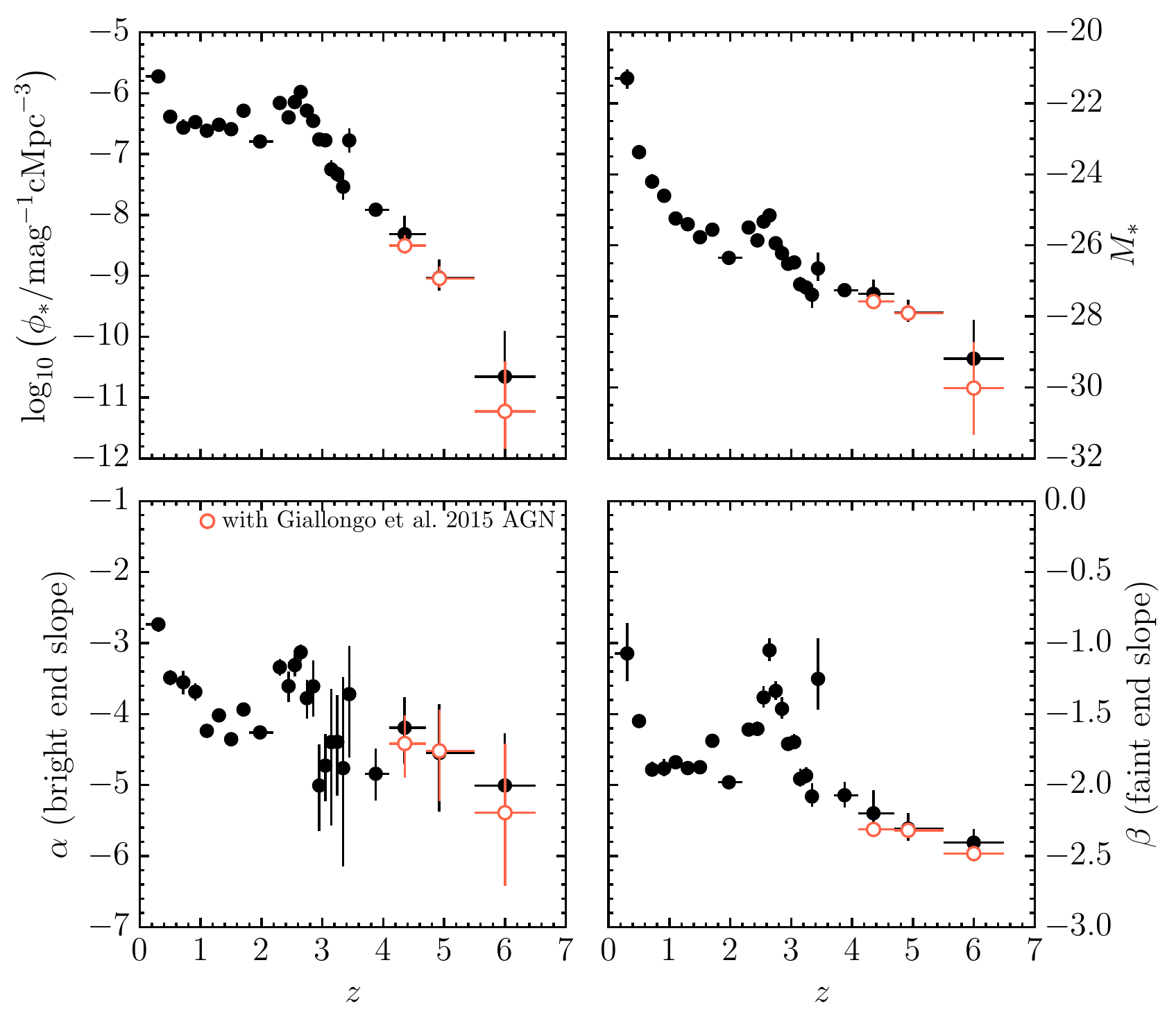}
  \end{center}
  \caption{Effect of the 19 AGN reported by \citet{2015AA...578A..83G}
    on the double power law luminosity function parameters in
    redshift bins from $z=0$ to $7$.  Black points show parameter
    values from Figure~\ref{fig:evoln}.  Red open circles show the
    parameter values obtained when the sample of G15 is added to the
    analysis.  In both cases, vertical error bars show one-sigma
    (68.26\%) uncertainties, and horizontal error bars show widths of
    the redshift bins. \label{fig:params_giallongo}}
\end{figure*}

\begin{figure*}
  \begin{center}
    % bins_withg.py
    % drawlf_giallongocompare.py 
    % giallongo_compare.py 
    \includegraphics[width=\textwidth]{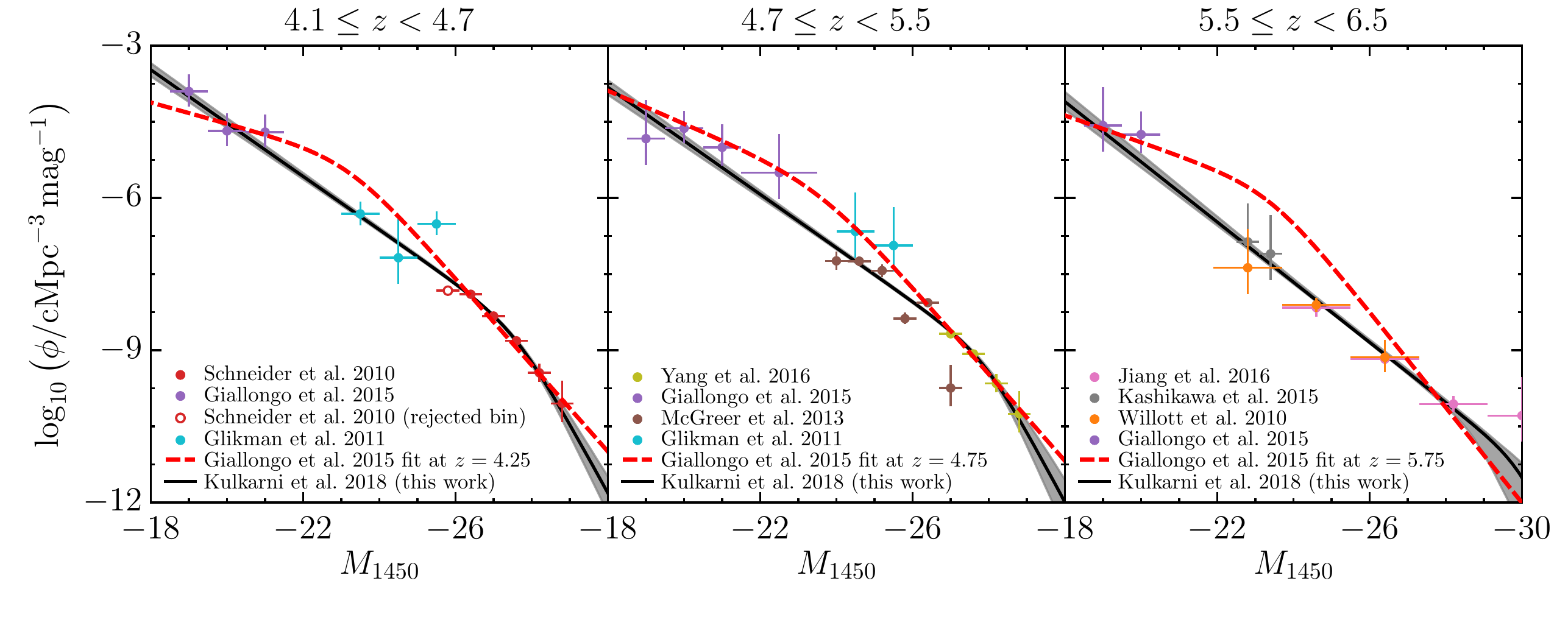}
  \end{center}
  \caption{Luminosity functions in three redshift bins at $z>4.1$.
    Black curves in each panel show the double power law,
    with the corresponding one-sigma (68.26\%) uncertainty shown by
    the grey shaded area.  There are 451, 270, and 69 AGN in each
    redshift bin from left to right, respectively.  These numbers are
    higher than those in Figure~\ref{fig:mosaic} because they include,
    respectively, 9, 7, and 3 AGN from G15.  The magnitude bins
    containing these AGN are shown in purple.  The red dashed curves
    show the double power law fits reported by G15 at $z=4.25, 4.75,$
    and $5.75$. \label{fig:lf_giallongo}}
\end{figure*}

In the double power law luminosity function models presented in
distinct redshift bins in Section~\ref{sec:bins}, we did not include
the 19 low-luminosity ($M_{1450}>-22.6$) AGN between redshifts $z=4.1$
and $6.3$ reported by \citep[][hereafter G15]{2015AA...578A..83G}.
While this was done in order to restrict our sample to quasars with
spectroscopic redshift determinations, it is instructive to consider
how our results are affected if the G15 AGN are added to the analysis.
In their work, G15 found a shallower faint-end slope ($\beta\sim -1.5$
to $-1.8$) for the luminosity function at $4.1 < z < 6.3$, relative to
our result from other AGN samples at these redshifts ($\beta\sim -2.0$
to $-2.5$).  Still, G15 derived a higher 912\,\AA\ emissivity than our
estimates (cf.\ Figure~\ref{fig:e912_2}), so that in their analysis
AGN can produce all ionizing photons necessary to keep hydrogen
ionized and explain the \lya data.  The black points in
Figure~\ref{fig:params_giallongo} show the parameters of the double
power law luminosity function from our analysis of
Section~\ref{sec:bins}.  The red open circles show the parameter
values obtained when the G15 sample is added to the analysis.  We find
that the two results are highly consistent, showing that the G15
sample is consistent with our double power law fit obtained from other
AGN samples are comparable redshifts.  This is surprising as the
integrated 912\,\AA\ emissivities in our model are smaller than those
derived by G15.  Figure~\ref{fig:lf_giallongo} provides an
explanation.  As seen in this figure, the double power law fits
favoured by G15 (red dashed curves) are quite different from our fits
(black curves).  The characteristic luminosity $M_*$ obtained by G15
is much fainter ($\sim -23$ at $z = 5$) than that resulting out of our
analysis ($\sim -29$ at $z = 5$).  Thus the 912\,\AA\ emissivities are
enhanced in G15 because of the increase contribution from
intermediate-luminosity AGN in their model.
Figure~\ref{fig:lf_giallongo} suggests that this is possibly because
of the inclusion of SDSS Stripe 82 data from
\citet{2013ApJ...768..105M} and the AGN samples of
\citet{2010AJ....139..906W} and \citet{2015ApJ...798...28K} in our
analysis.  Additionally, the homogenisation of data in our work may
also cause part of the difference.

\section{Tables of emissivities and photoionization rates}
\label{sec:tables}

Table~\ref{tab:emissivity_bins} shows the 912\,\AA\ and
1450\,\AA\ comoving emissivities obtained in various redshift bins
with the one-sigma (68.26\%) uncertainties.  Redshift bins severely
affected by systematic errors are also shown.  The result of the model
presented in Equation~(\ref{eqn:e912fit}), which describes the
emissivity evolution using a smooth function, is tabulated in
Table~\ref{tab:gamma2}, along with the hydrogen photoionization rate
computed in Section~\ref{sec:gammahi}.  In both tables, we show
results for our two integration limits of $M_{1450}<-18$ and
$M_{1450}<-21$.  These tables describe the curves shown in
Figures~\ref{fig:e912_2} and \ref{fig:gammapi}.  Note that the
photoionization rate calculation assumes an \HI\ column density
distribution given by \citet{2012ApJ...746..125H} and extrapolates
this to high redshifts.

\begin{table*}
  % gammapi.py and tabulate_emissivities.py. 
  \caption{
    Comoving emissivities at 912\,\AA\ and 1450\,\AA\ derived from
    our double power law luminosity function models in redshift bins
    (Table~\ref{tab:bins}) for two magnitude limits. The units are
    $10^{24}$\ erg\ s$^{-1}$\ Hz$^{-1}$\ cMpc$^{-3}$.
    Statistical uncertainties are one-sigma (68.26\%).
    }
  \label{tab:emissivity_bins}
  \begin{tabular}{ccc....}
    \hline
    $\langle z\rangle$ &
    $z_\mathrm{min}$ &
    $z_\mathrm{max}$ &
    \multicolumn{1}{c}{$\epsilon_{912}$} &
    \multicolumn{1}{c}{$\epsilon_{1450}$} &
    \multicolumn{1}{c}{$\epsilon_{912}$} &
    \multicolumn{1}{c}{$\epsilon_{1450}$} \\
    &
    &
    &
    \multicolumn{1}{c}{$(M_{1450}<-18)$} &
    \multicolumn{1}{c}{$(M_{1450}<-18)$} &
    \multicolumn{1}{c}{$(M_{1450}<-21)$} &
    \multicolumn{1}{c}{$(M_{1450}<-21)$} \\
    \hline
    0.31$^a$ & 0.10 & 0.40 & 0.41^{+0.01}_{-0.01} & 0.55^{+0.02}_{-0.02} & 0.27^{+0.01}_{-0.01} & 0.36^{+0.02}_{-0.01} \\
    0.50$^a$ & 0.40 & 0.60 & 0.72^{+0.01}_{-0.01} & 0.96^{+0.02}_{-0.02} & 0.53^{+0.01}_{-0.01} & 0.70^{+0.01}_{-0.01} \\
    0.72 & 0.60 & 0.80 & 2.03^{+0.08}_{-0.08} & 2.69^{+0.13}_{-0.11} & 1.22^{+0.02}_{-0.02} & 1.61^{+0.03}_{-0.03} \\
    0.91 & 0.80 & 1.00 & 3.64^{+0.23}_{-0.23} & 4.87^{+0.35}_{-0.37} & 2.33^{+0.06}_{-0.06} & 3.09^{+0.08}_{-0.08} \\
    1.10 & 1.00 & 1.20 & 4.47^{+0.16}_{-0.15} & 5.92^{+0.19}_{-0.19} & 3.18^{+0.05}_{-0.05} & 4.22^{+0.07}_{-0.07} \\
    1.30 & 1.20 & 1.40 & 7.33^{+0.38}_{-0.39} & 9.74^{+0.53}_{-0.53} & 5.10^{+0.12}_{-0.13} & 6.79^{+0.17}_{-0.18} \\
    1.50 & 1.40 & 1.60 & 8.84^{+0.35}_{-0.33} & 11.72^{+0.45}_{-0.48} & 6.32^{+0.13}_{-0.14} & 8.41^{+0.17}_{-0.17} \\
    1.71 & 1.60 & 1.80 & 9.05^{+0.26}_{-0.27} & 12.01^{+0.33}_{-0.32} & 7.51^{+0.13}_{-0.14} & 9.96^{+0.15}_{-0.15} \\
    1.98 & 1.80 & 2.20 & 14.27^{+0.53}_{-0.58} & 18.92^{+0.85}_{-0.79} & 9.40^{+0.19}_{-0.19} & 12.47^{+0.25}_{-0.24} \\
    2.30$^a$ & 2.20 & 2.40 & 10.20^{+0.25}_{-0.24} & 13.54^{+0.31}_{-0.31} & 8.88^{+0.11}_{-0.12} & 11.78^{+0.15}_{-0.17} \\
    2.45$^a$ & 2.40 & 2.50 & 8.05^{+0.23}_{-0.25} & 10.69^{+0.30}_{-0.30} & 7.16^{+0.14}_{-0.14} & 9.48^{+0.21}_{-0.19} \\
    2.55$^a$ & 2.50 & 2.60 & 6.60^{+0.20}_{-0.19} & 8.80^{+0.30}_{-0.28} & 6.22^{+0.13}_{-0.12} & 8.23^{+0.16}_{-0.16} \\
    2.65$^a$ & 2.60 & 2.70 & 6.56^{+0.16}_{-0.15} & 8.70^{+0.20}_{-0.21} & 6.44^{+0.16}_{-0.17} & 8.53^{+0.18}_{-0.22} \\
    2.75$^a$ & 2.70 & 2.80 & 7.47^{+0.22}_{-0.20} & 9.90^{+0.29}_{-0.26} & 7.19^{+0.18}_{-0.18} & 9.50^{+0.26}_{-0.26} \\
    2.85$^a$ & 2.80 & 2.90 & 7.90^{+0.30}_{-0.32} & 10.50^{+0.44}_{-0.43} & 7.50^{+0.28}_{-0.28} & 9.96^{+0.40}_{-0.38} \\
    2.95$^a$ & 2.90 & 3.00 & 7.96^{+0.37}_{-0.39} & 10.55^{+0.46}_{-0.47} & 6.85^{+0.20}_{-0.19} & 9.07^{+0.26}_{-0.25} \\
    3.05$^a$ & 3.00 & 3.10 & 7.25^{+0.42}_{-0.44} & 9.58^{+0.51}_{-0.56} & 6.22^{+0.20}_{-0.21} & 8.26^{+0.30}_{-0.29} \\
    3.15$^a$ & 3.10 & 3.20 & 9.91^{+1.14}_{-1.11} & 13.20^{+1.41}_{-1.32} & 6.99^{+0.35}_{-0.35} & 9.29^{+0.42}_{-0.42} \\
    3.25$^a$ & 3.20 & 3.30 & 8.35^{+0.88}_{-0.92} & 11.07^{+1.15}_{-1.26} & 6.10^{+0.36}_{-0.38} & 8.10^{+0.48}_{-0.53} \\
    3.34$^a$ & 3.30 & 3.40 & 12.19^{+2.77}_{-2.67} & 15.99^{+3.04}_{-3.34} & 7.30^{+0.79}_{-0.74} & 9.74^{+1.01}_{-1.10} \\
    3.44$^a$ & 3.40 & 3.50 & 4.40^{+0.51}_{-0.51} & 5.81^{+0.70}_{-0.64} & 4.28^{+0.45}_{-0.45} & 5.73^{+0.54}_{-0.61} \\
    3.88 & 3.70 & 4.10 & 4.43^{+1.37}_{-1.50} & 5.90^{+1.69}_{-1.95} & 2.53^{+0.45}_{-0.49} & 3.34^{+0.65}_{-0.65} \\
    4.35 & 4.10 & 4.70 & 4.17^{+2.21}_{-2.11} & 6.06^{+3.08}_{-3.01} & 1.77^{+0.43}_{-0.46} & 2.35^{+0.69}_{-0.71} \\
    4.92 & 4.70 & 5.50 & 2.69^{+0.93}_{-1.07} & 3.38^{+1.34}_{-1.20} & 0.97^{+0.17}_{-0.16} & 1.29^{+0.29}_{-0.26} \\
    6.00 & 5.50 & 6.50 & 0.66^{+0.26}_{-0.25} & 0.91^{+0.29}_{-0.31} & 0.21^{+0.05}_{-0.05} & 0.27^{+0.05}_{-0.05} \\
    \hline
  \end{tabular}
  \begin{minipage}{13.0cm}
    \textsuperscript{$a$}{Redshift bin considered to be severely
      affected by systematic errors (open circles in
      Figure~\ref{fig:e912_2}).  These bins should be avoided while
      considering the evolution of emissivities.}
  \end{minipage}
\end{table*}

\begin{table*}
  % rtg2.py and tabulate_emissivities_global.py.
  \caption{ Comoving emissivities at 912\,\AA\ and 1450\,\AA\ obtained
    by fitting Equation~\eqref{eqn:e912fit} to the emissivities in
    selected redshift bins from Table~\ref{tab:emissivity_bins}.
    Emissivities at $z<0.6$ and at $z>6.5$ are extrapolated assuming
    our best fits.  The derived \ion{H}{i} photoionization rates
    (Equation~\ref{eqn:gammapi}) are also given. Emissivity units are
    erg\ s$^{-1}$\ Hz$^{-1}$\ cMpc$^{-3}$, and photoionization rate
    units are s$^{-1}$. Statistical uncertainties are one-sigma
    (68.26\%) equal-tailed credibility intervals.  These values are
    shown in Figures~\ref{fig:e912_2} and \ref{fig:gammapi}.  See
    Sections~\ref{sec:e912} and \ref{sec:gammahi} for details.  }
  \label{tab:gamma2}
  \begin{tabular}{d......}
    \hline
    \multicolumn{1}{c}{$z$} &
    \multicolumn{1}{c}{$\log_{10}\epsilon_{1450}$} &
    \multicolumn{1}{c}{$\log_{10}\epsilon_{1450}$} &
    \multicolumn{1}{c}{$\log_{10}\epsilon_{912}$} &
    \multicolumn{1}{c}{$\log_{10}\epsilon_{912}$} & 
    \multicolumn{1}{c}{$\log_{10}\Gamma_\mathrm{HI}$} &
    \multicolumn{1}{c}{$\log_{10}\Gamma_\mathrm{HI}$} \\ 
    &
    \multicolumn{1}{c}{$(M_{1450}<-18)$} &
    \multicolumn{1}{c}{$(M_{1450}<-21)$} &
    \multicolumn{1}{c}{$(M_{1450}<-18)$} &
    \multicolumn{1}{c}{$(M_{1450}<-21)$} &
    \multicolumn{1}{c}{$(M_{1450}<-18)$} &
    \multicolumn{1}{c}{$(M_{1450}<-21)$} \\
    \hline
    0.0 & 23.13^{+0.33}_{-0.20} & 22.79^{+0.13}_{-0.15} & 23.01^{+0.33}_{-0.20} & 22.67^{+0.13}_{-0.15} & -13.66^{+0.08}_{-0.05} & -13.86^{+0.03}_{-0.03} \\
    0.1 & 23.39^{+0.25}_{-0.16} & 23.06^{+0.10}_{-0.12} & 23.27^{+0.25}_{-0.16} & 22.94^{+0.10}_{-0.12} & -13.45^{+0.06}_{-0.05} & -13.65^{+0.03}_{-0.02} \\
    0.2 & 23.61^{+0.18}_{-0.13} & 23.30^{+0.08}_{-0.09} & 23.49^{+0.18}_{-0.13} & 23.18^{+0.08}_{-0.09} & -13.27^{+0.05}_{-0.04} & -13.46^{+0.02}_{-0.02} \\
    0.3 & 23.81^{+0.13}_{-0.11} & 23.52^{+0.06}_{-0.06} & 23.69^{+0.13}_{-0.11} & 23.39^{+0.06}_{-0.06} & -13.10^{+0.04}_{-0.04} & -13.29^{+0.02}_{-0.02} \\
    0.4 & 23.99^{+0.09}_{-0.09} & 23.71^{+0.04}_{-0.04} & 23.86^{+0.09}_{-0.09} & 23.59^{+0.04}_{-0.04} & -12.95^{+0.04}_{-0.03} & -13.13^{+0.02}_{-0.02} \\
    0.5 & 24.14^{+0.07}_{-0.07} & 23.89^{+0.03}_{-0.03} & 24.02^{+0.07}_{-0.07} & 23.76^{+0.03}_{-0.03} & -12.82^{+0.03}_{-0.03} & -12.98^{+0.01}_{-0.01} \\
    0.6 & 24.28^{+0.05}_{-0.05} & 24.05^{+0.02}_{-0.02} & 24.16^{+0.05}_{-0.05} & 23.92^{+0.02}_{-0.02} & -12.69^{+0.03}_{-0.03} & -12.85^{+0.01}_{-0.01} \\
    0.7 & 24.41^{+0.04}_{-0.03} & 24.19^{+0.02}_{-0.02} & 24.28^{+0.04}_{-0.03} & 24.07^{+0.02}_{-0.02} & -12.58^{+0.02}_{-0.02} & -12.74^{+0.01}_{-0.01} \\
    0.8 & 24.52^{+0.03}_{-0.03} & 24.32^{+0.01}_{-0.01} & 24.40^{+0.03}_{-0.03} & 24.20^{+0.01}_{-0.01} & -12.48^{+0.02}_{-0.02} & -12.63^{+0.01}_{-0.01} \\
    0.9 & 24.62^{+0.02}_{-0.02} & 24.44^{+0.01}_{-0.01} & 24.50^{+0.02}_{-0.02} & 24.31^{+0.01}_{-0.01} & -12.39^{+0.02}_{-0.02} & -12.53^{+0.01}_{-0.01} \\
    1.0 & 24.71^{+0.02}_{-0.02} & 24.54^{+0.01}_{-0.01} & 24.59^{+0.02}_{-0.02} & 24.42^{+0.01}_{-0.01} & -12.31^{+0.02}_{-0.02} & -12.44^{+0.01}_{-0.01} \\
    1.1 & 24.79^{+0.02}_{-0.02} & 24.64^{+0.01}_{-0.01} & 24.67^{+0.02}_{-0.02} & 24.52^{+0.01}_{-0.01} & -12.24^{+0.02}_{-0.02} & -12.37^{+0.01}_{-0.01} \\
    1.2 & 24.87^{+0.02}_{-0.02} & 24.72^{+0.01}_{-0.01} & 24.75^{+0.02}_{-0.02} & 24.60^{+0.01}_{-0.01} & -12.17^{+0.02}_{-0.02} & -12.30^{+0.01}_{-0.01} \\
    1.3 & 24.93^{+0.02}_{-0.02} & 24.80^{+0.01}_{-0.01} & 24.81^{+0.02}_{-0.02} & 24.68^{+0.01}_{-0.01} & -12.12^{+0.02}_{-0.02} & -12.24^{+0.01}_{-0.01} \\
    1.4 & 24.99^{+0.02}_{-0.02} & 24.86^{+0.01}_{-0.01} & 24.87^{+0.02}_{-0.02} & 24.74^{+0.01}_{-0.01} & -12.08^{+0.02}_{-0.03} & -12.19^{+0.01}_{-0.01} \\
    1.5 & 25.04^{+0.02}_{-0.02} & 24.92^{+0.01}_{-0.01} & 24.92^{+0.02}_{-0.02} & 24.80^{+0.01}_{-0.01} & -12.04^{+0.02}_{-0.03} & -12.15^{+0.02}_{-0.01} \\
    1.6 & 25.08^{+0.02}_{-0.02} & 24.97^{+0.01}_{-0.01} & 24.96^{+0.02}_{-0.02} & 24.85^{+0.01}_{-0.01} & -12.01^{+0.03}_{-0.03} & -12.12^{+0.02}_{-0.02} \\
    1.7 & 25.12^{+0.02}_{-0.02} & 25.01^{+0.01}_{-0.01} & 25.00^{+0.02}_{-0.02} & 24.89^{+0.01}_{-0.01} & -11.99^{+0.03}_{-0.03} & -12.10^{+0.02}_{-0.02} \\
    1.8 & 25.15^{+0.02}_{-0.02} & 25.04^{+0.01}_{-0.01} & 25.03^{+0.02}_{-0.02} & 24.92^{+0.01}_{-0.01} & -11.97^{+0.03}_{-0.04} & -12.08^{+0.02}_{-0.02} \\
    1.9 & 25.17^{+0.02}_{-0.02} & 25.07^{+0.01}_{-0.01} & 25.05^{+0.02}_{-0.02} & 24.94^{+0.01}_{-0.01} & -11.96^{+0.04}_{-0.04} & -12.07^{+0.02}_{-0.02} \\
    2.0 & 25.19^{+0.03}_{-0.03} & 25.08^{+0.01}_{-0.01} & 25.07^{+0.03}_{-0.03} & 24.96^{+0.01}_{-0.01} & -11.95^{+0.04}_{-0.05} & -12.07^{+0.03}_{-0.03} \\
    2.1 & 25.20^{+0.03}_{-0.04} & 25.10^{+0.02}_{-0.02} & 25.08^{+0.03}_{-0.04} & 24.97^{+0.02}_{-0.02} & -11.95^{+0.04}_{-0.05} & -12.07^{+0.03}_{-0.03} \\
    2.2 & 25.21^{+0.03}_{-0.05} & 25.10^{+0.02}_{-0.02} & 25.09^{+0.03}_{-0.05} & 24.98^{+0.02}_{-0.02} & -11.95^{+0.05}_{-0.06} & -12.08^{+0.03}_{-0.03} \\
    2.3 & 25.22^{+0.04}_{-0.05} & 25.10^{+0.03}_{-0.03} & 25.09^{+0.04}_{-0.05} & 24.98^{+0.03}_{-0.03} & -11.96^{+0.05}_{-0.07} & -12.09^{+0.04}_{-0.04} \\
    2.4 & 25.22^{+0.05}_{-0.06} & 25.10^{+0.03}_{-0.03} & 25.09^{+0.05}_{-0.06} & 24.97^{+0.03}_{-0.03} & -11.97^{+0.06}_{-0.07} & -12.11^{+0.04}_{-0.04} \\
    2.5 & 25.21^{+0.05}_{-0.06} & 25.09^{+0.04}_{-0.04} & 25.09^{+0.05}_{-0.06} & 24.97^{+0.04}_{-0.04} & -11.99^{+0.06}_{-0.08} & -12.13^{+0.05}_{-0.04} \\
    2.6 & 25.21^{+0.05}_{-0.07} & 25.07^{+0.04}_{-0.04} & 25.08^{+0.05}_{-0.07} & 24.95^{+0.04}_{-0.04} & -12.01^{+0.07}_{-0.08} & -12.15^{+0.05}_{-0.04} \\
    2.7 & 25.19^{+0.06}_{-0.07} & 25.05^{+0.05}_{-0.04} & 25.07^{+0.06}_{-0.07} & 24.93^{+0.05}_{-0.04} & -12.03^{+0.07}_{-0.09} & -12.18^{+0.05}_{-0.05} \\
    2.8 & 25.18^{+0.07}_{-0.08} & 25.03^{+0.05}_{-0.05} & 25.06^{+0.06}_{-0.08} & 24.91^{+0.05}_{-0.05} & -12.05^{+0.08}_{-0.09} & -12.22^{+0.06}_{-0.05} \\
    2.9 & 25.16^{+0.07}_{-0.09} & 25.01^{+0.05}_{-0.05} & 25.04^{+0.07}_{-0.09} & 24.89^{+0.05}_{-0.05} & -12.08^{+0.08}_{-0.10} & -12.25^{+0.06}_{-0.05} \\
    3.0 & 25.14^{+0.08}_{-0.09} & 24.98^{+0.06}_{-0.05} & 25.02^{+0.08}_{-0.09} & 24.86^{+0.06}_{-0.05} & -12.11^{+0.09}_{-0.10} & -12.29^{+0.06}_{-0.05} \\
    3.1 & 25.12^{+0.08}_{-0.10} & 24.95^{+0.06}_{-0.05} & 25.00^{+0.08}_{-0.10} & 24.83^{+0.06}_{-0.05} & -12.14^{+0.09}_{-0.11} & -12.33^{+0.06}_{-0.05} \\
    3.2 & 25.09^{+0.09}_{-0.11} & 24.91^{+0.06}_{-0.05} & 24.97^{+0.09}_{-0.11} & 24.79^{+0.06}_{-0.05} & -12.18^{+0.10}_{-0.12} & -12.37^{+0.07}_{-0.06} \\
    3.3 & 25.06^{+0.09}_{-0.11} & 24.88^{+0.06}_{-0.06} & 24.94^{+0.09}_{-0.11} & 24.75^{+0.06}_{-0.06} & -12.22^{+0.10}_{-0.12} & -12.42^{+0.07}_{-0.06} \\
    3.4 & 25.03^{+0.10}_{-0.11} & 24.84^{+0.07}_{-0.05} & 24.91^{+0.10}_{-0.11} & 24.71^{+0.07}_{-0.05} & -12.26^{+0.11}_{-0.13} & -12.47^{+0.07}_{-0.06} \\
    3.5 & 25.00^{+0.10}_{-0.12} & 24.80^{+0.07}_{-0.06} & 24.88^{+0.10}_{-0.12} & 24.68^{+0.07}_{-0.06} & -12.30^{+0.11}_{-0.13} & -12.52^{+0.07}_{-0.07} \\
    3.6 & 24.97^{+0.11}_{-0.12} & 24.76^{+0.07}_{-0.07} & 24.84^{+0.11}_{-0.12} & 24.63^{+0.07}_{-0.07} & -12.34^{+0.11}_{-0.14} & -12.57^{+0.07}_{-0.07} \\
    3.7 & 24.93^{+0.11}_{-0.13} & 24.71^{+0.07}_{-0.07} & 24.81^{+0.11}_{-0.13} & 24.59^{+0.07}_{-0.07} & -12.38^{+0.11}_{-0.15} & -12.63^{+0.07}_{-0.08} \\
    3.8 & 24.90^{+0.11}_{-0.15} & 24.66^{+0.07}_{-0.07} & 24.78^{+0.11}_{-0.15} & 24.54^{+0.07}_{-0.07} & -12.43^{+0.12}_{-0.16} & -12.68^{+0.07}_{-0.08} \\
    3.9 & 24.86^{+0.11}_{-0.16} & 24.61^{+0.07}_{-0.08} & 24.74^{+0.11}_{-0.16} & 24.49^{+0.07}_{-0.08} & -12.48^{+0.12}_{-0.17} & -12.74^{+0.07}_{-0.08} \\
    4.0 & 24.83^{+0.11}_{-0.18} & 24.57^{+0.07}_{-0.08} & 24.71^{+0.11}_{-0.18} & 24.44^{+0.07}_{-0.08} & -12.52^{+0.12}_{-0.19} & -12.80^{+0.07}_{-0.09} \\
    4.1 & 24.79^{+0.12}_{-0.19} & 24.51^{+0.07}_{-0.09} & 24.67^{+0.12}_{-0.19} & 24.39^{+0.07}_{-0.09} & -12.58^{+0.13}_{-0.19} & -12.87^{+0.07}_{-0.09} \\
    4.2 & 24.74^{+0.13}_{-0.19} & 24.46^{+0.07}_{-0.09} & 24.62^{+0.13}_{-0.19} & 24.34^{+0.07}_{-0.09} & -12.63^{+0.14}_{-0.20} & -12.93^{+0.07}_{-0.09} \\
    4.3 & 24.70^{+0.14}_{-0.20} & 24.40^{+0.07}_{-0.09} & 24.57^{+0.14}_{-0.20} & 24.28^{+0.07}_{-0.09} & -12.69^{+0.15}_{-0.21} & -12.99^{+0.08}_{-0.10} \\
    4.4 & 24.65^{+0.15}_{-0.21} & 24.35^{+0.08}_{-0.10} & 24.53^{+0.15}_{-0.21} & 24.22^{+0.08}_{-0.10} & -12.75^{+0.15}_{-0.22} & -13.06^{+0.08}_{-0.11} \\
    4.5 & 24.60^{+0.15}_{-0.22} & 24.29^{+0.08}_{-0.10} & 24.48^{+0.15}_{-0.22} & 24.17^{+0.08}_{-0.10} & -12.80^{+0.16}_{-0.23} & -13.13^{+0.08}_{-0.12} \\
    4.6 & 24.56^{+0.15}_{-0.22} & 24.24^{+0.08}_{-0.12} & 24.43^{+0.15}_{-0.22} & 24.11^{+0.08}_{-0.12} & -12.86^{+0.16}_{-0.24} & -13.20^{+0.08}_{-0.13} \\
    4.7 & 24.51^{+0.16}_{-0.23} & 24.17^{+0.09}_{-0.13} & 24.39^{+0.16}_{-0.23} & 24.05^{+0.09}_{-0.13} & -12.92^{+0.17}_{-0.25} & -13.27^{+0.09}_{-0.14} \\
    4.8 & 24.46^{+0.17}_{-0.25} & 24.11^{+0.09}_{-0.14} & 24.34^{+0.17}_{-0.25} & 23.99^{+0.09}_{-0.14} & -12.98^{+0.17}_{-0.26} & -13.34^{+0.09}_{-0.15} \\
    4.9 & 24.41^{+0.18}_{-0.26} & 24.05^{+0.09}_{-0.15} & 24.29^{+0.18}_{-0.26} & 23.93^{+0.09}_{-0.15} & -13.04^{+0.18}_{-0.27} & -13.41^{+0.09}_{-0.16} \\
    5.0 & 24.36^{+0.17}_{-0.28} & 23.99^{+0.09}_{-0.16} & 24.24^{+0.17}_{-0.28} & 23.87^{+0.09}_{-0.16} & -13.10^{+0.18}_{-0.29} & -13.49^{+0.09}_{-0.17} \\
    \hline
  \end{tabular}
\end{table*}

\begin{table*}
  \contcaption{}
  \begin{tabular}{d......}
    \hline
    \multicolumn{1}{c}{$z$} &    
    \multicolumn{1}{c}{$\log_{10}\epsilon_{1450}$} &
    \multicolumn{1}{c}{$\log_{10}\epsilon_{1450}$} &
    \multicolumn{1}{c}{$\log_{10}\epsilon_{912}$} &
    \multicolumn{1}{c}{$\log_{10}\epsilon_{912}$} & 
    \multicolumn{1}{c}{$\log_{10}\Gamma_\mathrm{HI}$} &
    \multicolumn{1}{c}{$\log_{10}\Gamma_\mathrm{HI}$} \\ 
    &
    \multicolumn{1}{c}{$(M_{1450}<-18)$} &
    \multicolumn{1}{c}{$(M_{1450}<-21)$} &
    \multicolumn{1}{c}{$(M_{1450}<-18)$} &
    \multicolumn{1}{c}{$(M_{1450}<-21)$} &
    \multicolumn{1}{c}{$(M_{1450}<-18)$} &
    \multicolumn{1}{c}{$(M_{1450}<-21)$} \\
    \hline
    5.1 & 24.31^{+0.18}_{-0.29} & 23.93^{+0.09}_{-0.17} & 24.18^{+0.18}_{-0.29} & 23.80^{+0.09}_{-0.17} & -13.17^{+0.19}_{-0.30} & -13.56^{+0.10}_{-0.17} \\
    5.2 & 24.25^{+0.19}_{-0.30} & 23.86^{+0.10}_{-0.17} & 24.13^{+0.19}_{-0.30} & 23.74^{+0.10}_{-0.17} & -13.24^{+0.20}_{-0.31} & -13.64^{+0.11}_{-0.18} \\
    5.3 & 24.19^{+0.20}_{-0.32} & 23.79^{+0.11}_{-0.19} & 24.07^{+0.20}_{-0.32} & 23.67^{+0.11}_{-0.19} & -13.31^{+0.21}_{-0.33} & -13.72^{+0.12}_{-0.20} \\
    5.4 & 24.14^{+0.22}_{-0.33} & 23.73^{+0.12}_{-0.20} & 24.01^{+0.22}_{-0.33} & 23.60^{+0.12}_{-0.20} & -13.38^{+0.23}_{-0.34} & -13.80^{+0.13}_{-0.21} \\
    5.5 & 24.08^{+0.23}_{-0.34} & 23.66^{+0.13}_{-0.21} & 23.95^{+0.23}_{-0.34} & 23.53^{+0.13}_{-0.21} & -13.47^{+0.24}_{-0.36} & -13.89^{+0.14}_{-0.22} \\
    5.6 & 24.01^{+0.24}_{-0.36} & 23.59^{+0.14}_{-0.23} & 23.89^{+0.24}_{-0.36} & 23.47^{+0.14}_{-0.23} & -13.56^{+0.25}_{-0.37} & -14.00^{+0.14}_{-0.24} \\
    5.7 & 23.96^{+0.25}_{-0.38} & 23.52^{+0.15}_{-0.25} & 23.84^{+0.25}_{-0.38} & 23.40^{+0.15}_{-0.25} & -13.66^{+0.26}_{-0.39} & -14.11^{+0.15}_{-0.25} \\
    5.8 & 23.90^{+0.26}_{-0.39} & 23.45^{+0.16}_{-0.25} & 23.78^{+0.26}_{-0.39} & 23.32^{+0.16}_{-0.25} & -13.76^{+0.27}_{-0.40} & -14.22^{+0.16}_{-0.26} \\
    5.9 & 23.83^{+0.28}_{-0.40} & 23.37^{+0.17}_{-0.26} & 23.71^{+0.28}_{-0.40} & 23.25^{+0.17}_{-0.26} & -13.88^{+0.28}_{-0.41} & -14.34^{+0.18}_{-0.26} \\
    6.0 & 23.77^{+0.28}_{-0.42} & 23.30^{+0.18}_{-0.27} & 23.65^{+0.28}_{-0.42} & 23.17^{+0.18}_{-0.27} & -13.99^{+0.29}_{-0.43} & -14.47^{+0.19}_{-0.28} \\
    6.1 & 23.71^{+0.30}_{-0.43} & 23.23^{+0.19}_{-0.29} & 23.59^{+0.30}_{-0.43} & 23.11^{+0.19}_{-0.29} & -14.10^{+0.30}_{-0.44} & -14.59^{+0.19}_{-0.29} \\
    6.2 & 23.65^{+0.31}_{-0.45} & 23.15^{+0.20}_{-0.30} & 23.52^{+0.31}_{-0.45} & 23.03^{+0.20}_{-0.30} & -14.22^{+0.32}_{-0.46} & -14.72^{+0.21}_{-0.30} \\
    6.3 & 23.59^{+0.33}_{-0.47} & 23.07^{+0.21}_{-0.31} & 23.47^{+0.33}_{-0.47} & 22.95^{+0.21}_{-0.31} & -14.33^{+0.33}_{-0.48} & -14.85^{+0.22}_{-0.32} \\
    6.4 & 23.53^{+0.34}_{-0.50} & 23.00^{+0.22}_{-0.33} & 23.41^{+0.34}_{-0.50} & 22.88^{+0.22}_{-0.33} & -14.46^{+0.34}_{-0.50} & -14.99^{+0.23}_{-0.33} \\
    6.5 & 23.47^{+0.35}_{-0.51} & 22.91^{+0.24}_{-0.34} & 23.34^{+0.35}_{-0.51} & 22.79^{+0.24}_{-0.34} & -14.59^{+0.36}_{-0.52} & -15.14^{+0.25}_{-0.34} \\
    6.6 & 23.40^{+0.37}_{-0.53} & 22.83^{+0.26}_{-0.35} & 23.28^{+0.37}_{-0.53} & 22.71^{+0.26}_{-0.35} & -14.72^{+0.38}_{-0.54} & -15.29^{+0.27}_{-0.35} \\
    6.7 & 23.33^{+0.39}_{-0.55} & 22.76^{+0.27}_{-0.37} & 23.21^{+0.39}_{-0.55} & 22.64^{+0.27}_{-0.37} & -14.86^{+0.40}_{-0.55} & -15.44^{+0.27}_{-0.38} \\
    6.8 & 23.26^{+0.42}_{-0.56} & 22.68^{+0.28}_{-0.40} & 23.14^{+0.42}_{-0.56} & 22.56^{+0.28}_{-0.40} & -15.01^{+0.43}_{-0.56} & -15.59^{+0.28}_{-0.40} \\
    6.9 & 23.19^{+0.44}_{-0.58} & 22.60^{+0.29}_{-0.41} & 23.07^{+0.44}_{-0.58} & 22.48^{+0.29}_{-0.41} & -15.16^{+0.45}_{-0.58} & -15.75^{+0.30}_{-0.41} \\
    7.0 & 23.12^{+0.46}_{-0.59} & 22.52^{+0.31}_{-0.42} & 23.00^{+0.46}_{-0.59} & 22.40^{+0.31}_{-0.42} & -15.31^{+0.46}_{-0.60} & -15.91^{+0.31}_{-0.43} \\
    7.1 & 23.05^{+0.48}_{-0.61} & 22.45^{+0.32}_{-0.44} & 22.93^{+0.48}_{-0.61} & 22.32^{+0.32}_{-0.44} & -15.47^{+0.48}_{-0.61} & -16.07^{+0.32}_{-0.45} \\
    7.2 & 22.98^{+0.50}_{-0.62} & 22.37^{+0.33}_{-0.47} & 22.85^{+0.50}_{-0.62} & 22.24^{+0.33}_{-0.47} & -15.62^{+0.50}_{-0.62} & -16.24^{+0.33}_{-0.47} \\
    7.3 & 22.91^{+0.52}_{-0.64} & 22.29^{+0.34}_{-0.48} & 22.78^{+0.52}_{-0.64} & 22.17^{+0.34}_{-0.48} & -15.78^{+0.52}_{-0.64} & -16.40^{+0.34}_{-0.49} \\
    7.4 & 22.84^{+0.53}_{-0.66} & 22.21^{+0.35}_{-0.50} & 22.72^{+0.53}_{-0.66} & 22.09^{+0.35}_{-0.50} & -15.94^{+0.54}_{-0.66} & -16.57^{+0.35}_{-0.51} \\
    7.5 & 22.77^{+0.55}_{-0.68} & 22.13^{+0.36}_{-0.52} & 22.65^{+0.55}_{-0.68} & 22.01^{+0.36}_{-0.52} & -16.09^{+0.56}_{-0.68} & -16.73^{+0.37}_{-0.53} \\
    7.6 & 22.70^{+0.57}_{-0.69} & 22.05^{+0.38}_{-0.54} & 22.57^{+0.57}_{-0.69} & 21.93^{+0.38}_{-0.54} & -16.25^{+0.58}_{-0.70} & -16.90^{+0.38}_{-0.55} \\
    7.7 & 22.63^{+0.59}_{-0.71} & 21.97^{+0.39}_{-0.56} & 22.51^{+0.59}_{-0.71} & 21.85^{+0.39}_{-0.56} & -16.40^{+0.60}_{-0.72} & -17.07^{+0.40}_{-0.56} \\
    7.8 & 22.56^{+0.61}_{-0.73} & 21.89^{+0.41}_{-0.58} & 22.44^{+0.61}_{-0.73} & 21.76^{+0.41}_{-0.58} & -16.56^{+0.61}_{-0.74} & -17.23^{+0.41}_{-0.58} \\
    7.9 & 22.49^{+0.62}_{-0.76} & 21.80^{+0.43}_{-0.60} & 22.37^{+0.62}_{-0.76} & 21.68^{+0.43}_{-0.60} & -16.71^{+0.62}_{-0.76} & -17.40^{+0.43}_{-0.61} \\
    8.0 & 22.41^{+0.63}_{-0.77} & 21.72^{+0.44}_{-0.63} & 22.29^{+0.63}_{-0.77} & 21.60^{+0.44}_{-0.63} & -16.86^{+0.64}_{-0.78} & -17.56^{+0.44}_{-0.63} \\
    8.1 & 22.34^{+0.65}_{-0.80} & 21.63^{+0.45}_{-0.65} & 22.22^{+0.65}_{-0.80} & 21.51^{+0.45}_{-0.65} & -17.01^{+0.66}_{-0.80} & -17.72^{+0.46}_{-0.65} \\
    8.2 & 22.26^{+0.68}_{-0.81} & 21.55^{+0.47}_{-0.67} & 22.14^{+0.68}_{-0.81} & 21.43^{+0.47}_{-0.67} & -17.17^{+0.68}_{-0.82} & -17.88^{+0.47}_{-0.68} \\
    8.3 & 22.19^{+0.70}_{-0.83} & 21.46^{+0.48}_{-0.70} & 22.06^{+0.70}_{-0.83} & 21.34^{+0.48}_{-0.70} & -17.31^{+0.71}_{-0.84} & -18.04^{+0.49}_{-0.70} \\
    8.4 & 22.12^{+0.72}_{-0.86} & 21.38^{+0.50}_{-0.72} & 21.99^{+0.72}_{-0.86} & 21.25^{+0.50}_{-0.72} & -17.46^{+0.72}_{-0.86} & -18.20^{+0.51}_{-0.72} \\
    8.5 & 22.04^{+0.73}_{-0.87} & 21.29^{+0.52}_{-0.74} & 21.92^{+0.73}_{-0.87} & 21.16^{+0.52}_{-0.74} & -17.61^{+0.74}_{-0.88} & -18.36^{+0.53}_{-0.74} \\
    8.6 & 21.96^{+0.76}_{-0.89} & 21.20^{+0.54}_{-0.76} & 21.84^{+0.76}_{-0.89} & 21.07^{+0.54}_{-0.76} & -17.75^{+0.77}_{-0.89} & -18.52^{+0.54}_{-0.76} \\
    8.7 & 21.88^{+0.79}_{-0.90} & 21.11^{+0.55}_{-0.78} & 21.76^{+0.79}_{-0.90} & 20.99^{+0.55}_{-0.78} & -17.90^{+0.79}_{-0.90} & -18.67^{+0.56}_{-0.79} \\
    8.8 & 21.81^{+0.81}_{-0.91} & 21.03^{+0.57}_{-0.81} & 21.69^{+0.81}_{-0.91} & 20.91^{+0.57}_{-0.81} & -18.04^{+0.82}_{-0.92} & -18.82^{+0.57}_{-0.81} \\
    8.9 & 21.73^{+0.83}_{-0.93} & 20.95^{+0.58}_{-0.83} & 21.61^{+0.83}_{-0.93} & 20.82^{+0.58}_{-0.83} & -18.18^{+0.84}_{-0.94} & -18.97^{+0.59}_{-0.84} \\
    9.0 & 21.65^{+0.85}_{-0.94} & 20.86^{+0.60}_{-0.85} & 21.53^{+0.85}_{-0.94} & 20.74^{+0.60}_{-0.85} & -18.33^{+0.86}_{-0.95} & -19.12^{+0.60}_{-0.86} \\
    9.1 & 21.57^{+0.88}_{-0.95} & 20.77^{+0.61}_{-0.87} & 21.45^{+0.88}_{-0.95} & 20.65^{+0.61}_{-0.87} & -18.47^{+0.89}_{-0.96} & -19.28^{+0.62}_{-0.88} \\
    9.2 & 21.50^{+0.90}_{-0.97} & 20.68^{+0.63}_{-0.89} & 21.37^{+0.90}_{-0.97} & 20.56^{+0.63}_{-0.89} & -18.61^{+0.91}_{-0.98} & -19.43^{+0.63}_{-0.90} \\
    9.3 & 21.42^{+0.92}_{-0.99} & 20.59^{+0.64}_{-0.92} & 21.30^{+0.92}_{-0.99} & 20.47^{+0.64}_{-0.92} & -18.75^{+0.93}_{-1.00} & -19.58^{+0.65}_{-0.92} \\
    9.4 & 21.34^{+0.94}_{-1.01} & 20.50^{+0.66}_{-0.94} & 21.22^{+0.94}_{-1.01} & 20.38^{+0.65}_{-0.94} & -18.89^{+0.95}_{-1.02} & -19.72^{+0.66}_{-0.95} \\
    9.5 & 21.26^{+0.97}_{-1.03} & 20.42^{+0.67}_{-0.96} & 21.14^{+0.97}_{-1.03} & 20.29^{+0.67}_{-0.96} & -19.02^{+0.98}_{-1.04} & -19.87^{+0.68}_{-0.97} \\
    9.6 & 21.18^{+0.99}_{-1.05} & 20.33^{+0.68}_{-0.98} & 21.06^{+0.99}_{-1.05} & 20.20^{+0.68}_{-0.98} & -19.16^{+1.00}_{-1.05} & -20.02^{+0.69}_{-0.99} \\
    9.7 & 21.10^{+1.02}_{-1.06} & 20.24^{+0.70}_{-1.01} & 20.97^{+1.02}_{-1.06} & 20.12^{+0.70}_{-1.01} & -19.30^{+1.03}_{-1.07} & -20.16^{+0.70}_{-1.02} \\
    9.8 & 21.01^{+1.04}_{-1.08} & 20.15^{+0.71}_{-1.03} & 20.89^{+1.04}_{-1.08} & 20.03^{+0.71}_{-1.03} & -19.44^{+1.06}_{-1.09} & -20.31^{+0.71}_{-1.04} \\
    9.9 & 20.93^{+1.07}_{-1.10} & 20.06^{+0.72}_{-1.06} & 20.81^{+1.07}_{-1.10} & 19.94^{+0.72}_{-1.06} & -19.58^{+1.08}_{-1.11} & -20.45^{+0.73}_{-1.07} \\
    10.0 & 20.85^{+1.10}_{-1.12} & 19.97^{+0.73}_{-1.08} & 20.72^{+1.10}_{-1.12} & 19.85^{+0.73}_{-1.08} & -19.72^{+1.11}_{-1.12} & -20.60^{+0.74}_{-1.09} \\
    \hline
  \end{tabular}
\end{table*}

\begin{table*}
  \contcaption{}
  \begin{tabular}{d......}
    \hline
    \multicolumn{1}{c}{$z$} &    
    \multicolumn{1}{c}{$\log_{10}\epsilon_{1450}$} &
    \multicolumn{1}{c}{$\log_{10}\epsilon_{1450}$} &
    \multicolumn{1}{c}{$\log_{10}\epsilon_{912}$} &
    \multicolumn{1}{c}{$\log_{10}\epsilon_{912}$} & 
    \multicolumn{1}{c}{$\log_{10}\Gamma_\mathrm{HI}$} &
    \multicolumn{1}{c}{$\log_{10}\Gamma_\mathrm{HI}$} \\ 
    &
    \multicolumn{1}{c}{$(M_{1450}<-18)$} &
    \multicolumn{1}{c}{$(M_{1450}<-21)$} &
    \multicolumn{1}{c}{$(M_{1450}<-18)$} &
    \multicolumn{1}{c}{$(M_{1450}<-21)$} &
    \multicolumn{1}{c}{$(M_{1450}<-18)$} &
    \multicolumn{1}{c}{$(M_{1450}<-21)$} \\
    \hline
    10.1 & 20.76^{+1.12}_{-1.13} & 19.88^{+0.75}_{-1.10} & 20.64^{+1.12}_{-1.13} & 19.76^{+0.75}_{-1.10} & -19.86^{+1.14}_{-1.14} & -20.74^{+0.75}_{-1.11} \\
    10.2 & 20.68^{+1.15}_{-1.15} & 19.80^{+0.76}_{-1.13} & 20.56^{+1.15}_{-1.15} & 19.67^{+0.76}_{-1.13} & -20.00^{+1.16}_{-1.16} & -20.88^{+0.76}_{-1.14} \\
    10.3 & 20.60^{+1.18}_{-1.17} & 19.71^{+0.77}_{-1.15} & 20.47^{+1.18}_{-1.17} & 19.58^{+0.77}_{-1.15} & -20.13^{+1.19}_{-1.18} & -21.03^{+0.78}_{-1.16} \\
    10.4 & 20.52^{+1.20}_{-1.20} & 19.62^{+0.78}_{-1.17} & 20.39^{+1.20}_{-1.20} & 19.49^{+0.78}_{-1.17} & -20.26^{+1.21}_{-1.21} & -21.17^{+0.79}_{-1.18} \\
    10.5 & 20.43^{+1.22}_{-1.22} & 19.52^{+0.80}_{-1.19} & 20.31^{+1.22}_{-1.22} & 19.40^{+0.80}_{-1.19} & -20.40^{+1.23}_{-1.23} & -21.31^{+0.80}_{-1.20} \\
    10.6 & 20.36^{+1.24}_{-1.24} & 19.43^{+0.81}_{-1.21} & 20.23^{+1.24}_{-1.24} & 19.31^{+0.81}_{-1.21} & -20.53^{+1.26}_{-1.25} & -21.46^{+0.82}_{-1.22} \\
    10.7 & 20.27^{+1.27}_{-1.26} & 19.34^{+0.83}_{-1.23} & 20.15^{+1.27}_{-1.26} & 19.22^{+0.83}_{-1.23} & -20.66^{+1.28}_{-1.27} & -21.60^{+0.83}_{-1.24} \\
    10.8 & 20.19^{+1.30}_{-1.28} & 19.24^{+0.84}_{-1.25} & 20.06^{+1.30}_{-1.28} & 19.12^{+0.84}_{-1.25} & -20.80^{+1.31}_{-1.29} & -21.75^{+0.85}_{-1.26} \\
    10.9 & 20.10^{+1.33}_{-1.30} & 19.15^{+0.86}_{-1.27} & 19.98^{+1.33}_{-1.30} & 19.02^{+0.86}_{-1.27} & -20.93^{+1.34}_{-1.31} & -21.89^{+0.87}_{-1.28} \\
    11.0 & 20.01^{+1.36}_{-1.32} & 19.05^{+0.88}_{-1.29} & 19.89^{+1.36}_{-1.32} & 18.93^{+0.88}_{-1.29} & -21.07^{+1.37}_{-1.33} & -22.04^{+0.89}_{-1.31} \\
    11.1 & 19.93^{+1.39}_{-1.34} & 18.96^{+0.90}_{-1.32} & 19.80^{+1.39}_{-1.34} & 18.84^{+0.90}_{-1.32} & -21.21^{+1.40}_{-1.35} & -22.18^{+0.90}_{-1.33} \\
    11.2 & 19.84^{+1.41}_{-1.36} & 18.86^{+0.91}_{-1.34} & 19.72^{+1.41}_{-1.36} & 18.74^{+0.91}_{-1.34} & -21.34^{+1.43}_{-1.37} & -22.32^{+0.92}_{-1.35} \\
    11.3 & 19.75^{+1.44}_{-1.38} & 18.77^{+0.93}_{-1.36} & 19.63^{+1.44}_{-1.38} & 18.65^{+0.93}_{-1.36} & -21.47^{+1.46}_{-1.39} & -22.46^{+0.94}_{-1.37} \\
    11.4 & 19.67^{+1.47}_{-1.39} & 18.67^{+0.95}_{-1.39} & 19.54^{+1.47}_{-1.39} & 18.55^{+0.95}_{-1.39} & -21.61^{+1.49}_{-1.41} & -22.61^{+0.95}_{-1.40} \\
    11.5 & 19.58^{+1.50}_{-1.41} & 18.58^{+0.96}_{-1.41} & 19.46^{+1.50}_{-1.41} & 18.46^{+0.96}_{-1.41} & -21.74^{+1.51}_{-1.42} & -22.75^{+0.97}_{-1.42} \\
    11.6 & 19.49^{+1.52}_{-1.44} & 18.48^{+0.98}_{-1.43} & 19.37^{+1.52}_{-1.44} & 18.36^{+0.98}_{-1.43} & -21.88^{+1.54}_{-1.45} & -22.89^{+0.99}_{-1.44} \\
    11.7 & 19.41^{+1.55}_{-1.46} & 18.39^{+0.99}_{-1.45} & 19.28^{+1.55}_{-1.46} & 18.27^{+0.99}_{-1.45} & -22.01^{+1.57}_{-1.47} & -23.03^{+1.00}_{-1.46} \\
    11.8 & 19.32^{+1.58}_{-1.48} & 18.29^{+1.01}_{-1.47} & 19.20^{+1.58}_{-1.48} & 18.17^{+1.01}_{-1.47} & -22.14^{+1.59}_{-1.49} & -23.17^{+1.02}_{-1.48} \\
    11.9 & 19.23^{+1.60}_{-1.50} & 18.20^{+1.02}_{-1.49} & 19.11^{+1.60}_{-1.50} & 18.08^{+1.02}_{-1.49} & -22.27^{+1.62}_{-1.52} & -23.31^{+1.03}_{-1.50} \\
    12.0 & 19.15^{+1.63}_{-1.53} & 18.11^{+1.04}_{-1.51} & 19.02^{+1.63}_{-1.53} & 17.98^{+1.04}_{-1.51} & -22.40^{+1.65}_{-1.54} & -23.45^{+1.05}_{-1.52} \\
    12.1 & 19.06^{+1.66}_{-1.55} & 18.01^{+1.05}_{-1.53} & 18.94^{+1.66}_{-1.55} & 17.89^{+1.05}_{-1.53} & -22.54^{+1.67}_{-1.56} & -23.59^{+1.06}_{-1.55} \\
    12.2 & 18.97^{+1.68}_{-1.57} & 17.92^{+1.07}_{-1.56} & 18.85^{+1.68}_{-1.57} & 17.79^{+1.07}_{-1.56} & -22.67^{+1.70}_{-1.59} & -23.73^{+1.08}_{-1.57} \\
    12.3 & 18.89^{+1.71}_{-1.60} & 17.82^{+1.08}_{-1.58} & 18.76^{+1.71}_{-1.60} & 17.70^{+1.08}_{-1.58} & -22.80^{+1.73}_{-1.61} & -23.87^{+1.09}_{-1.60} \\
    12.4 & 18.80^{+1.74}_{-1.62} & 17.73^{+1.10}_{-1.61} & 18.68^{+1.74}_{-1.62} & 17.61^{+1.10}_{-1.61} & -22.93^{+1.76}_{-1.63} & -24.00^{+1.11}_{-1.63} \\
    12.5 & 18.71^{+1.76}_{-1.64} & 17.63^{+1.11}_{-1.64} & 18.59^{+1.76}_{-1.64} & 17.51^{+1.11}_{-1.64} & -23.06^{+1.79}_{-1.65} & -24.14^{+1.12}_{-1.66} \\
    12.6 & 18.62^{+1.80}_{-1.65} & 17.54^{+1.13}_{-1.67} & 18.49^{+1.80}_{-1.65} & 17.41^{+1.13}_{-1.67} & -23.20^{+1.82}_{-1.66} & -24.28^{+1.14}_{-1.68} \\
    12.7 & 18.52^{+1.83}_{-1.66} & 17.44^{+1.15}_{-1.69} & 18.40^{+1.83}_{-1.66} & 17.32^{+1.15}_{-1.69} & -23.34^{+1.85}_{-1.67} & -24.42^{+1.15}_{-1.71} \\
    12.8 & 18.43^{+1.87}_{-1.68} & 17.34^{+1.16}_{-1.72} & 18.31^{+1.87}_{-1.68} & 17.22^{+1.16}_{-1.72} & -23.47^{+1.88}_{-1.69} & -24.56^{+1.17}_{-1.74} \\
    12.9 & 18.34^{+1.89}_{-1.70} & 17.24^{+1.18}_{-1.75} & 18.22^{+1.89}_{-1.70} & 17.12^{+1.18}_{-1.75} & -23.60^{+1.91}_{-1.71} & -24.70^{+1.19}_{-1.77} \\
    13.0 & 18.25^{+1.92}_{-1.72} & 17.15^{+1.19}_{-1.78} & 18.13^{+1.92}_{-1.72} & 17.02^{+1.19}_{-1.78} & -23.73^{+1.94}_{-1.73} & -24.84^{+1.20}_{-1.80} \\
    13.1 & 18.17^{+1.94}_{-1.74} & 17.05^{+1.21}_{-1.81} & 18.04^{+1.94}_{-1.74} & 16.93^{+1.21}_{-1.81} & -23.86^{+1.96}_{-1.75} & -24.98^{+1.22}_{-1.82} \\
    13.2 & 18.08^{+1.97}_{-1.76} & 16.95^{+1.23}_{-1.84} & 17.96^{+1.97}_{-1.76} & 16.83^{+1.23}_{-1.84} & -23.99^{+1.99}_{-1.77} & -25.12^{+1.24}_{-1.85} \\
    13.3 & 17.99^{+2.00}_{-1.78} & 16.85^{+1.24}_{-1.87} & 17.86^{+2.00}_{-1.78} & 16.73^{+1.24}_{-1.87} & -24.12^{+2.02}_{-1.79} & -25.26^{+1.25}_{-1.88} \\
    13.4 & 17.89^{+2.03}_{-1.80} & 16.75^{+1.26}_{-1.89} & 17.77^{+2.03}_{-1.80} & 16.63^{+1.26}_{-1.89} & -24.25^{+2.05}_{-1.81} & -25.40^{+1.27}_{-1.90} \\
    13.5 & 17.80^{+2.06}_{-1.82} & 16.66^{+1.28}_{-1.91} & 17.68^{+2.06}_{-1.82} & 16.53^{+1.28}_{-1.91} & -24.38^{+2.08}_{-1.83} & -25.54^{+1.29}_{-1.93} \\
    13.6 & 17.71^{+2.09}_{-1.84} & 16.56^{+1.29}_{-1.94} & 17.59^{+2.09}_{-1.84} & 16.44^{+1.29}_{-1.94} & -24.51^{+2.11}_{-1.86} & -25.68^{+1.31}_{-1.95} \\
    13.7 & 17.63^{+2.11}_{-1.87} & 16.46^{+1.31}_{-1.96} & 17.50^{+2.11}_{-1.87} & 16.34^{+1.31}_{-1.96} & -24.64^{+2.13}_{-1.88} & -25.81^{+1.32}_{-1.97} \\
    13.8 & 17.54^{+2.14}_{-1.89} & 16.36^{+1.33}_{-1.98} & 17.42^{+2.14}_{-1.89} & 16.24^{+1.33}_{-1.98} & -24.77^{+2.16}_{-1.91} & -25.95^{+1.34}_{-1.99} \\
    13.9 & 17.45^{+2.16}_{-1.92} & 16.26^{+1.35}_{-2.00} & 17.33^{+2.16}_{-1.92} & 16.14^{+1.35}_{-2.00} & -24.89^{+2.18}_{-1.94} & -26.09^{+1.36}_{-2.02} \\
    14.0 & 17.37^{+2.19}_{-1.95} & 16.16^{+1.37}_{-2.03} & 17.24^{+2.19}_{-1.95} & 16.04^{+1.37}_{-2.03} & -25.02^{+2.20}_{-1.96} & -26.23^{+1.38}_{-2.04} \\
    14.1 & 17.29^{+2.21}_{-1.98} & 16.06^{+1.39}_{-2.05} & 17.16^{+2.21}_{-1.98} & 15.94^{+1.39}_{-2.05} & -25.14^{+2.22}_{-2.00} & -26.37^{+1.40}_{-2.06} \\
    14.2 & 17.21^{+2.22}_{-2.01} & 15.96^{+1.40}_{-2.07} & 17.08^{+2.22}_{-2.01} & 15.84^{+1.40}_{-2.07} & -25.26^{+2.24}_{-2.03} & -26.51^{+1.42}_{-2.09} \\
    14.3 & 17.12^{+2.24}_{-2.04} & 15.86^{+1.42}_{-2.10} & 17.00^{+2.24}_{-2.04} & 15.74^{+1.42}_{-2.10} & -25.38^{+2.26}_{-2.05} & -26.65^{+1.43}_{-2.11} \\
    14.4 & 17.04^{+2.27}_{-2.07} & 15.76^{+1.44}_{-2.13} & 16.91^{+2.27}_{-2.07} & 15.64^{+1.44}_{-2.13} & -25.51^{+2.28}_{-2.08} & -26.79^{+1.45}_{-2.14} \\
    14.5 & 16.95^{+2.29}_{-2.09} & 15.66^{+1.46}_{-2.16} & 16.83^{+2.29}_{-2.09} & 15.54^{+1.46}_{-2.16} & -25.64^{+2.31}_{-2.11} & -26.93^{+1.47}_{-2.17} \\
    14.6 & 16.87^{+2.32}_{-2.12} & 15.56^{+1.48}_{-2.18} & 16.74^{+2.32}_{-2.12} & 15.44^{+1.48}_{-2.18} & -25.77^{+2.33}_{-2.13} & -27.07^{+1.49}_{-2.19} \\
    14.7 & 16.78^{+2.35}_{-2.14} & 15.46^{+1.50}_{-2.21} & 16.65^{+2.35}_{-2.14} & 15.34^{+1.50}_{-2.21} & -25.90^{+2.36}_{-2.15} & -27.22^{+1.50}_{-2.22} \\
    14.8 & 16.68^{+2.38}_{-2.16} & 15.36^{+1.51}_{-2.23} & 16.56^{+2.38}_{-2.16} & 15.24^{+1.51}_{-2.23} & -26.05^{+2.39}_{-2.17} & -27.37^{+1.52}_{-2.24} \\
    14.9 & 16.59^{+2.41}_{-2.18} & 15.26^{+1.53}_{-2.26} & 16.47^{+2.41}_{-2.18} & 15.14^{+1.53}_{-2.26} & -26.21^{+2.42}_{-2.19} & -27.54^{+1.53}_{-2.26} \\
    15.0 & 16.50^{+2.45}_{-2.20} & 15.16^{+1.55}_{-2.29} & 16.37^{+2.45}_{-2.20} & 15.04^{+1.55}_{-2.29} & -26.44^{+2.45}_{-2.20} & -27.78^{+1.55}_{-2.29} \\
    \hline
  \end{tabular}
\end{table*}

\section{Code and data}
\label{sec:code}

We make the code and data used in this work publicly available at
\url{https://github.com/gkulkarni/QLF}.  This includes homogenised AGN
catalogues and selection functions and the code used for developing
and analysing luminosity function models.  The evolution of the
luminosity function can be derived using any of the three global
models that we presented in Section~\ref{sec:global}.  However, as we
discussed in Section~\ref{sec:global}, none of these models should be
extrapolated to higher redshifts.  Alternatively, our binned
luminosity function estimates from Section~\ref{sec:bins} can be used.
For applications that require extrapolation to higher redshifts, such
as calculations of photoionization fluxes and emissivities, the fits
presented in Equations~(\ref{eqn:e912_21}) and (\ref{eqn:e912_18}) can
be used, or a similar approach of fitting functional forms to
quantities suitably derived from the binned luminosity functions
should be adopted.

\bibliographystyle{mnras}
\bibliography{refs}

\bsp
\label{lastpage}
\end{document}